\documentclass[%
 aip,
 amsmath,amssymb,
 reprint,%
]{revtex4-1}

\usepackage{graphicx}
\usepackage{dcolumn}
\usepackage{multirow}
\usepackage{bm}

\usepackage[utf8]{inputenc}
\usepackage[T1]{fontenc}
\usepackage{mathptmx}
\usepackage{etoolbox}
\usepackage{url}
\usepackage[hidelinks]{hyperref}
\urldef\faboURL\url{https://github.com/LXYan2333/fabo}

\usepackage{tikz}
\usetikzlibrary{positioning, arrows.meta, shapes.geometric}

\usepackage{algorithm}
\makeatletter
\def\ftype@algorithm{8}
\makeatother
\usepackage{algpseudocode}
\algrenewcommand\algorithmicrequire{\textbf{Input:}}
\algrenewcommand\algorithmicensure{\textbf{Output:}}

\usepackage[acronym]{glossaries-extra}
\glsdisablehyper
\setabbreviationstyle[acronym]{long-short}
\newacronym[
  longplural={Gaussian-type orbitals},
  shortplural={GTOs}
]{gto}{GTO}{Gaussian-type orbital}

\newacronym[
  longplural={electron-repulsion integrals},
  shortplural={ERIs}
]{eri}{ERI}{electron-repulsion integral}

\newacronym[
  longplural={graphics processing units},
  shortplural={GPUs}
]{gpu}{GPU}{graphics processing unit}

\newacronym{simd}{SIMD}{single-instruction, multiple-data}

\newacronym{ilp}{ILP}{instruction-level parallelism}

\newacronym{ooo}{OoO}{out-of-order}

\newacronym{fma}{FMA}{fused multiply-add}

\newacronym{flop}{FLOP}{floating-point operation}

\newacronym{dag}{DAG}{directed acyclic graph}

\newacronym{avx512}{AVX-512}{Intel Advanced Vector Extensions 512}

\newacronym[
  longplural={central processing units},
  shortplural={CPUs}
]{cpu}{CPU}{central processing unit}

\newacronym{ieee754}{IEEE 754}{Institute of Electrical and Electronics Engineers Standard for Floating-Point Arithmetic}

\newacronym{gmp}{GMP}{GNU Multiple Precision Arithmetic Library}

\newacronym{fabo}{\texttt{FABO}}{\texttt{Fast and Accurate BOys}}

\newacronym{gds}{GDS}{Gather Data Sampling}

\newacronym{scf}{SCF}{self-consistent field}

\newacronym{rhf}{RHF}{restricted Hartree--Fock}

\newacronym{vsr}{VSR}{Vikhamar-Sandberg--Repisky}

\newacronym[longplural={units in the last place}]{ulp}{ULP}{unit in the last place}

\newglossaryentry{mpfr}{name={MPFR},description={a multiple-precision floating-point library supporting directed rounding}}

\newacronym{cuda}{CUDA}{Compute Unified Device Architecture}

\newacronym{mkl}{MKL}{Math Kernel Library}

\newacronym{sm}{SM}{streaming multiprocessor}

\makeatletter
\def\@email#1#2{%
 \endgroup
 \patchcmd{\titleblock@produce}
  {\frontmatter@RRAPformat}
  {\frontmatter@RRAPformat{\produce@RRAP{*#1\href{mailto:#2}{#2}}}\frontmatter@RRAPformat}
  {}{}
}%
\makeatother
\begin{document}

\preprint{AIP/123-QED}

\title{Fast and Accurate Boys Function Evaluation Using Nonuniform Gridded Taylor Expansions}
\author{Longxiang Yan}
\author{Wei Wu}
\author{Peifeng Su}
\email{supi@xmu.edu.cn}
\affiliation{The State Key Laboratory of Physical Chemistry of Solid Surfaces, Fujian Provincial Key Laboratory of Theoretical and Computational Chemistry, and College of Chemistry and Chemical Engineering, Xiamen University, Xiamen, Fujian 361005, China}
\date[]{}

\begin{abstract}
Efficient Boys function evaluation is important for Gaussian integral calculations on both \glspl{cpu} and \glspl{gpu}. Gridded Taylor expansion provides a simple evaluation procedure, but a uniformly spaced reference table requires substantial storage over a wide argument range. We address this limitation using a nonuniform grid whose spacing increases with the argument. For large arguments, we introduce a normalized auxiliary function whose scaling and derivative properties allow it to be evaluated using the same Taylor procedure as the main-domain Boys functions. The main-domain and large-argument reference tables are constructed jointly, with polynomial degrees and grid parameters selected under accuracy and storage constraints. Low-part corrections to the reference values are included to improve the accuracy of the Taylor evaluation. The resulting method is implemented in the \gls{fabo} library and evaluated in QUICK and NWChem--Simint. \gls{fabo} gives the smallest maximum and mean errors among the tested implementations for all three reported error measures, with the largest sampled error being approximately 1.067 \glspl{ulp}. In application benchmarks, \gls{fabo} reduces the Fock-build time relative to the original Boys evaluators in both applications and is competitive with the other tested implementations.
\end{abstract}
\maketitle

{\small\noindent\textbf{License.} Except where otherwise noted, this article is licensed under the \href{https://creativecommons.org/licenses/by-nc/4.0/}{Creative Commons Attribution-NonCommercial 4.0 International License} (CC BY-NC 4.0).\par}

\glsresetall


\section{Introduction}
\label{sec:introduction}

The evaluation of molecular integrals over \glspl{gto} is a fundamental component of electronic structure calculations. In particular, \glspl{eri} are repeatedly evaluated in Hartree--Fock and density functional theory calculations. Efficient integral evaluation is therefore important for calculations on both \glspl{cpu} and \glspl{gpu}.\cite{Pritchard2016Simint,Cruzeiro2021AMBERQUICK}

The Boys functions,
\begin{equation}
F_n(x)=\int_0^1 t^{2n}e^{-xt^2}\,\mathrm{d}t,
\label{eq:boys-definition}
\end{equation}
are central to many algorithms for evaluating Gaussian integrals.\cite{McMurchie1978Integrals,HeadGordon1988Integrals} They provide the fundamental quantities from which higher-angular-momentum integrals are generated through recurrence relations. Their arguments and requested orders depend on the target molecular geometry, basis set, and integral evaluation algorithm. A Boys function evaluator must therefore handle a range of arguments and output orders efficiently while providing sufficient accuracy for the subsequent integral calculation.

A variety of approaches have been developed for Boys function evaluation. These include series expansions, polynomial or rational approximations, interpolation from tabulated values, and recurrence relations between different orders. The choice of algorithm involves a balance among approximation accuracy, floating-point work, and storage requirements. Recent work has also considered how these choices affect vectorized and \gls{gpu} execution. For example, Vikhamar-Sandberg and Repisky use rational minimax approximations together with order recurrences, while Laikov develops a global approximation for vectorized computation.\cite{VikhamarSandberg2026Boys,Laikov2025Boys} Other approaches use tabulated values and local polynomial evaluation, often combined with different formulas for other argument ranges.\cite{Tsuji2025BoysGPU,Asadchev2025LibintX}

Among these approaches, gridded Taylor evaluation has long been used
in molecular integral calculations\cite{McMurchie1978Integrals}
and remains attractive. It has a simple computational structure and can be evaluated primarily with \glspl{fma}, which are efficient on modern processors. Within the interpolation region, the same polynomial evaluation procedure can be used for different arguments without argument-dependent branches, and the Taylor truncation error can be easily analyzed using the remainder.

Despite these advantages, extending a uniformly spaced reference table over a wide argument range requires substantial storage. Existing implementations therefore often restrict gridded Taylor evaluation to part of the argument range and use recurrence relations or asymptotic formulas for larger arguments.\cite{Tsuji2025BoysGPU,Ito2026GANSU} These different regions require different evaluation procedures, and the accuracy of the recurrence-generated values can deteriorate as rounding errors accumulate. When arguments from different regions occur within the same \gls{gpu} warp or \gls{cpu} \gls{simd} vector, the different procedures can also increase the work required to process them, which leads to worse performance.

To address these limitations, we develop a nonuniform gridded Taylor method for Boys-function evaluation and implement it in a new open-source library called \gls{fabo}.\footnote{\faboURL} In the main region, we use a nonuniform grid whose spacing increases with the argument, reducing the reference storage required over a wide range. For large arguments, a normalized reference table is constructed using the scaling identity of an auxiliary function. The main and large-argument tables share the same Taylor evaluation procedure, so that inputs from both regions can follow the same computational structure. Low-part corrections to the reference values associated with the requested output orders improve the numerical accuracy of the evaluation. The table parameters and polynomial degrees are selected jointly, prioritizing reduced polynomial work before reducing storage under the prescribed accuracy and storage requirements. Together, these designs provide an accurate Boys function evaluator with compact reference tables and a shared evaluation procedure over a wide argument range.

This paper is structured as follows. Section~\ref{sec:numerical-formulation} presents the nonuniform gridded Taylor method and its numerical target. Section~\ref{sec:wide-range} describes the normalized large-argument representation and the joint construction of the reference tables. Section~\ref{sec:implementation} discusses the implementation in integral kernels. Section~\ref{sec:experimental-methodology} describes the evaluation protocols, and Sec.~\ref{sec:results-discussion} presents the accuracy and performance results. Finally, Sec.~\ref{sec:conclusions} summarizes the conclusions.

\section{Nonuniform Gridded Taylor Evaluation}
\label{sec:numerical-formulation}

\subsection{Taylor Representation and Numerical Target}
\label{subsec:gridded-taylor}

Gaussian integral calculations commonly require a sequence of Boys function values $F_0(x),F_1(x),\ldots,F_N(x)$ for the same argument $x$. We evaluate these values using Taylor expansions around a set of reference points. For each argument, a reference point $r_i$ is selected from the grid, and each requested Boys function is approximated by a polynomial in the displacement from $r_i$. The choice of grid affects both the polynomial degrees required for a given accuracy and the amount of reference data.

The coefficients of these polynomials can be obtained directly from Boys function values. Differentiating Eq.~\eqref{eq:boys-definition} gives
\begin{equation}
\frac{\mathrm{d}^{j}F_n(x)}{\mathrm{d}x^{j}}
=
(-1)^jF_{n+j}(x),
\qquad j=0,1,\ldots,
\label{eq:boys-derivative}
\end{equation}
so the derivatives of $F_n$ at $r_i$ are given by higher-order Boys functions at the same point. To allow for calculations that require only part of the sequence, we consider a general interval of requested orders $a\leq n\leq b$. Defining $h=r_i-x$, we write
\begin{equation}
F_n(x)=T_{n,p_n}(x;r_i)+R_{n,p_n}(x;r_i),
\label{eq:taylor-decomposition}
\end{equation}
where $R_{n,p_n}$ is the truncation remainder.
The Taylor polynomial of degree $p_n$ for order $n$ is
\begin{equation}
T_{n,p_n}(x;r_i)
=\sum_{j=0}^{p_n}\frac{F_{n+j}(r_i)}{j!}h^j.
\label{eq:gridded-taylor}
\end{equation}
The choice $h=r_i-x$ incorporates the alternating signs
in Eq.~\eqref{eq:boys-derivative}.

For a requested output range $[a,b]$, we refer to the complete set of numerical data and parameters used to evaluate the corresponding Boys functions as a \emph{numerical model}. A model comprises the main-domain and normalized large-argument reference tables, their grid parameters and switching boundary, the polynomial degree $p_n$ assigned to each requested order, and the associated low-part corrections. Each model is constructed independently for its requested output range and is used by the evaluator to compute $F_a(x),\ldots,F_b(x)$ over the supported argument range.

We use an error of one \gls{ulp} as the numerical target for constructing the reference grid and selecting the polynomial degrees. The \gls{ulp} measures the spacing between adjacent representable floating-point numbers at a given magnitude and therefore expresses the error relative to the available floating-point resolution.\cite{Goldberg1991FloatingPoint} Let $\widehat F_n(x)$ denote the computed binary64 result and let $\operatorname{RN}_{64}$ denote rounding to binary64 using round-to-nearest, ties-to-even.\cite{IEEE7542019} The target is expressed as
\begin{equation}
\frac{\left|\widehat F_n(x)-F_n(x)\right|}
{\operatorname{ulp}\!\left(\operatorname{RN}_{64}(F_n(x))\right)}
\leq 1,
\qquad a\leq n\leq b.
\label{eq:boys-accuracy-target}
\end{equation}
This target is less restrictive than correct rounding, which would require $\widehat F_n(x)=\operatorname{RN}_{64}(F_n(x))$. 

The error in Eq.~\eqref{eq:boys-accuracy-target} includes more than the Taylor remainder. It also reflects the representation of the reference data and rounding during floating-point evaluation. In the large-argument region introduced in Sec.~\ref{sec:wide-range}, it additionally includes the error from approximating $F_n$ by the scaled auxiliary function. These contributions combine in the computed result, so a small truncation error alone does not ensure that the final error remains below one \gls{ulp}.

Testing every binary64 input for every candidate grid and degree combination would make the parameter search impractical. We therefore evaluate the complete floating-point results at a selected set of qualification inputs. These inputs sample the reference cells, including regions where the mathematical approximation errors are expected to be large. A candidate model is accepted only if it satisfies Eq.~\eqref{eq:boys-accuracy-target} at the prescribed qualification inputs. The construction of these inputs and the numerical qualification procedure are described in the supplementary material (Sec.~S2.3).

Because qualification uses a finite set of inputs, passing these tests does not guarantee that Eq.~\eqref{eq:boys-accuracy-target} holds for every binary64 input. In particular, floating-point rounding errors need not be largest where the mathematical approximation error is largest. This behavior is observed in the accuracy results in Sec.~\ref{subsec:function-accuracy-results}.

\subsection{Mathematical Basis for Nonuniform Reference Spacing}
\label{subsec:nonuniform-spacing}

Gridded Taylor evaluation offers a simple computational structure: once the reference values are available, the local polynomial can be evaluated using multiplications and \glspl{fma}, without evaluating an exponential, error function, or reciprocal square root at the input argument. For a fixed polynomial degree, selecting a different reference point changes the coefficients but the polynomial evaluation sequence remains unchanged. The Taylor remainder also provides a direct connection between reference spacing and approximation accuracy. These properties make gridded Taylor evaluation attractive, provided that the reference data can be stored and accessed economically.

A natural implementation uses equally spaced reference points, $r_i=i\Delta$. With a fixed set of reference orders, covering $0\leq x\leq X$ then requires approximately $X/\Delta$ reference rows. A spacing fine enough to meet the accuracy target near the origin is consequently retained throughout the table, and extending the supported range increases storage linearly. One way to limit this storage is to use gridded Taylor evaluation only over part of the argument range and apply other methods elsewhere. For example, Tsuji et al. use uniformly spaced reference values for small arguments, together with recurrence and large-argument methods for the remaining regions.\cite{Tsuji2025BoysGPU} Here, we instead examine whether the reference spacing can increase with $x$, allowing gridded Taylor evaluation to cover a wider range without a proportional increase in storage.

The Boys functions provide a mathematical basis for such a grid. For normal-valued results, a fixed \gls{ulp} target corresponds to a relative-error scale that varies by at most a factor of two with the significand.\cite{Goldberg1991FloatingPoint} We therefore examine the Taylor remainder relative to $F_n(x)$. The required derivative bounds follow directly from the integral definition. For $y>0$ and a nonnegative integer $k$, integrating the derivative of $t^{2k+1}e^{-yt^2}$ gives
\begin{equation}
\begin{aligned}
e^{-y}
&=\int_0^1
\frac{\mathrm{d}}{\mathrm{d}t}
\left(t^{2k+1}e^{-yt^2}\right)\,\mathrm{d}t\\
&=(2k+1)F_k(y)-2yF_{k+1}(y).
\end{aligned}
\label{eq:boys-grid-integration-identity}
\end{equation}
Since the Boys functions and the exponential are positive, rearrangement yields
\begin{equation}
\frac{F_{k+1}(y)}{F_k(y)}
\leq\frac{k+\tfrac12}{y}.
\label{eq:boys-adjacent-ratio-bound}
\end{equation}
Applying this inequality successively at orders $n,n+1,\ldots,n+m-1$ bounds the higher derivatives relative to the function itself:
\begin{equation}
\frac{F_{n+m}(y)}{F_n(y)}
\leq
\frac{1}{y^m}
\prod_{j=0}^{m-1}\left(n+j+\frac12\right),
\qquad m\geq1.
\label{eq:boys-relative-derivative-bound}
\end{equation}
Each additional derivative thus introduces a factor bounded by a constant divided by the argument.

To apply this observation to the Taylor approximation, set $m=p_n+1$, $d=|x-r_i|$, and $y=\min(x,r_i)>0$. Taylor's theorem places the remainder derivative at a point between $x$ and $r_i$. Because $F_{n+m}$ decreases with its argument,
\begin{equation}
\left|R_{n,p_n}(x;r_i)\right|
\leq
\frac{d^m}{m!}F_{n+m}(y).
\label{eq:taylor-remainder-bound}
\end{equation}
The bound in Eq.~\eqref{eq:boys-relative-derivative-bound} compares this derivative with $F_n(y)$, whereas the desired error scale is $F_n(x)$. To express this bound relative to $F_n(x)$, we also need to bound the ratio $F_n(y)/F_n(x)$. Using $F_n'=-F_{n+1}$ and integrating Eq.~\eqref{eq:boys-adjacent-ratio-bound} gives
\begin{equation}
\ln\frac{F_n(y)}{F_n(x)}
=
\int_y^x\frac{F_{n+1}(v)}{F_n(v)}\,\mathrm{d}v
\leq
\left(n+\frac12\right)\ln\frac{x}{y}.
\label{eq:boys-local-ratio-bound}
\end{equation}
Combining these results and writing $\rho=d/y$, with $x/y\leq1+\rho$, yields
\begin{equation}
\frac{\left|R_{n,p_n}(x;r_i)\right|}{F_n(x)}
\leq
(1+\rho)^{n+\frac12}
\frac{\rho^m}{m!}
\prod_{j=0}^{m-1}\left(n+j+\frac12\right).
\label{eq:relative-taylor-remainder}
\end{equation}

For fixed order and degree, the right-hand side of Eq.~\eqref{eq:relative-taylor-remainder} depends on the relative displacement $\rho$. Doubling both $x$ and $r_i$, for example, doubles their separation but leaves this bound unchanged. Away from the origin, reference intervals can therefore widen approximately in proportion to $x$ while retaining the same truncation-error bound. Near zero, the relative-displacement description is unsuitable, but the truncation error can instead be bounded in terms of the absolute displacement $d$. Since
\begin{equation}
F_{n+m}(y)\leq F_{n+m}(0)=\frac{1}{2n+2m+1},
\end{equation}
Eq.~\eqref{eq:taylor-remainder-bound} gives
\begin{equation}
|R_{n,p_n}(x;r_i)|\leq \frac{d^m}{m!(2n+2m+1)}.
\end{equation}
Because $F_n(x)$ remains bounded away from zero in a neighborhood of the origin, a sufficiently small absolute grid spacing also controls the relative truncation error there. Together with the relative-displacement bound away from the origin, this motivates a grid with fine, finite spacing near zero and progressively wider intervals at larger arguments.

Such a grid changes how storage grows with the supported range. Once the spacing is proportional to the argument, each doubling of the upper limit requires approximately the same number of additional reference points. The reference count therefore grows logarithmically with the upper limit in this region, rather than linearly as for a uniform grid with fixed spacing. This provides a route to extending gridded Taylor evaluation over a broad domain without retaining the near-origin resolution throughout the table.

\subsection{An Expanding Grid from a Floating-Point Offset}
\label{subsec:expanding-grid}

The expanding spacing reduces the number of reference points, but an efficient evaluator must also locate the appropriate reference row inexpensively for every input. A search through stored reference coordinates would introduce additional work into each Boys evaluation. We therefore seek a grid that combines the spacing pattern motivated above with a reference index obtainable directly from the input representation.

Binary floating-point numbers provide such a structure: their spacing is uniform within each power-of-two interval and doubles in the next interval.\cite{IEEE7542019} A positive normal value has the form $2^\ell(1+f)$, where $\ell$ is an integer exponent and $0\leq f<1$. Retaining $b_g$ fraction bits gives $Q=2^{b_g}$ lattice points in each interval $[2^\ell,2^{\ell+1})$:
\begin{equation}
z_{\ell,j}
=
2^\ell\left(1+\frac{j}{Q}\right),
\qquad j=0,\ldots,Q-1.
\label{eq:shifted-floating-point-lattice}
\end{equation}
Their spacing is
\begin{equation}
\Delta_\ell=2^{\ell-b_g}.
\label{eq:expanding-grid-spacing}
\end{equation}
The exponent and retained fraction bits therefore describe both the expanding grid and the position of a point within it.

Applying this lattice directly to $x$ would, however, retain progressively finer spacing as $x$ approaches zero. We avoid that unnecessary refinement by introducing a positive offset $s$ and applying the lattice to $x+s$. Choose $s$ to be a lattice point and translate the retained points $z_i\geq s$ back to the original coordinate:
\begin{equation}
r_i^\star=z_i-s.
\label{eq:offset-reference-grid}
\end{equation}
The point $z_i=s$ then gives a reference at the origin. Near zero, $x+s$ lies in the same power-of-two interval as $s$, producing a finite uniform spacing. Farther away, the spacing doubles whenever $x+s$ crosses the next power of two. For illustration, $s=1$ and $b_g=2$ give spacings of $1/4$ on $[0,1]$, $1/2$ on $[1,3]$, and $1$ on $[3,7]$. The offset sets the near-origin scale, while the number of retained fraction bits controls the grid density.

The same representation makes the grid inexpensive to index. At runtime, the shifted coordinate is formed by
\begin{equation}
z=\operatorname{RN}_{64}(x+s).
\label{eq:additive-grid-index-coordinate}
\end{equation}
Integer rounding and shifts of its binary encoding select a nearby lattice point and identify the corresponding reference row. The selected lattice value $z_i$ is recovered from those bits, and the reference coordinate is
\begin{equation}
r_i=\operatorname{RN}_{64}(z_i-s).
\label{eq:rounded-offset-reference}
\end{equation}
Reference-point selection therefore requires neither a search over stored coordinates nor a logarithm or floating-point scaling multiplication: the indexing coordinate is obtained with one floating-point addition.

With $Q$ intervals per complete power-of-two range of the shifted coordinate, this construction realizes the approximately logarithmic reference-count growth described in Sec.~\ref{subsec:nonuniform-spacing}, while allowing the row to be selected without a search over reference coordinates.

The numerical construction uses the actual floating-point mapping rather than idealized cell boundaries. Reference coefficients are generated at the reconstructed coordinates $r_i$, and the grid and Taylor degrees are accepted using the final-output criterion in Eq.~\eqref{eq:boys-accuracy-target}. The detailed index construction and qualification procedure are provided in the supplementary material (Secs.~S1.2 and~S2.3).

Fig.~\ref{fig:nonuniform-grid-remainder} compares the Taylor truncation errors for $F_0$ using the nonuniform grid constructed for \gls{fabo} evaluations that request only $F_0$ and a uniform grid with the same number of reference points and polynomial degree. At each interior boundary, the larger error of the two neighboring Taylor polynomials is shown. With uniform spacing, the relative remainder is much smaller than one \gls{ulp} over much of the large-argument region, but exceeds one \gls{ulp} near the origin. Thus, under the same storage budget, reference points are used in regions where the resulting truncation error is already well below the target, while the spacing remains too coarse where a finer grid is needed. The nonuniform grid redistributes the reference points toward the small-argument region and permits the spacing to increase with $x$. As a result, the remainder at every tested cell boundary remains below one \gls{ulp}, and the largest relative remainder is reduced from $3.12\times10^{-15}$ to $6.48\times10^{-17}$. This comparison concerns the mathematical Taylor remainder and excludes rounding of the reference coefficients and runtime arithmetic. Grid parameters and computational details are provided in the supplementary material (Sec.~S3.1).

\begin{figure}[htbp]
\centering
\includegraphics[width=\linewidth]{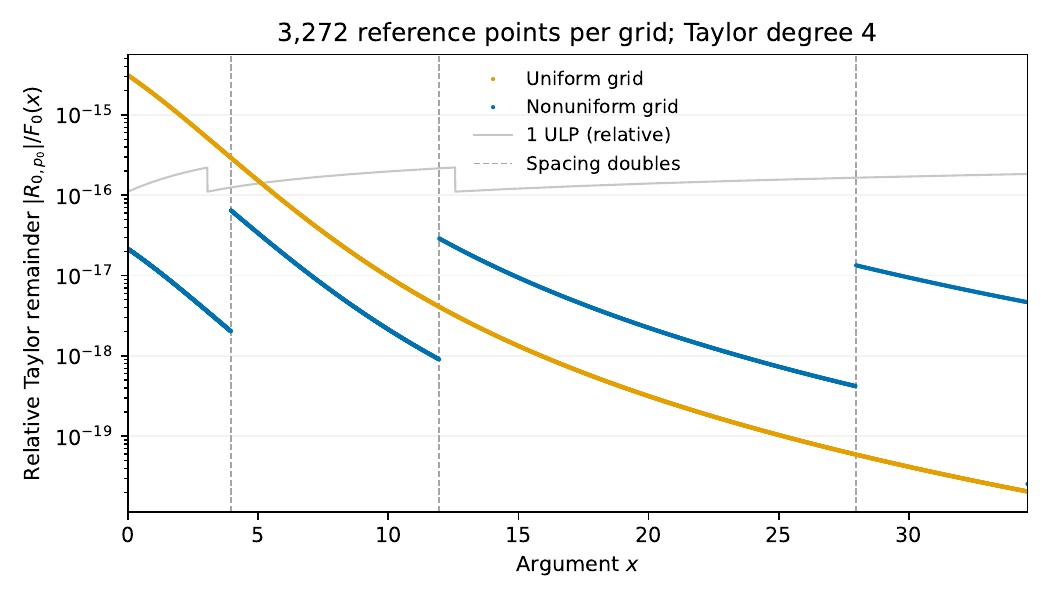}
\caption{Relative Taylor truncation errors for $F_0$ at reference-cell boundaries using uniform and nonuniform grids with the same number of reference points and polynomial degree.}
\label{fig:nonuniform-grid-remainder}
\end{figure}

\subsection{Reference Tables and Polynomial Evaluation}
\label{subsec:reference-tables}

The Taylor expansions of different output orders require overlapping sets of reference values. For example, $F_{n+1}(r_i)$ is both the constant term in the expansion of $F_{n+1}(x)$ and the coefficient of $h$ in the expansion of $F_n(x)$. We therefore store a single sequence of Boys values at each reference point, rather than a separate set of factorial-weighted coefficients for every output. The factorial factors are incorporated during polynomial evaluation, allowing the same stored value to serve different positions in different expansions.

To avoid evaluating unnecessary terms, the polynomial degree $p_n$ is selected separately for each output order according to the numerical acceptance criterion. For a given order, that degree is fixed across the reference grid, so selecting a different reference point changes the coefficients but not the polynomial evaluation sequence. For the requested interval $[a,b]$, the highest reference order needed is
\begin{equation}
H=\max_{a\leq n\leq b}(n+p_n).
\label{eq:highest-reference-order}
\end{equation}
Each reference row consequently stores
\begin{equation}
G_{i,k}=\operatorname{RN}_{64}\!\left(F_k(r_i)\right),
\qquad k=a,\ldots,H.
\label{eq:rounded-reference-value}
\end{equation}
The orders above $b$ supply Taylor coefficients rather than additional outputs. With $N_r$ reference points, the reference values alone occupy $8N_r(H-a+1)$ bytes. Their reuse across output orders avoids duplicating the overlapping coefficient sequences.

The factorial factors in Eq.~\eqref{eq:gridded-taylor} can be applied without forming powers of $h$ or storing a separate factorial-scaled table. In exact arithmetic, a factorial form of Horner's rule gives, for $p_n\geq1$,
\begin{equation}
T_{n,p_n}(x;r_i)=F_n(r_i)+h\,v_{n,1}
\label{eq:factorial-horner}
\end{equation}
where the intermediate values are defined by
\begin{subequations}
\label{eq:factorial-horner-intermediates}
\begin{align}
v_{n,p_n}
&=F_{n+p_n}(r_i),
\label{eq:factorial-horner-init}\\
v_{n,j}
&=F_{n+j}(r_i)+\frac{h}{j+1}v_{n,j+1},
\qquad j=p_n-1,\ldots,1.
\label{eq:factorial-horner-recursion}
\end{align}
\end{subequations}
For example, the cubic polynomial is written as
\begin{widetext}
\begin{equation}
T_{n,3}(x;r_i)
=
F_n(r_i)+h\left[
F_{n+1}(r_i)+\frac{h}{2}\left(
F_{n+2}(r_i)+\frac{h}{3}F_{n+3}(r_i)
\right)\right].
\label{eq:factorial-horner-example}
\end{equation}
\end{widetext}
The successive factors $h$, $h/2$, and $h/3$ generate the required factorial denominators while preserving the use of unscaled reference values.

In the floating-point evaluator, the reference values in the nonconstant Horner chain are replaced by $G_{i,k}$, and each multiply-add stage is evaluated with an \gls{fma}. Factors corresponding to $h/j$ are formed by multiplying the computed displacement by precomputed binary64 reciprocals of the positive integers $j$, rather than by runtime division. Since these factors depend on the displacement and Taylor position but not on the output order, they can be shared by the requested polynomials. Each output nevertheless has its own Horner accumulator and is evaluated from its reference coefficients, without propagating another computed output through an order recurrence at $x$.

The final combination of the constant coefficient and the nonconstant contribution is kept separate from the Horner chain. Although $G_{i,n}$ is correctly rounded, its difference from the true value $F_n(r_i)$ enters this combination without a displacement factor. That error can remain important even when the Taylor truncation error is small. The next subsection introduces a compact low-part correction that retains additional information about the constant coefficient in the final output assembly.

\subsection{Low-Part Corrections for Taylor Evaluation}
\label{subsec:output-low-parts}

Rounding the reference values to binary64 discards information that can affect the final Taylor result. For the constant coefficient, this discarded part is
\begin{equation}
\rho_{i,n}=F_n(r_i)-G_{i,n},
\label{eq:reference-low-part}
\end{equation}
where $G_{i,n}$ is the correctly rounded reference value defined in Eq.~\eqref{eq:rounded-reference-value}. The representation error in $F_{n+j}(r_i)$ is suppressed by the factor $h^j/j!$, whereas the constant-coefficient residual is not suppressed by any power of the displacement $h$. Moving the input closer to the reference point therefore does not suppress this contribution. We retain additional information about the constant coefficient and incorporate it into the final output assembly.

The correction is combined with the nonconstant Taylor contribution inside a \gls{fma}, before adding the high part of the constant coefficient. This ordering avoids first adding a small residual to $G_{i,n}$, which would simply round away the information we intend to preserve. Storing the residual as another binary64 value would require eight additional bytes per output and reference point. Instead, we use a one-byte approximation, concentrating the extra precision on the coefficient whose error is not attenuated by the displacement.

To reserve floating-point execution resources for the Taylor polynomials, we also seek to decode this byte without integer-to-floating-point conversion or floating-point scaling. A natural approach is to place the stored bits directly into a floating-point representation, deriving its scale from the exponent of $G_{i,n}$. However, a normal binary floating-point number has a significand of the form $1.f$, with an implicit leading one.\cite{IEEE7542019} The residual $\rho_{i,n}$ can have either sign, vanish, or be arbitrarily small. It therefore cannot be represented simply by filling the fraction bits of a normal number at a fixed scale.

We accommodate the implicit leading one by changing the division between the high part and its correction. For a positive normal reference value, let $u_{i,n}=\operatorname{ulp}(G_{i,n})$. We define the adjusted high part as the next representable binary64 value above $G_{i,n}$:
\begin{equation}
G_{i,n}^{+}=\operatorname{nextUp}(G_{i,n})=G_{i,n}+u_{i,n}.
\label{eq:next-up-reference}
\end{equation}
The corresponding correction is defined as the difference between the adjusted high part and the exact reference value:
\begin{equation}
\Delta_{i,n}=G_{i,n}^{+}-F_n(r_i)=u_{i,n}-\rho_{i,n}.
\label{eq:next-up-correction}
\end{equation}
Correct rounding gives $|\rho_{i,n}|\leq u_{i,n}/2$, and hence
\begin{equation}
\frac12u_{i,n}\leq\Delta_{i,n}\leq\frac32u_{i,n}.
\label{eq:next-up-correction-range}
\end{equation}
The signed residual near zero has thus been replaced by a positive correction confined to a narrow range. Its nonzero leading significand bit is now useful rather than an obstacle. Subtracting this correction compensates for the one-\gls{ulp} increase in the high part.

The byte stores one exponent bit and seven fraction bits. Writing its unsigned code as $c=128e_c+f_c$, with $e_c\in\{0,1\}$ and $0\leq f_c<128$, the represented correction is
\begin{equation}
\widehat{\Delta}_{i,n}
=
u_{i,n}\,2^{e_c-1}
\left(1+\frac{f_c}{128}\right).
\label{eq:output-correction-encoding}
\end{equation}
Codes from 0 through 192 cover the interval in Eq.~\eqref{eq:next-up-correction-range}. In particular, code 128 represents $u_{i,n}$ and therefore handles an exactly zero original residual without requiring a zero-valued decoded correction. During table construction, $\Delta_{i,n}$ is rounded to the nearest encoded value, with ties resolved to even, giving
\begin{equation}
\left|\widehat{\Delta}_{i,n}-\Delta_{i,n}\right|
\leq\frac{u_{i,n}}{256}.
\label{eq:output-correction-quantization}
\end{equation}
The arithmetic in Eq.~\eqref{eq:output-correction-encoding} describes the represented value, not the runtime decoder: the binary64 encoding is assembled from the original exponent of $G_{i,n}$ and the correction byte using integer and bitwise operations. The next larger high part is likewise obtained by incrementing the encoding of the positive reference value.

Let $\widehat v_{n,1}$ denote the nonconstant Horner accumulator from Eq.~\eqref{eq:factorial-horner}, evaluated using the original reference coefficients. We write $\operatorname{FMA}_{64}(a,b,c)=\operatorname{RN}_{64}(ab+c)$ to denote the \gls{fma} defined by the \gls{ieee754}, where the product and sum are evaluated without intermediate rounding.\cite{IEEE7542019} The corrected output is
\begin{equation}
\widehat F_n(x)
=
\operatorname{RN}_{64}\!\left[
G_{i,n}^{+}
+
\operatorname{FMA}_{64}
\!\left(h,\widehat v_{n,1},-\widehat{\Delta}_{i,n}\right)
\right].
\label{eq:taylor-low-part-correction}
\end{equation}
This retains low-part information through the final fused accumulation while requiring only one additional byte per output and reference point, with no floating-point arithmetic to construct the decoded correction. The bit-level decoder and construction checks ensuring a finite shifted high part and a normal decoded correction are given in the supplementary material (Sec.~S1.3).

\section{Large-Argument Evaluation Using a Scaled Asymptotic Function}
\label{sec:wide-range}

Although the expanding grid introduced in the preceding section reduces the growth of reference storage, extending gridded Taylor evaluation to all finite, nonnegative binary64 arguments would still require a substantial table of reference values. Such a table may fit in memory but can exceed the cache budget available for efficient evaluation within integral kernels. We therefore introduce a scaled asymptotic function that can be tabulated over a fixed normalized interval and evaluated using the same Taylor procedure as the original Boys functions.

\subsection{Limitations of Conventional Large-Argument Evaluation}
\label{subsec:large-argument-limitations}

Conventional Boys implementations commonly use asymptotic approximations and recurrence relations for sufficiently large arguments.\cite{Asadchev2025LibintX,Tsuji2025BoysGPU,VikhamarSandberg2026Boys} The asymptotic approximation can be derived directly from the integral definition. For $n=0$, substituting $y=\sqrt{x}\,t$ into Eq.~\eqref{eq:boys-definition} gives
\begin{equation}
\begin{aligned}
F_0(x)
&=\int_0^1 e^{-xt^2}\,\mathrm{d}t\\
&=\frac{1}{\sqrt{x}}\int_0^{\sqrt{x}}e^{-y^2}\,\mathrm{d}y\\
&=\frac{\sqrt{\pi}}{2\sqrt{x}}
  \operatorname{erf}(\sqrt{x}).
\end{aligned}
\label{eq:boys-zero-erf}
\end{equation}
Since $\operatorname{erf}(\sqrt{x})$ approaches unity as $x$ increases, the leading asymptotic approximation is
\begin{equation}
F_0(x)\approx\frac{\sqrt{\pi}}{2\sqrt{x}}.
\label{eq:asymptotic-f0}
\end{equation}

Higher orders can be obtained through the upward recurrence. Rearranging Eq.~\eqref{eq:boys-grid-integration-identity} gives
\begin{equation}
F_{n+1}(x)
=
\frac{(2n+1)F_n(x)-e^{-x}}{2x},
\qquad x>0.
\label{eq:upward-recurrence}
\end{equation}
Although this recurrence provides an inexpensive way to generate successive orders, its numerical accuracy can deteriorate as the recurrence proceeds. This is particularly important when the required accuracy is substantially tighter than the typical error tolerated by conventional applications.

To assess the error introduced by upward recurrence, we start from correctly rounded binary64 values of $F_0(x)$ and $e^{-x}$. For each argument, we compute $t=\operatorname{RN}_{64}(2x)$, $\rho=\operatorname{RN}_{64}(1/t)$, and $q=\operatorname{RN}_{64}(e^{-x}\rho)$. The recurrence is then evaluated as
\begin{equation}
\widehat F_n=\operatorname{FMA}_{64}(a_n,\widehat F_{n-1},-q),
\end{equation}
where $a_n=\operatorname{RN}_{64}((2n-1)\rho)$ and $\widehat F_0=\operatorname{RN}_{64}(F_0(x))$. The rounded exponential and reciprocal are reused at every step.

For each requested maximum order $N=1,\ldots,12$ and tolerance of 1, 2, 4, 8, or 16 \glspl{ulp}, we scan the arguments $x_i=\operatorname{RN}_{64}(i/10^6)$, $i=1,\ldots,10^9$, and identify the first window of one million consecutive samples in which every computed value $\widehat F_1,\ldots,\widehat F_N$ meets the tolerance. The reported boundary is the first argument in that window. Errors are measured against high-precision mathematical reference values, using the binary64 spacing at the correctly rounded reference to define the \gls{ulp}. As discussed in Sec.~\ref{subsec:gridded-taylor}, satisfying an error tolerance at sampled inputs does not guarantee that it is met at other inputs.

\begin{figure}[tbp]
\centering
\includegraphics[width=\columnwidth]{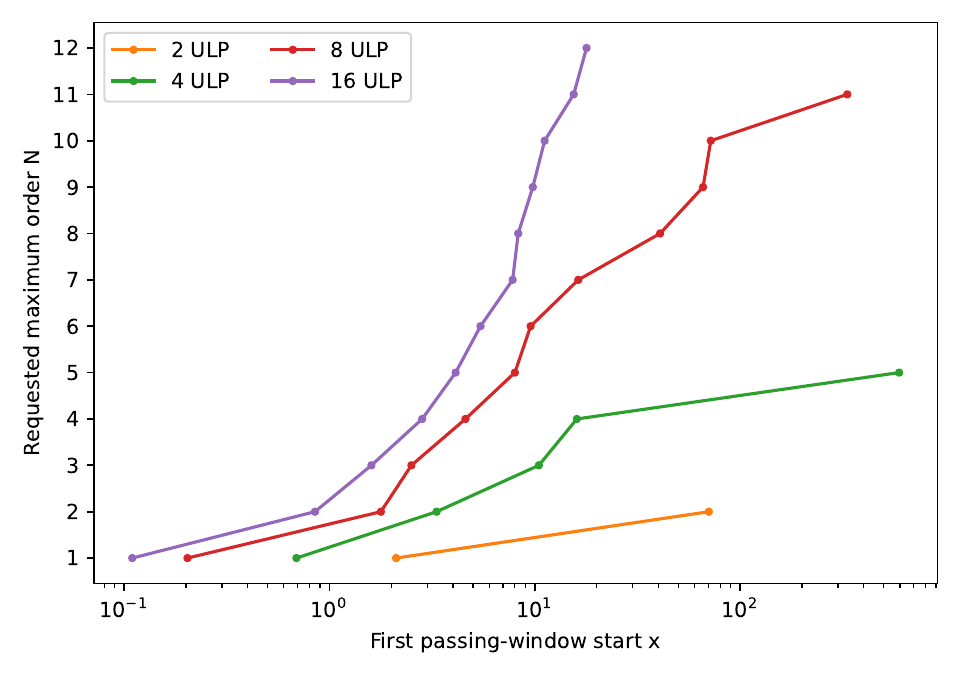}
\caption{First passing-window start for binary64 upward recurrence at different requested maximum orders and error tolerances. No such window was found for the 1-\gls{ulp} tolerance.}
\label{fig:upward-recurrence-limits}
\end{figure}

Fig.~\ref{fig:upward-recurrence-limits} shows that the first passing window generally shifts to larger arguments as the requested maximum order increases. Tightening the error tolerance has a similar effect, and for several combinations of order and tolerance, no complete passing window is found within the already large scanned range. In particular, none of the tested order ranges produces a passing window at the one-\gls{ulp} tolerance, even with correctly rounded starting values. These results illustrate the difficulty of maintaining high accuracy with upward recurrence, especially when several orders are required. 

Besides the numerical limitation, conventional hybrid evaluators generally employ different computational procedures in the small- and large-argument regions. When arguments from both regions occur within the same \gls{gpu} warp, the corresponding branches may execute under different thread masks, increasing the amount of serialized work. Different lanes of a \gls{cpu} \gls{simd} vector can encounter the same problem. The two regions also require different coefficient-access and evaluation procedures, making it difficult to share the main computational work.

These observations motivate an alternative to conventional asymptotic recurrence: a compact large-argument representation that retains the local Taylor evaluation structure of the original Boys functions.

\subsection{Large-Argument Scaling and Normalization}
\label{subsec:large-argument-model}

The asymptotic approximation in Eq.~\eqref{eq:asymptotic-f0} can be understood as extending the upper integration limit of the Boys integral from unity to infinity. We define the corresponding family of functions as
\begin{equation}
A_n(x)=\int_0^\infty t^{2n}e^{-xt^2}\,\mathrm{d}t,
\qquad x>0.
\label{eq:asymptotic-model-definition}
\end{equation}
For $n=0$, the Gaussian integral gives
\begin{equation}
A_0(x)=\frac{\sqrt{\pi}}{2\sqrt{x}},
\label{eq:asymptotic-model-zero}
\end{equation}
which is precisely the leading asymptotic approximation of $F_0(x)$.

The higher orders follow from integration by parts. Since
\begin{equation}
\frac{\mathrm{d}}{\mathrm{d}t}
\left(t^{2n+1}e^{-xt^2}\right)
=
(2n+1)t^{2n}e^{-xt^2}
-
2xt^{2n+2}e^{-xt^2},
\end{equation}
integrating over $[0,\infty)$ and using the vanishing boundary terms yields
\begin{equation}
A_{n+1}(x)
=
\frac{2n+1}{2x}A_n(x).
\label{eq:large-argument-recurrence}
\end{equation}
Thus, the entire asymptotic family can be generated from Eq.~\eqref{eq:asymptotic-model-zero} without evaluating an error function or an exponential. By comparing Eqs.~\eqref{eq:boys-definition} and~\eqref{eq:asymptotic-model-definition}, the difference between the model and the original Boys function is exactly
\begin{equation}
A_n(x)-F_n(x)
=
\int_1^\infty t^{2n}e^{-xt^2}\,\mathrm{d}t.
\label{eq:asymptotic-model-error}
\end{equation}
The model therefore overestimates the Boys function, with a discrepancy determined by the additional tail introduced by extending the upper integration limit to infinity. Whether this approximation is sufficiently accurate depends on both the argument and the requested order, and is considered in the numerical qualification of the complete evaluator.

An important property of this model is its scaling under a fourfold change in the argument. From Eq.~\eqref{eq:asymptotic-model-zero},
\begin{equation}
A_0(4x)=\frac{1}{2}A_0(x).
\end{equation}
Applying the recurrence in Eq.~\eqref{eq:large-argument-recurrence} successively gives
\begin{equation}
A_n(4x)=2^{-(2n+1)}A_n(x).
\label{eq:large-argument-fourfold-scaling}
\end{equation}
The scaling factor is therefore an integer power of two for every order. For example, increasing the argument by a factor of four halves the zero-order value and divides the first-order value by eight.

We exploit this identity to map large arguments onto a fixed interval. Any $x\geq1$ can be uniquely expressed as
\begin{equation}
x=4^{k_x}u
\label{eq:large-argument-normalization}
\end{equation}
where $k_x\in\mathbb{Z}_{\geq0}$ and $1\leq u<4$. Repeated application of Eq.~\eqref{eq:large-argument-fourfold-scaling} then gives
\begin{equation}
A_n(x)
=
2^{-(2n+1)k_x}A_n(u).
\label{eq:large-argument-scaling}
\end{equation}

This normalization separates the argument magnitude from the function evaluation. Regardless of how large the original input becomes, the reference grid for the asymptotic model only needs to cover $[1,4)$. Moreover, for a normal binary64 input, $k_x$ and $u$ can be obtained directly from its exponent and significand: each increment of two in the binary exponent corresponds to one increment of $k_x$. No expensive logarithm or floating-point division is required.

The large-argument model thus provides a compact representation over an arbitrarily wide input range. More importantly, as shown next, its derivative structure is identical to that of the original Boys functions, allowing both domains to use the same Taylor evaluation procedure.

\subsection{A Shared Taylor Evaluation Path}
\label{subsec:shared-taylor-path}

The normalized large-argument representation can be evaluated using the same Taylor representation as the original Boys functions. To see this, differentiating Eq.~\eqref{eq:asymptotic-model-definition} under the integral sign gives:
\begin{equation}
\frac{\mathrm{d}A_n(x)}{\mathrm{d}x}
=
-\int_0^\infty t^{2n+2}e^{-xt^2}\,\mathrm{d}t
=
-A_{n+1}(x).
\label{eq:asymptotic-model-derivative}
\end{equation}
Repeated differentiation therefore yields
\begin{equation}
\frac{\mathrm{d}^{j}A_n(x)}{\mathrm{d}x^{j}}
=
(-1)^j A_{n+j}(x),
\qquad j=0,1,\ldots.
\label{eq:asymptotic-model-derivatives}
\end{equation}
This is identical to the derivative relation of the original Boys functions in Eq.~\eqref{eq:boys-derivative}. Consequently, both function families have the same Taylor expansion structure.

Let $r_i$ denote a reference point selected from the main-domain grid, and let $u_i$ denote a reference point from the normalized large-argument grid. For an input $x$ in the main domain, define the displacement $h=r_i-x$. For a large-argument input, first obtain its normalized coordinate $u$ using Eq.~\eqref{eq:large-argument-normalization}, and define $h=u_i-u$. The corresponding Taylor polynomials are
\begin{subequations}
\label{eq:shared-domain-taylor}
\begin{align}
T_{n,p_n}^{(F)}(x;r_i)
&=
\sum_{j=0}^{p_n}
\frac{F_{n+j}(r_i)}{j!}h^j,
\label{eq:shared-domain-taylor-main}\\
T_{n,p_n}^{(A)}(u;u_i)
&=
\sum_{j=0}^{p_n}
\frac{A_{n+j}(u_i)}{j!}h^j.
\label{eq:shared-domain-taylor-tail}
\end{align}
\end{subequations}
The two polynomials differ in their reference coefficients and displacements, but have identical algebraic forms. In particular, both can be evaluated using the factorial Horner scheme and the low-part correction introduced in Secs.~\ref{subsec:reference-tables} and~\ref{subsec:output-low-parts}.

Mathematically, the two domains do not need to use the same polynomial degree. We impose this constraint deliberately: for each output order $n$, a common degree $p_n$ is used in both domains. Different output orders may still use different degrees. This ensures that the two evaluations require the same number of polynomial stages and the same sequence of coefficient accesses for each output order, allowing them to share the polynomial evaluation procedure without domain-dependent changes to its arithmetic.

The common degree vector also determines the reference orders required by both domains. For a requested output interval $[a,b]$, the highest required reference order is
\begin{equation}
H=\max_{a\leq n\leq b}(n+p_n).
\label{eq:shared-highest-reference-order}
\end{equation}
Each main-domain reference row therefore stores the sequence of rounded values $F_k(r_i)$ for $a\leq k\leq H$, while each normalized large-argument row stores the corresponding sequence $A_k(u_i)$. Both use the same logical row layout, with the reference coefficients and output low-part corrections occupying corresponding positions. The numerical values are domain-specific, but their interpretation and use during evaluation are identical.

The resulting evaluation separates domain-dependent preparation from the common polynomial calculation. The main-domain input selects its reference row and displacement directly from $x$. A large-argument input is first normalized, after which its reference row and displacement are obtained from $u$; the associated integer scaling exponent $k_x$ is retained. Once the reference row and displacement have been prepared, both domains follow the same coefficient-loading, correction-decoding, and Horner evaluation sequence. The main-domain polynomial directly approximates $F_n(x)$, whereas the normalized polynomial approximates $A_n(u)$. For the latter, the corresponding Boys value is obtained using Eq.~\eqref{eq:large-argument-scaling} after polynomial evaluation.

This construction allows two different representations of the Boys functions to share the same Taylor computation. The core runtime evaluation procedure is summarized in Alg.~\ref{alg:fabo-evaluation}. $x_{\mathrm{switching}}$ denotes the model-dependent switching boundary between the main domain and the normalized large-argument domain, and $C_{i,m}$ denotes the ordinary rounded reference value $\operatorname{RN}_{64}(F_m(r_i))$ in the main domain or $\operatorname{RN}_{64}(A_m(u_i))$ in the normalized large-argument domain.

The remaining problem is to construct compatible reference grids and a common degree vector that satisfy the numerical target while respecting a shared storage budget. We address this joint construction and its numerical qualification next.

\begin{algorithm}[tbp]
\caption{Evaluation of $F_0(x),\ldots,F_N(x)$.}
\label{alg:fabo-evaluation}
\begin{algorithmic}[1]
\Require Finite binary64 argument $x\geq 0$ (including $-0$), maximum output order $N$
\Ensure $\widehat F_0(x),\ldots,\widehat F_N(x)$

\If{$x \leq x_{\mathrm{switching}}$}
    \Comment{Main-domain evaluation}
    \State $z \gets \operatorname{RN}_{64}(x+s)$
    \State $i \gets \Call{MainGridIndex}{z}$
    \State $r_i \gets \Call{MainReferencePoint}{z,i}$
    \State $h \gets \operatorname{RN}_{64}(r_i-x)$
\Else
    \Comment{Normalized large-argument evaluation}
    \State $(k_x,u) \gets \Call{Normalize}{x}$
    \State $i \gets \Call{TailGridIndex}{u}$
    \State $u_i \gets \Call{TailReferencePoint}{i}$
    \State $h \gets \operatorname{RN}_{64}(u_i-u)$
\EndIf

\For{$n=0,\ldots,N$}
    \State $v \gets C_{i,n+p_n}$
    \For{$j=p_n-1,\ldots,1$}
        \State $\alpha_j \gets \operatorname{RN}_{64}(h\times \operatorname{RN}_{64}(1/(j+1)))$
        \State $v \gets \operatorname{FMA}_{64}(\alpha_j,v,C_{i,n+j})$
    \EndFor
    \State $G^+_{i,n} \gets \operatorname{nextUp}(C_{i,n})$
    \State Decode $\widehat{\Delta}_{i,n}$ using Eq.~\eqref{eq:output-correction-encoding}
    \State $q_n \gets \operatorname{FMA}_{64}(h,v,-\widehat{\Delta}_{i,n})$
    \State $y_n \gets \operatorname{RN}_{64}(G^+_{i,n}+q_n)$
    \If{$x > x_{\mathrm{switching}}$}
        \State $\widehat F_n(x) \gets
        \Call{ScaleByPowerOfTwo}{y_n,-(2n+1)k_x}$
    \Else
        \State $\widehat F_n(x) \gets y_n$
    \EndIf
\EndFor

\State \Return $\widehat F_0(x),\ldots,\widehat F_N(x)$
\end{algorithmic}
\end{algorithm}

\subsection{Joint Reference-Table Construction and Numerical Qualification}
\label{subsec:joint-reference-tables}

The main-domain and normalized large-argument tables are constructed jointly because they share the same Taylor degrees and reference-row layout. Their grid parameters can differ, but the choice of polynomial degrees affects the accuracy and storage requirements of both tables. We therefore search for a compatible pair of tables under a combined storage budget.

The search prioritizes reducing the Taylor degrees. For a candidate large-argument grid, a common degree ceiling is considered for the requested output orders. Numerical qualification of the large-argument representation determines a switching boundary, which also determines the extent of the main domain.

After allocating storage for the large-argument table, the remaining budget is available for the main table. Its grid density is controlled by the offset $s$ in Eq.~\eqref{eq:additive-grid-index-coordinate}: for a fixed number of retained significand bits, increasing $s$ generally makes the grid over the input interval coarser and reduces the number of reference points. We therefore seek an offset as small as the remaining budget permits, using as much of the available storage as possible to obtain a fine main grid. This grid reduces the Taylor displacement and can allow lower polynomial degrees.

With the two grids established, the degree for each output order is reduced where numerical qualification permits. The same degree is retained for each order in both argument regions, preserving the shared Taylor evaluation procedure. After the degrees have been selected, we increase $s$ to reduce the main-table footprint while retaining those degrees and satisfying the numerical qualification. In this way, the search first uses the available storage to reduce polynomial work, and then reduces storage without introducing additional polynomial stages.

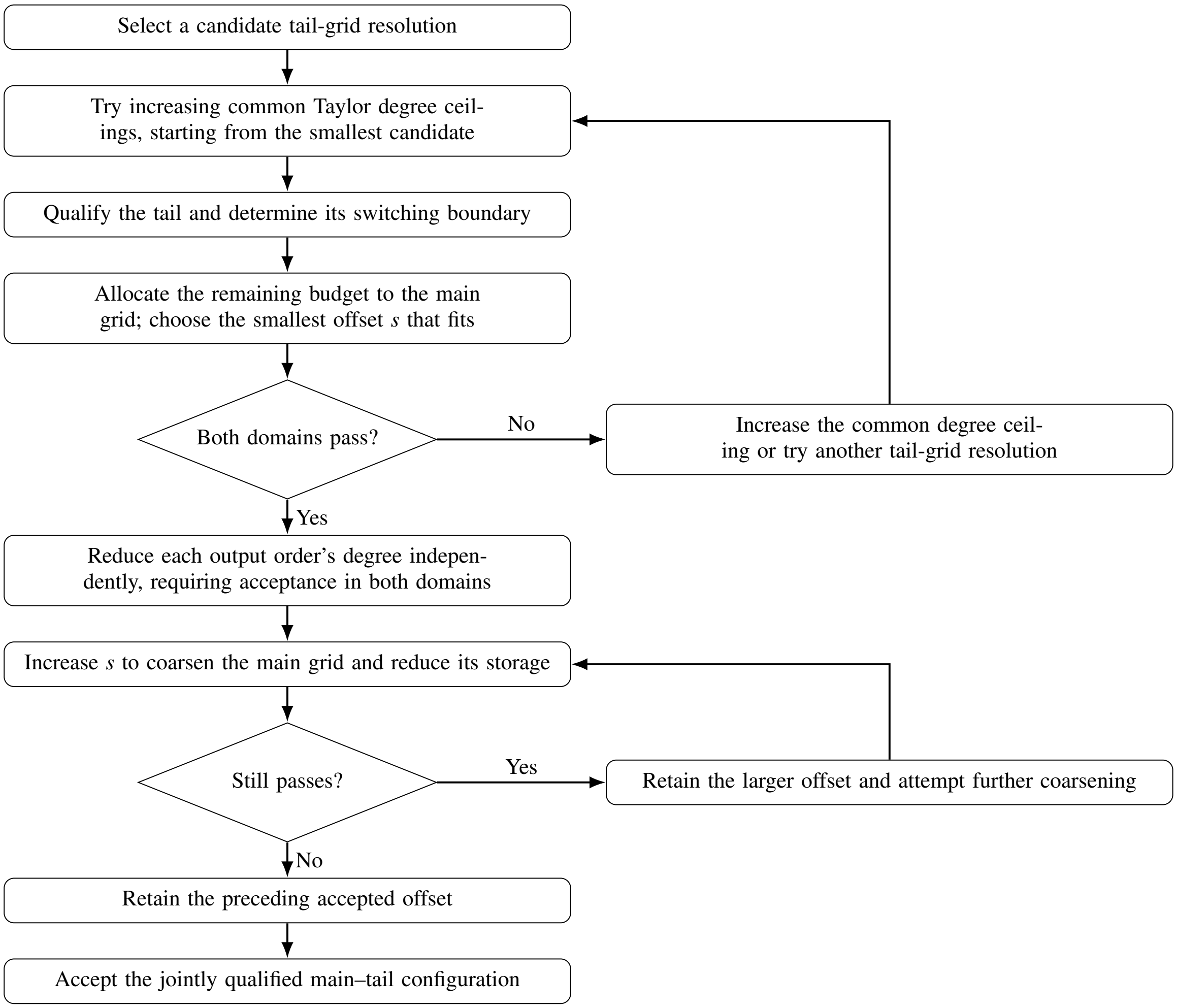
\begin{figure*}[t]
\centering
\resizebox{0.85\textwidth}{!}{%
\begin{tikzpicture}[
  node distance=5mm and 12mm,
  every node/.style={font=\small},
  box/.style={
    draw,
    rounded corners,
    align=center,
    inner sep=5pt,
    text width=0.43\textwidth
  },
  decision/.style={
    draw,
    diamond,
    aspect=2.5,
    align=center,
    inner sep=2pt,
    text width=0.17\textwidth
  },
  line/.style={-Latex, thick}
]

\node[box] (start) {
  Select a candidate tail-grid resolution
};

\node[box, below=of start] (degree) {
  Try increasing common Taylor degree ceilings,
  starting from the smallest candidate
};

\node[box, below=of degree] (tail) {
  Qualify the tail and determine its
  switching boundary
};

\node[box, below=of tail] (main) {
  Allocate the remaining budget to the main grid;
  choose the smallest offset $s$ that fits
};

\node[decision, below=of main] (common) {
  Both domains pass?
};

\node[box, right=of common, xshift=12mm] (increase) {
  Increase the common degree ceiling
  or try another tail-grid resolution
};

\node[box, below=of common] (individual) {
  Reduce each output order's degree
  independently, requiring acceptance
  in both domains
};

\node[box, below=of individual] (offset) {
  Increase $s$ to coarsen the main grid
  and reduce its storage
};

\node[decision, below=of offset] (shrink) {
  Still passes?
};

\node[box, right=of shrink, xshift=12mm] (continue) {
  Retain the larger offset and
  attempt further coarsening
};

\node[box, below=of shrink] (retain) {
  Retain the preceding accepted offset
};

\node[box, below=of retain] (finish) {
  Accept the jointly qualified
  main--tail configuration
};

\draw[line] (start) -- (degree);
\draw[line] (degree) -- (tail);
\draw[line] (tail) -- (main);
\draw[line] (main) -- (common);

\draw[line] (common.east) -- node[above] {No} (increase.west);
\draw[line] (increase.north) |- (degree.east);

\draw[line] (common.south) -- node[right] {Yes} (individual);
\draw[line] (individual) -- (offset);
\draw[line] (offset) -- (shrink);

\draw[line] (shrink.east) -- node[above] {Yes} (continue.west);
\draw[line] (continue.north) |- (offset.east);

\draw[line] (shrink.south) -- node[right] {No} (retain);
\draw[line] (retain) -- (finish);

\end{tikzpicture}%
}
\caption{Illustrative workflow for the joint construction of the main-domain and normalized large-argument reference tables.}
\label{fig:paired-search-workflow}
\end{figure*}

Fig.~\ref{fig:paired-search-workflow} illustrates the main stages of the joint construction. Both tables store reference coefficients and output low-part corrections in the same logical row layout. Their combined storage, including row padding, is subject to the 1 MiB budget. Numerical qualification is based on the complete computed Boys values at selected inputs and accounts for the combined effects of approximation, reference-data representation, floating-point evaluation, and large-argument scaling. The detailed parameter search, reference construction, qualification procedure, and selected model parameters are described in the supplementary material (Sec.~S2 and Table~S1).

\section{Implementation in Integral Kernels}
\label{sec:implementation}

We implemented the proposed method in \gls{fabo}, with \gls{cuda} and \gls{cpu} \gls{simd} backends. To assess its application-level Fock-build performance in established quantum-chemistry applications, we integrated the library into QUICK for \gls{gpu} calculations and into NWChem through its Simint integral engine for \gls{cpu} calculations.\cite{Cruzeiro2021AMBERQUICK,nwchem,Pritchard2016Simint} These integrations replace Boys evaluation within the selected integral paths while retaining the surrounding screening, integral recurrences, contraction, and Fock accumulation. The implementation therefore follows the applications' existing request patterns rather than requiring arguments to be collected into a separate Boys-evaluation batch.

\subsection{GPU Thread Mapping and Order Specialization}
\label{subsec:gpu-thread-mapping}

Within QUICK's two-electron integral kernels, each participating thread supplies an argument $x$ and requests the prefix $F_0(x),\ldots,F_N(x)$ for its subsequent integral calculation. We therefore evaluate one argument and its requested outputs within the same thread. The evaluator is inlined into the integral kernel and writes directly into QUICK's existing order-strided workspace.

Although the requested endpoint $N$ can vary between threads, the integral call site and build configuration establish a common upper bound. For example, ordinary SP and SPD requests have ceilings of 4 and 8, respectively, whereas the SPD and SPDF paths in an \texttt{ENABLEF} build can require order 12. We use the appropriate bound to select the numerical model at compile time, fixing the reference-table parameters, coefficient offsets, and Taylor degrees. The actual endpoint remains a per-thread runtime input, so selecting a model that supports higher orders does not require every call to evaluate them.

To combine this specialization with request-dependent work, the generated evaluator contains an unrolled sequence of output stages in ascending order. Each stage has a fixed polynomial degree and fixed reference-data offsets. After writing its result, the stage returns from the Boys evaluator if the requested endpoint has been reached; otherwise, execution continues to the next stage. Compile-time restrictions on the requested interval remove excluded stages, while the runtime endpoint determines where each participating thread stops.

Early termination should also avoid data accesses that serve only unrequested outputs. Preloading the entire reference row would retain these accesses even when later polynomial stages are skipped. Conversely, loading each polynomial's coefficients independently would repeatedly fetch their overlapping reference values. We therefore schedule each coefficient load at its first use and retain the value for subsequent polynomials. The coefficients are represented by named scalar variables, so their accesses and reuse are explicit in the generated stages rather than expressed through a dynamically indexed per-thread array.

For example, the selected model for outputs $F_0$ through $F_4$ uses degree four for both $F_0$ and $F_1$. The first stage requires reference orders 0 through 4, whereas the second introduces only order 5. A request ending at $F_1$ returns before the later stages and their additional coefficient loads. More generally, the loading schedule follows the union of the coefficient ranges needed by the outputs reached so far; it does not assume that advancing one output always requires exactly one new coefficient.

Termination is local to each thread. Threads requesting higher orders continue through the remaining stages without requiring a warp-wide maximum-order reduction or full-warp participation. This accommodates the mixed requests and partial participation arising within QUICK's screened integral calculation. The same generated output stages operate on either a main-domain or a normalized large-argument reference row; the preparation and access of those rows are described next.

\subsection{GPU Reference Access and Shared-Path Execution}
\label{subsec:gpu-reference-access}

The common Taylor evaluator requires both numerical domains to provide reference data in the same format. We therefore store the main-domain and normalized large-argument reference rows in a single contiguous allocation. Main-domain rows precede the large-argument rows, whose starting position is specified by the compile-time offset \texttt{tail\_row\_offset}. Both regions use the same row size and logical arrangement of reference coefficients and output corrections. Consequently, either domain can be accessed through a single table pointer and a common set of coefficient offsets.

Within each row, output corrections are interleaved with groups of reference coefficients. A correction word precedes each group of up to eight corresponding output orders and reference coefficients; higher-order coefficients required only for Taylor expansion follow without additional correction words. This layout improves cache locality for short requests, reducing the likelihood that accessing these data requires additional cache lines. We also preserve the same layout in both domains, allowing the generated evaluator to access either reference sequence using identical offsets.

The two domains differ in how their reference coordinates are prepared. For an input in the main domain, the evaluator forms the shifted coordinate $x+s$, selects the corresponding grid point, reconstructs its reference coordinate, and calculates the Taylor displacement relative to $x$. For a large-argument input, the evaluator first normalizes the argument according to Eq.~\eqref{eq:large-argument-normalization}, obtaining $u$ and the corresponding scaling exponent $k_x$. It then selects a reference point from the normalized grid and calculates the displacement relative to $u$. The selected tail row is addressed by adding its compile-time offset to the local row index.

Once the reference row and Taylor displacement have been prepared, both domains enter the same evaluation procedure. They follow identical sequences of reference-coefficient loads, correction decoding, and factorial-Horner operations. The resulting Taylor polynomial directly approximates $F_n(x)$ for main-domain inputs, whereas for large-argument inputs it approximates $A_n(u)$. Thus, the domain-dependent operations are confined to reference-coordinate preparation, while the coefficient access and polynomial evaluation are shared.

The large-argument result requires one additional step after the common Taylor evaluation. Using the scaling identity in Eq.~\eqref{eq:large-argument-scaling}, the normalized result is converted back to the original argument:
\begin{equation}
\widehat F_n(x)
=
2^{-(2n+1)k_x}T_n,
\label{eq:gpu-output-scaling}
\end{equation}
where $T_n$ is the computed Taylor result for the normalized large-argument model. Main-domain results require no such rescaling and can be stored directly. In the implementation, the two cases are handled through a common output-scaling interface, with the main-domain case reducing to an identity operation.

This organization confines the main/tail distinction to coordinate preparation and final output assembly, allowing the computationally substantial coefficient-loading and polynomial-evaluation stages to be shared. It does not require neighbouring threads to select the same reference row or argument domain. Threads with different requested output endpoints can still terminate at different stages, as described in Sec.~\ref{subsec:gpu-thread-mapping}, but neither domain selection nor reference access requires a warp-wide operation.

\subsection{\glsfmtshort{cpu} \glsfmtshort{simd} Extension}
\label{subsec:cpu-simd}

The \gls{cpu} implementation follows the existing NWChem--Simint organization of integral evaluation. Simint's generated kernels vectorize over primitive-integral tasks, with each \gls{simd} lane supplying its own Boys argument. Within a single vector call, all lanes request the same output-order range, which is known at compile time. However, successive calls need not request the same range, since NWChem may combine different groups of shell quartets. We therefore specialize the evaluator for the order range of each generated integral kernel rather than assuming a common order across an entire batch.

The Highway-based \gls{simd} evaluator assigns one argument to each lane and returns the requested Boys values directly into Simint's existing vector workspace.\cite{Highway} It reuses the same numerical models, combined reference table, and row layout as the \gls{gpu} implementation. Since different lanes can select different reference points and even different argument domains, their logical row indices are determined independently before gathering the required reference coefficients and correction words. Once these data have been gathered, the lanes execute the same vector Taylor evaluation, including the low-part correction and final scaling.

The \gls{cpu} backend therefore reuses the same numerical construction and combined reference table as the \gls{cuda} implementation, with lane-dependent reference selection followed by shared vector Taylor evaluation. This provides a \gls{simd} implementation of the proposed method within a conventional \gls{cpu} integral engine.

\section{Experimental Methodology}
\label{sec:experimental-methodology}

We evaluated the proposed method through function-level accuracy tests and
application-level benchmarks. The primary performance comparisons use QUICK
on \gls{gpu}, while NWChem--Simint provides a complementary
\gls{cpu} \gls{simd} evaluation.\cite{Cruzeiro2021AMBERQUICK,nwchem,Pritchard2016Simint} We also compared the effects of the Boys-function implementations on molecular calculation results. This section describes the compared implementations and computing platforms, followed by the workloads, timing procedure, and accuracy assessments.

\subsection{Applications, Platforms, and Compared Implementations}
\label{subsec:comparison-design}

For the \gls{gpu} benchmarks, we compared \gls{fabo} with QUICK's original Boys evaluator and evaluators adapted from LibintX, \gls{vsr}, and GANSU.\cite{Cruzeiro2021AMBERQUICK,Asadchev2025LibintX,VikhamarSandberg2026Boys,Ito2026GANSU} LibintX and GANSU were adapted at the interface and output-layout level while retaining their numerical evaluation procedures, whereas the authors' OpenACC implementation of \gls{vsr} was ported to \gls{cuda}. Each backend was compiled into a separate QUICK executable, leaving the surrounding screening, integral recurrences, contraction, and Fock accumulation unchanged.

For the \gls{cpu} benchmarks, we compared \gls{fabo} with Simint's original Boys evaluator, LibintX, \gls{vsr}, and Laikov's approximation within NWChem using the Simint integral engine.\cite{Pritchard2016Simint,Asadchev2025LibintX,VikhamarSandberg2026Boys,Laikov2025Boys} LibintX retained its native \gls{simd} implementation, whereas \gls{vsr} and Laikov were evaluated through Highway-based \gls{simd} ports.\cite{Highway} The Boys backends were substituted in Simint's generated integral kernels and linked into separate NWChem executables.

The QUICK calculations used an NVIDIA GeForce RTX 5090 D v2 \gls{gpu}. \gls{cuda} code was compiled with NVCC 13.4.92 for \texttt{sm\_120}, while retaining QUICK's \texttt{-O2} and \texttt{--use\_fast\_math} options. We did not enable its f-calculation path, which is disabled by default. The NWChem--Simint calculations ran on an Intel Xeon Gold 6526Y processor on a single thread, using the \gls{avx512} \gls{simd} instruction set. All \gls{cpu} integral variants were compiled with GCC 14.4.0 using \texttt{-O3}, \texttt{-ffast-math}, and \texttt{-march=sapphirerapids}. \gls{cpu} linear algebra used single-threaded Intel \gls{mkl} with automatic instruction-set dispatch. These settings describe the timing builds; numerical-validation builds are distinguished in Sec.~\ref{subsec:accuracy-protocol}. Complete hardware specifications, software versions, source revisions, and build settings are provided in the supplementary material (Sec.~S5.1 and Table~S3).

\subsection{Workloads and Model Selection}
\label{subsec:workloads-model-selection}

The application benchmarks used \gls{rhf} calculations on water, uracil, benzene, and pentane dimers, corresponding to systems 1, 17, 24, and 34 of S66, respectively.\cite{Rezac2011S66} We used the equilibrium-separation geometries and the 6-31G*,\cite{Hehre1972Basis,Hariharan1973Polarization} def2-SVP,\cite{Weigend2005Basis} and cc-pVDZ\cite{Dunning1989Basis} basis sets. Each benchmark evaluates a single density update from a fixed initial density.

To obtain sufficiently long execution times, we constructed linear stacks of translated copies of each dimer without changing its internal geometry. Stack sizes were selected separately for QUICK and NWChem. The resulting \gls{gpu} stacks contained 128, 8, 16, and 16 dimers for water, uracil, benzene, and pentane, respectively; the \gls{cpu} workloads contained eight water dimers and one dimer of each other system. Once selected, each geometry and stack count were held fixed across all three basis sets and all Boys backends. The complete geometries and sizing records are provided in the supplementary material (Sec.~S5.2 and Table~S4).

\gls{fabo}'s numerical models were generated using the joint construction in Sec.~\ref{subsec:joint-reference-tables}. We constructed a separate model for each output range $[0,N]$, $0\leq N\leq24$, subject to a sampled one-\gls{ulp} acceptance criterion and a combined main--tail table budget of 1MiB per model. This budget includes stored coefficients, low-part corrections, and row padding.

In QUICK, the s/p and s/p/d integral-kernel families use the $[0,4]$ and $[0,8]$ models, respectively, and each thread evaluates only the orders requested by its integral calculation. In NWChem--Simint, the maximum requested Boys order is known for each generated integral kernel, which uses the corresponding $[0,N]$ model. Both applications use the same numerically selected model data; only their evaluation interfaces and execution backends differ.

\subsection{Performance Measurements}
\label{subsec:timing-protocol}

Each measurement performed one Fock build and one orbital/density update from a saved initial state. QUICK used its default initial SAD guess, while NWChem used its atomic initial guess. Within each application and molecular case, the same saved state was replayed for every Boys backend and repetition. Initialization, initial-guess preparation, fixture loading, and Boys-table allocation and upload were completed before timing.

For QUICK, the Fock-build interval enclosed \texttt{scf\_operator(.false.)}, which performs the full, rather than density-difference, Fock construction. This includes density-dependent screening preparation, the associated host--device transfers, two-electron integral evaluation through \texttt{gpu\_get\_cshell\_eri}, and Fock-matrix assembly. The one-electron matrix was prepared beforehand and reused. Elapsed wall time was measured with \texttt{clock\_gettime(CLOCK\_MONOTONIC)}, with \texttt{cudaDeviceSynchronize()} before the starting timestamp and before the ending timestamp, so the measurement includes completion of \gls{gpu} work.

For NWChem--Simint, the timing interval enclosed \texttt{fock\_2e(...)}, measuring the two-electron Coulomb and exchange contribution to the Fock build.  These elapsed wall times were measured using NWChem's \texttt{util\_wallsec()}. Thus, the \gls{cpu} timing isolates the two-electron build, whereas the QUICK timing includes the surrounding Fock-construction operations. Matrix serialization and diagnostic output were performed after timing.

Each backend was measured five times per molecular case, with a separate process for each repetition and a cyclic rotation of backend order across repetitions. We report the arithmetic mean of the five elapsed times. Individual measurements and their dispersion are retained in the supplementary material (Sec.~S6 and Tables~S6 and~S7).

\subsection{Function and Molecular Accuracy}
\label{subsec:accuracy-protocol}

Function-level accuracy was assessed for each compared implementation and each requested output range $[0,N]$, $0\leq N\leq12$. The upper order was chosen to cover the maximum Boys-function order required by the \gls{rhf} calculations in QUICK when the \texttt{ENABLEF} option is enabled. The assessment used a uniform grid covering $[0,1000]$ with spacing $10^{-6}$. High-precision references were evaluated at the sampled binary64 arguments, without rounding the reference values to binary64 before computing errors.

Maximum and arithmetic-mean absolute, relative and \gls{ulp} errors were recorded for each requested range, pooling all sampled arguments and all returned orders with equal weight. Strict-arithmetic and fast-math validation results are reported separately.

Molecular results were compared with those obtained using a backend that supplies correctly rounded binary64 Boys values. For each case, we evaluated the absolute difference in the energy and the maximum and mean absolute elementwise differences in the Fock and updated density matrices. The results summarized across cases are the largest casewise values of these quantities.

To compare the effects of different Boys evaluators under identical conditions, all calculations start from the same initial density matrix for each molecular system. This is important because differences in the Boys evaluation can propagate through successive \gls{scf} iterations, leading to different density matrices and potentially different density-dependent screening decisions. Consequently, the set of integrals evaluated and the associated computational workload may also differ between calculations. By fixing the initial density matrix and performing only one Fock build followed by a single diagonalization and density update, we avoid these differences in the preceding \gls{scf} iteration histories. This protocol provides a controlled assessment of how Boys-function evaluation errors affect the Fock matrix and subsequent density update within the same application workflow.

For QUICK, the order of floating-point accumulation can vary between runs because of dynamic \gls{gpu} scheduling. Since floating-point addition is not associative, this variation can lead to differences in the calculated results even when the same Boys evaluator is used. We therefore averaged the energy and each matrix element over five runs for each backend, including the reference, before calculating the differences. For the single-threaded NWChem calculations, repeated outputs were checked for reproducibility and compared with the reference for the same saved initial state.

\section{Results and Discussion}
\label{sec:results-discussion}

We first examine the accuracy of the Boys-function implementations and the reference storage required by \gls{fabo}. We then compare their performance within QUICK and NWChem--Simint.\cite{Cruzeiro2021AMBERQUICK,nwchem,Pritchard2016Simint} Finally, we assess the differences in molecular calculation results obtained with each implementation relative to the correctly rounded Boys reference.

\subsection{Boys-Function Accuracy and Reference Storage}
\label{subsec:function-accuracy-results}

\begin{figure*}[tbp]
\centering
\includegraphics[width=\textwidth]{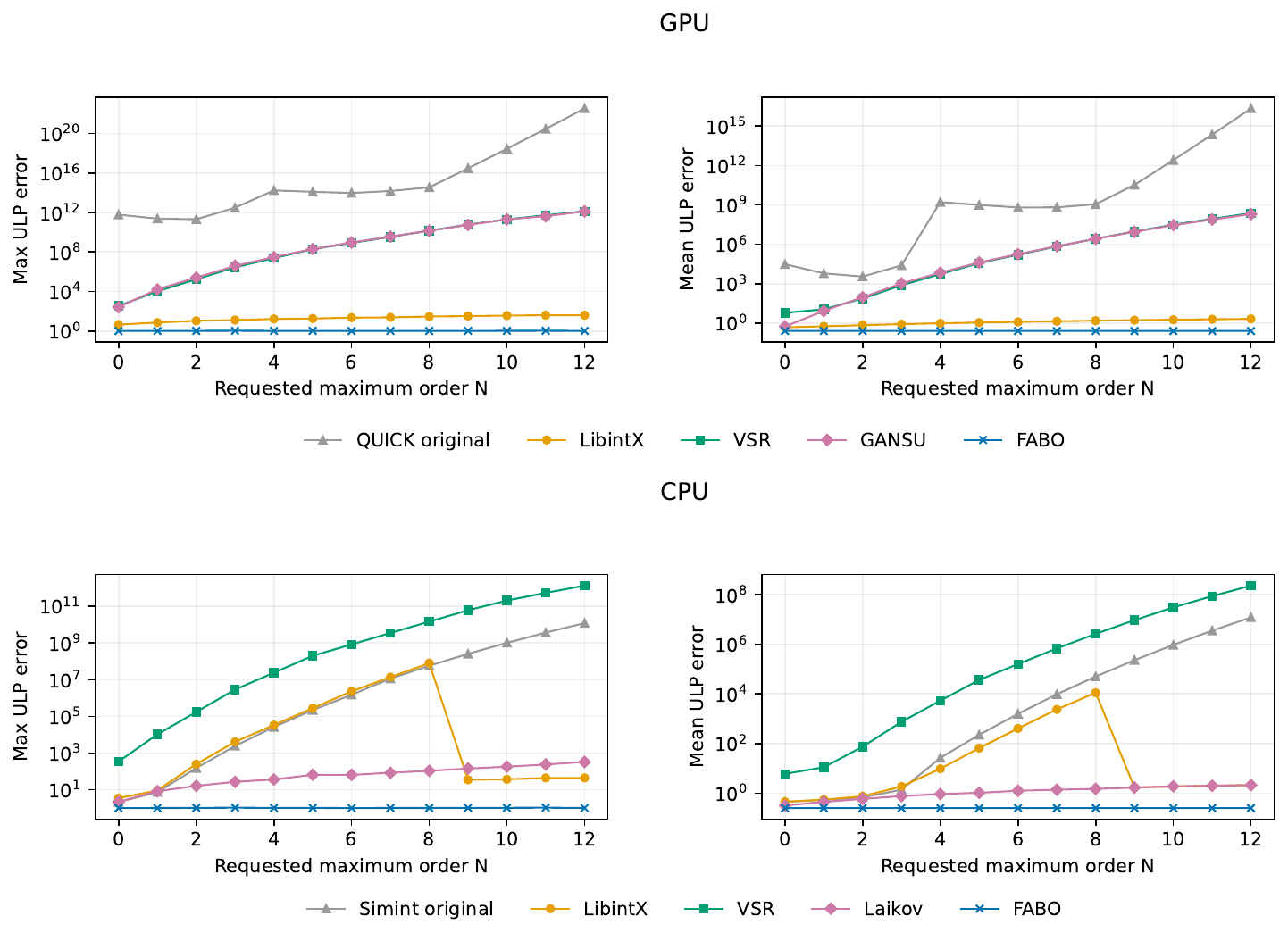}
\caption{Maximum and mean sampled \gls{ulp} errors for each requested output range $[0,N]$ under strict arithmetic, shown for the \gls{gpu} and \gls{cpu} implementations. Each range was evaluated on the $10^{-6}$ grid over $[0,1000]$.}
\label{fig:function-accuracy}
\end{figure*}

Fig.~\ref{fig:function-accuracy} compares the maximum and mean \gls{ulp} errors for each requested output range under strict arithmetic. Table~\ref{tab:function-accuracy} summarizes the absolute, relative, and \gls{ulp} errors over all requested ranges, including both strict-arithmetic and fast-math results. Errors are measured against unrounded high-precision references. For each mean, all returned values are weighted equally. The complete statistics and the inputs at which the maxima occur are provided in the accompanying data (see the supplementary material, Sec.~S4 and Table~S2).

\begin{table*}[tbp]
\caption{Maximum and mean absolute, relative, and \gls{ulp} errors for each Boys evaluator, pooled over all requested ranges $[0,N]$, $N=0,\ldots,12$, and sampled arguments on the $10^{-6}$ grid over $[0,1000]$.}
\label{tab:function-accuracy}
\begingroup
\setlength{\tabcolsep}{3pt}
\begin{ruledtabular}
\begin{tabular}{llrrrrrr}
Platform / mode & Boys backend & \multicolumn{2}{c}{Absolute error} & \multicolumn{2}{c}{Relative error} & \multicolumn{2}{c}{ULP error} \\
 & & Max & Mean & Max & Mean & Max & Mean \\
\hline
\multirow{5}{*}{GPU / strict} & QUICK original & $2.36\times10^{5}$ & $2.21\times10^{-2}$ & $6.49\times10^{6}$ & $6.11\times10^{-1}$ & $3.41\times10^{22}$ & $3.19\times10^{15}$ \\
 & LibintX & $4.30\times10^{-16}$ & $6.70\times10^{-19}$ & $5.06\times10^{-15}$ & $2.34\times10^{-16}$ & $3.93\times10^{1}$ & $1.52\times10^{0}$ \\
 & VSR & $4.37\times10^{-14}$ & $4.52\times10^{-17}$ & $1.97\times10^{-4}$ & $7.75\times10^{-9}$ & $1.30\times10^{12}$ & $4.97\times10^{7}$ \\
 & GANSU & $7.34\times10^{-15}$ & $6.04\times10^{-18}$ & $1.67\times10^{-4}$ & $6.63\times10^{-9}$ & $1.30\times10^{12}$ & $4.32\times10^{7}$ \\
 & FABO & $7.67\times10^{-17}$ & $3.27\times10^{-19}$ & $1.87\times10^{-16}$ & $4.01\times10^{-17}$ & $1.07\times10^{0}$ & $2.50\times10^{-1}$ \\
\hline
\multirow{5}{*}{GPU / fast math} & QUICK original & $2.34\times10^{5}$ & $2.21\times10^{-2}$ & $6.43\times10^{6}$ & $6.11\times10^{-1}$ & $3.38\times10^{22}$ & $3.19\times10^{15}$ \\
 & LibintX & $4.30\times10^{-16}$ & $6.70\times10^{-19}$ & $5.06\times10^{-15}$ & $2.34\times10^{-16}$ & $3.94\times10^{1}$ & $1.52\times10^{0}$ \\
 & VSR & $4.37\times10^{-14}$ & $4.52\times10^{-17}$ & $1.97\times10^{-4}$ & $7.75\times10^{-9}$ & $1.30\times10^{12}$ & $4.97\times10^{7}$ \\
 & GANSU & $7.34\times10^{-15}$ & $6.04\times10^{-18}$ & $1.67\times10^{-4}$ & $6.63\times10^{-9}$ & $1.30\times10^{12}$ & $4.32\times10^{7}$ \\
 & FABO & $7.67\times10^{-17}$ & $3.27\times10^{-19}$ & $1.87\times10^{-16}$ & $4.01\times10^{-17}$ & $1.07\times10^{0}$ & $2.50\times10^{-1}$ \\
\hline
\multirow{5}{*}{CPU / strict} & Simint original & $2.00\times10^{-7}$ & $4.33\times10^{-12}$ & $1.37\times10^{-6}$ & $3.74\times10^{-10}$ & $1.19\times10^{10}$ & $2.35\times10^{6}$ \\
 & LibintX & $2.91\times10^{-16}$ & $6.12\times10^{-19}$ & $9.73\times10^{-9}$ & $2.01\times10^{-13}$ & $7.80\times10^{7}$ & $1.35\times10^{3}$ \\
 & VSR & $4.37\times10^{-14}$ & $4.52\times10^{-17}$ & $1.97\times10^{-4}$ & $7.75\times10^{-9}$ & $1.30\times10^{12}$ & $4.97\times10^{7}$ \\
 & Laikov & $1.57\times10^{-15}$ & $1.07\times10^{-18}$ & $3.95\times10^{-14}$ & $2.35\times10^{-16}$ & $3.29\times10^{2}$ & $1.52\times10^{0}$ \\
 & FABO & $7.67\times10^{-17}$ & $3.27\times10^{-19}$ & $1.87\times10^{-16}$ & $4.01\times10^{-17}$ & $1.07\times10^{0}$ & $2.50\times10^{-1}$ \\
\hline
\multirow{5}{*}{CPU / fast math} & Simint original & $2.00\times10^{-7}$ & $4.33\times10^{-12}$ & $1.37\times10^{-6}$ & $3.74\times10^{-10}$ & $1.19\times10^{10}$ & $2.35\times10^{6}$ \\
 & LibintX & $2.90\times10^{-16}$ & $6.05\times10^{-19}$ & $9.73\times10^{-9}$ & $2.01\times10^{-13}$ & $7.80\times10^{7}$ & $1.35\times10^{3}$ \\
 & VSR & $4.37\times10^{-14}$ & $4.52\times10^{-17}$ & $1.97\times10^{-4}$ & $7.75\times10^{-9}$ & $1.30\times10^{12}$ & $4.97\times10^{7}$ \\
 & Laikov & $1.57\times10^{-15}$ & $1.09\times10^{-18}$ & $3.95\times10^{-14}$ & $2.35\times10^{-16}$ & $3.29\times10^{2}$ & $1.52\times10^{0}$ \\
 & FABO & $7.67\times10^{-17}$ & $3.27\times10^{-19}$ & $1.87\times10^{-16}$ & $4.01\times10^{-17}$ & $1.07\times10^{0}$ & $2.50\times10^{-1}$ \\
\end{tabular}
\end{ruledtabular}
\endgroup
\end{table*}

The comparison of the three error measures provides a consistent assessment of the tested implementations: \gls{fabo} gives the smallest maximum and mean errors in absolute, relative, and \gls{ulp} terms on both platforms and under both arithmetic modes. Thus, its accuracy advantage is not specific to the choice of error measure. The extent to which these function-level differences affect subsequent molecular calculations is examined separately in Sec.~\ref{subsec:molecular-accuracy-results}.

Although \gls{fabo} achieves the smallest errors among the tested implementations, its largest sampled \gls{ulp} error is approximately 1.067, slightly exceeding the one-\gls{ulp} target used during construction. This does not contradict the construction results, since the target was enforced only at the prescribed qualification inputs, whereas the present dense-grid scan evaluates additional arguments. As discussed in Sec.~\ref{subsec:gridded-taylor}, passing the qualification tests does not guarantee that the target is satisfied for every binary64 input. The observed exceedance therefore illustrates the limitation of finite numerical qualification rather than a failure to meet the prescribed construction criterion. The corresponding input $x$ and output order are reported in the supplementary material (Table~S2).

The errors of the other implementations vary with the requested order and the approximation used in each argument region. The original QUICK evaluator shows particularly pronounced error growth immediately after its switch to upward recurrence at $x=0.1$. At such small arguments, rounding errors are rapidly amplified as the recurrence proceeds to higher orders. For example, under strict arithmetic, the computed $F_{12}(0.100019)$ is approximately $-2.36\times10^5$. LibintX shows markedly different error behavior on the \gls{cpu} and \gls{gpu}, with a much larger maximum error on the \gls{cpu}. This difference arises because the asymptotic switching points differ: for the $[0,8]$ output range, the \gls{cpu} implementation switches at $x=36$, where its error increases, whereas the \gls{gpu} sequence implementation switches at $x=117$, resulting in a smaller approximation error.

The fast-math results are generally close to their strict-arithmetic counterparts, although a few error statistics improve slightly. We attribute these improvements to the compiler's ability to replace separate multiplication and addition operations with an \gls{fma}, thereby reducing the number of rounding steps.\cite{IEEE7542019}

\begin{figure}[tbp]
\centering
\includegraphics[width=\columnwidth]{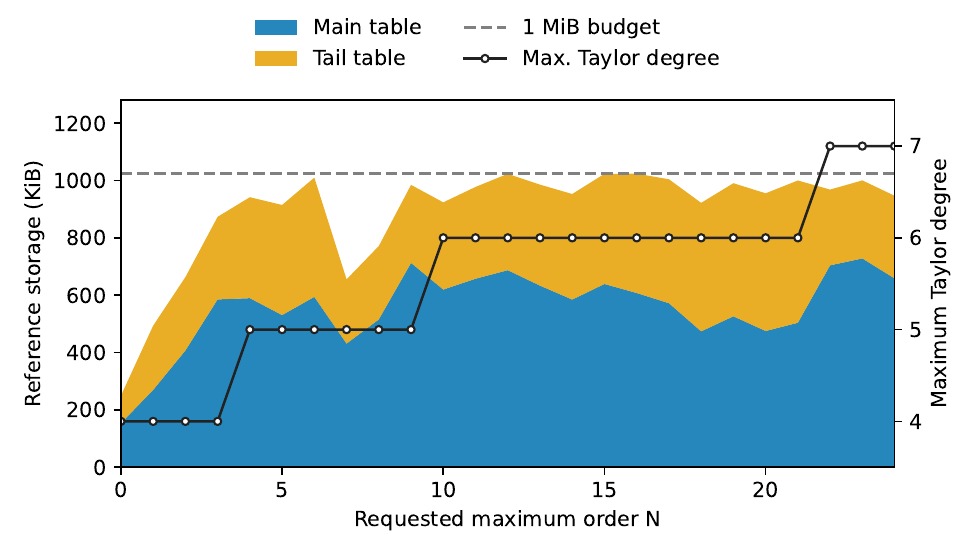}
\caption{Reference storage and maximum Taylor degree for each independently
constructed numerical model covering the output range $[0,N]$.}
\label{fig:reference-storage}
\end{figure}

Fig.~\ref{fig:reference-storage} shows the reference storage for the constructed models. All combined main and tail tables fit within the 1 MiB budget assigned to each model. The stacked areas show the main- and tail-table sizes, including stored coefficients, low-part corrections, and row padding. The storage does not increase monotonically with $N$, because the grid and polynomial degrees are selected separately for each requested range, and the grid parameters and polynomial degrees are jointly optimized under the same storage budget. Increasing $N$ adds more requested Boys functions and can increase the number of coefficients stored in each reference row. However, a higher Taylor degree can also permit a coarser grid with fewer reference points, compensating for the additional coefficients. The maximum Taylor degree, shown on the right axis of Fig.~\ref{fig:reference-storage}, increases in steps from four to seven over the tested range. The resulting storage variation reflects the trade-off between the number of coefficients per reference row and the number of reference rows. Although this 1 MiB budget may appear unfavorable for cache utilization, its effect on overall performance must be assessed in end-to-end application benchmarks, as discussed in Secs.~\ref{subsec:gpu-performance} and~\ref{subsec:cpu-performance}.

The accuracy and storage results illustrate the trade-offs underlying the \gls{fabo} construction. The nonuniform grid redistributes reference points toward regions where finer spacing is needed, reducing the maximum Taylor remainder without increasing the number of reference points, as shown in Fig.~\ref{fig:nonuniform-grid-remainder}. The low-part corrections introduce additional storage, but this cost is included in the combined table budget and allows the final evaluation to retain information lost when the reference coefficients are rounded to binary64. The storage results in Fig.~\ref{fig:reference-storage} further show that the main- and tail-table sizes depend jointly on the requested output orders, polynomial degrees, and grid parameters. Their nonmonotonic variation with $N$ reflects the independent optimization of each model under the same storage constraint. Importantly, the compact representation does not require separate polynomial evaluation procedures for the two argument regions. Once the reference data and displacements have been prepared, both regions use the same coefficient-loading, correction-decoding, and Taylor evaluation sequence. This shared computational structure allows the storage reduction and numerical accuracy improvements to be incorporated without introducing additional polynomial stages for large arguments. However, the effect of these design choices on execution time also depends on the surrounding integral calculations, including screening, contraction, and Fock-matrix assembly. Their practical performance is therefore assessed through the application-level Fock-build benchmarks in the following sections.

\subsection{\glsfmtshort{gpu} Performance in QUICK}
\label{subsec:gpu-performance}

\begin{figure*}[tbp]
\centering
\begin{minipage}{0.49\textwidth}
\centering
\includegraphics[width=\linewidth]{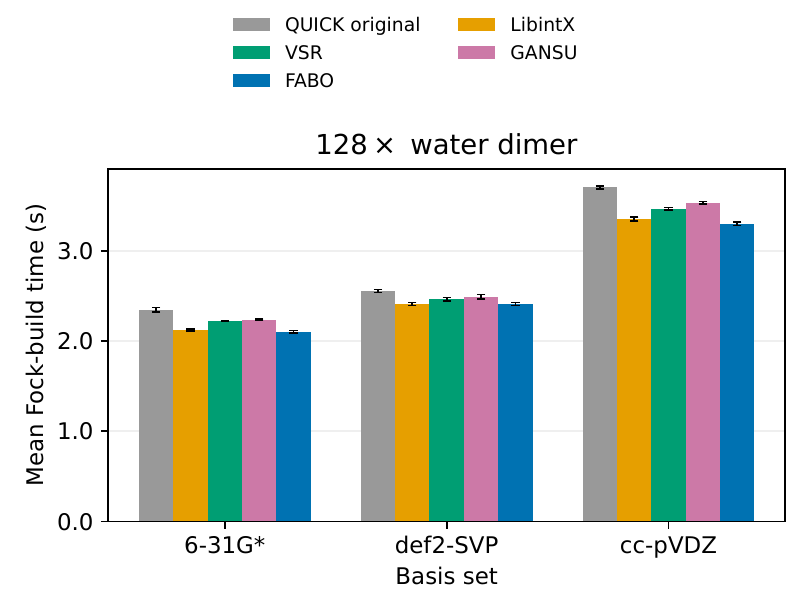}
\par (a)
\end{minipage}\hfill
\begin{minipage}{0.49\textwidth}
\centering
\includegraphics[width=\linewidth]{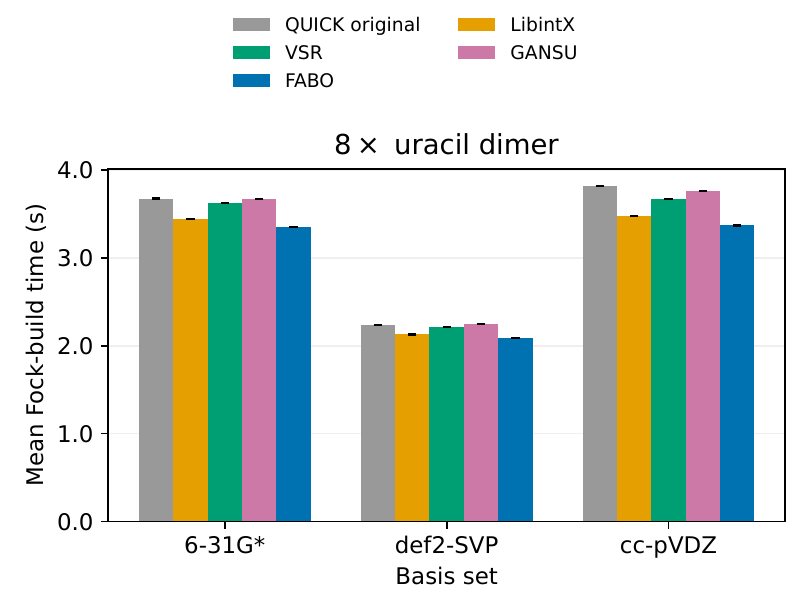}
\par (b)
\end{minipage}
\par\medskip
\begin{minipage}{0.49\textwidth}
\centering
\includegraphics[width=\linewidth]{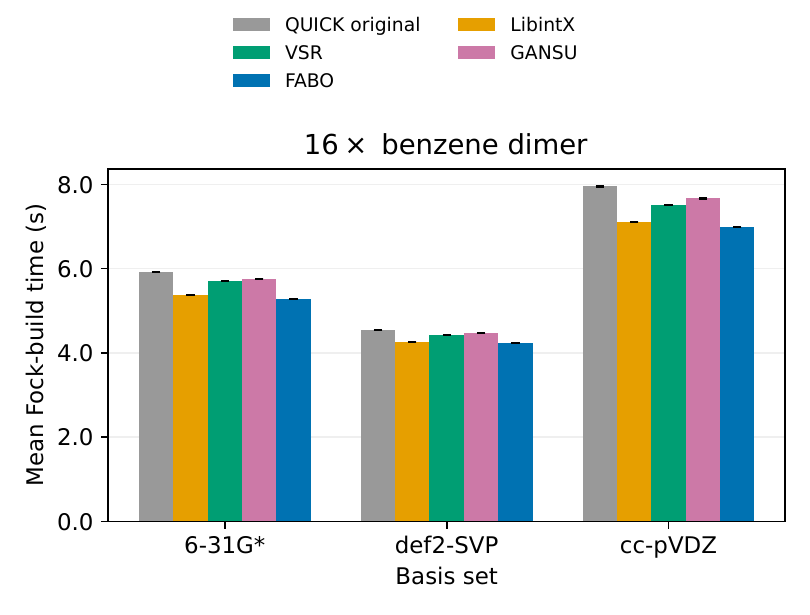}
\par (c)
\end{minipage}\hfill
\begin{minipage}{0.49\textwidth}
\centering
\includegraphics[width=\linewidth]{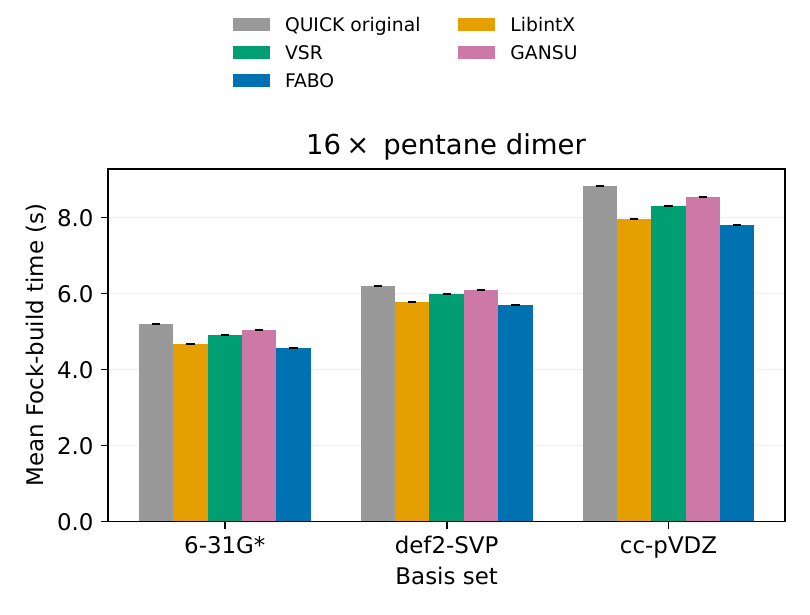}
\par (d)
\end{minipage}
\caption{Fock-build times in QUICK for stacks of (a) water, (b) uracil, (c) benzene, and (d) pentane dimers. Bars show the mean of five measurements, and error bars show one sample standard deviation. The panels use independent vertical scales.}
\label{fig:gpu-fock-performance}
\end{figure*}

Fig.~\ref{fig:gpu-fock-performance} compares the Fock-build times obtained with the five Boys evaluators in QUICK. The corresponding numerical values are listed in Table~\ref{tab:gpu-fock-times}. \gls{fabo} gives the lowest mean Fock-build time in all twelve tested cases and consistently improves upon QUICK's original Boys evaluator. The improvement is observed across all four molecular systems and three basis sets, although its magnitude varies with the workload.

\begin{table*}[tbp]
\caption{Mean Fock-build times in seconds for the QUICK benchmarks shown in Fig.~\ref{fig:gpu-fock-performance}.}
\label{tab:gpu-fock-times}
\begin{ruledtabular}
\begin{tabular}{llrrrrr}
Molecule & Basis & QUICK original & LibintX & VSR & GANSU & FABO \\
\hline
 & 6-31G* & 2.347 & 2.122 & 2.221 & 2.237 & 2.103 \\
$128\times$ water dimer & def2-SVP & 2.553 & 2.410 & 2.463 & 2.490 & 2.408 \\
 & cc-pVDZ & 3.703 & 3.350 & 3.462 & 3.530 & 3.297 \\
\hline
 & 6-31G* & 3.674 & 3.439 & 3.624 & 3.669 & 3.348 \\
$8\times$ uracil dimer & def2-SVP & 2.237 & 2.128 & 2.212 & 2.251 & 2.091 \\
 & cc-pVDZ & 3.818 & 3.473 & 3.667 & 3.760 & 3.370 \\
\hline
 & 6-31G* & 5.918 & 5.380 & 5.701 & 5.765 & 5.290 \\
$16\times$ benzene dimer & def2-SVP & 4.540 & 4.264 & 4.424 & 4.473 & 4.234 \\
 & cc-pVDZ & 7.955 & 7.113 & 7.503 & 7.665 & 6.984 \\
\hline
 & 6-31G* & 5.192 & 4.668 & 4.916 & 5.037 & 4.567 \\
$16\times$ pentane dimer & def2-SVP & 6.192 & 5.765 & 5.977 & 6.082 & 5.699 \\
 & cc-pVDZ & 8.823 & 7.947 & 8.298 & 8.535 & 7.800 \\
\end{tabular}
\end{ruledtabular}
\end{table*}

The difference between function-level accuracy and application-level performance is worth noting. As shown in Sec.~\ref{subsec:function-accuracy-results}, \gls{fabo} substantially reduces the Boys-function evaluation errors compared with the other tested implementations. However, the measured Fock-build time includes not only Boys evaluation but also screening, integral recurrences, contraction, and Fock-matrix assembly. According to Amdahl's law~\cite{Amdahl1967}, if Boys evaluation accounts for a fraction $f$ of the original execution time and is accelerated by a factor $S_{\mathrm{B}}$, the overall speedup is bounded by
\begin{equation}
S_{\mathrm{total}}
=
\frac{1}{(1-f)+f/S_{\mathrm{B}}}.
\label{eq:gpu-amdahl}
\end{equation}
This expression illustrates why improvements in Boys evaluation do not translate directly into equivalent improvements in the complete Fock build.

One may therefore ask why the time spent on Boys evaluation is not measured separately. The main reason is that its performance cannot be assessed independently of the surrounding integral kernel. The evaluator is inlined into the \gls{gpu} integral kernels, where its instructions are interleaved with integral recurrences and other operations. In particular, the register requirements of the complete kernel determine how many warps can reside concurrently on each streaming multiprocessor (SM), thereby affecting the ability to hide instruction and memory latency. Measuring the evaluator separately would change this resource usage and the opportunities for overlapping its operations with the surrounding calculation. Reference-table accesses also compete with other kernel data for cache capacity, so their performance depends on the memory-access pattern of the complete integral kernel. Consequently, an isolated measurement would not reliably represent the contribution of Boys evaluation to the application-level Fock-build time.

Among the alternative evaluators, LibintX gives the closest timings to \gls{fabo}. The geometric-mean speedup of \gls{fabo} over LibintX is approximately $1.016$, corresponding to a $1.62\%$ reduction in mean execution time across the twelve cases. The difference is particularly small for the water stack with def2-SVP, where the mean times are $2.408$ and $2.410$~s for \gls{fabo} and LibintX, respectively. This difference is much smaller than the run-to-run variation of either implementation, so their performance in this case should be regarded as comparable. The corresponding geometric-mean speedups over \gls{vsr} and GANSU are $1.062$ and $1.080$, respectively. Relative to QUICK's original evaluator, \gls{fabo} achieves a geometric-mean speedup of $1.107$, corresponding to a $9.64\%$ reduction in Fock-build time.

These results demonstrate that the numerical improvements introduced by \gls{fabo} can be incorporated into QUICK without sacrificing application-level performance. In particular, the nonuniform reference grid and the normalized large-argument representation provide compact reference tables, while the low-part corrections improve the accuracy of the final Taylor evaluation. Despite these additional numerical features and the associated reference-data accesses, the resulting evaluator remains competitive with established \gls{gpu} implementations.

\subsection{\glsfmtshort{cpu} Performance in NWChem--Simint}
\label{subsec:cpu-performance}

\begin{figure*}[tbp]
\centering
\begin{minipage}{0.49\textwidth}
\centering
\includegraphics[width=\linewidth]{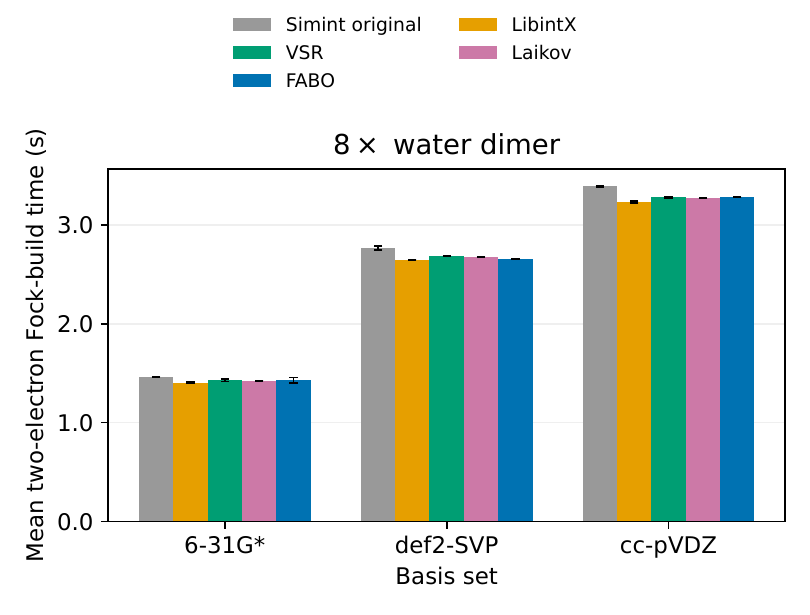}
\par (a)
\end{minipage}\hfill
\begin{minipage}{0.49\textwidth}
\centering
\includegraphics[width=\linewidth]{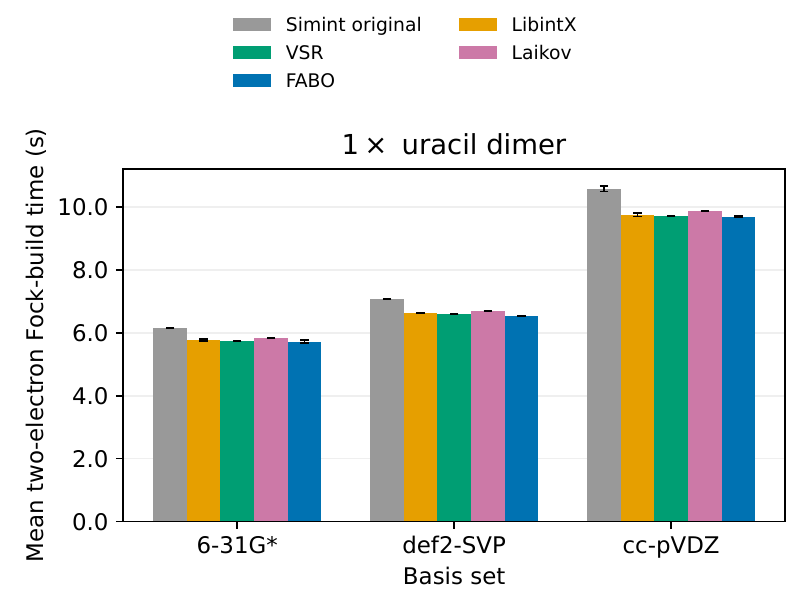}
\par (b)
\end{minipage}
\par\medskip
\begin{minipage}{0.49\textwidth}
\centering
\includegraphics[width=\linewidth]{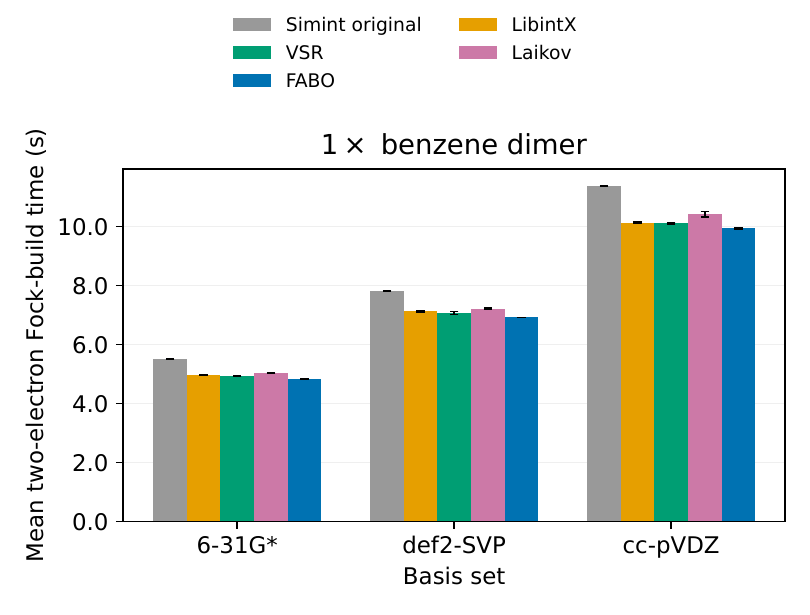}
\par (c)
\end{minipage}\hfill
\begin{minipage}{0.49\textwidth}
\centering
\includegraphics[width=\linewidth]{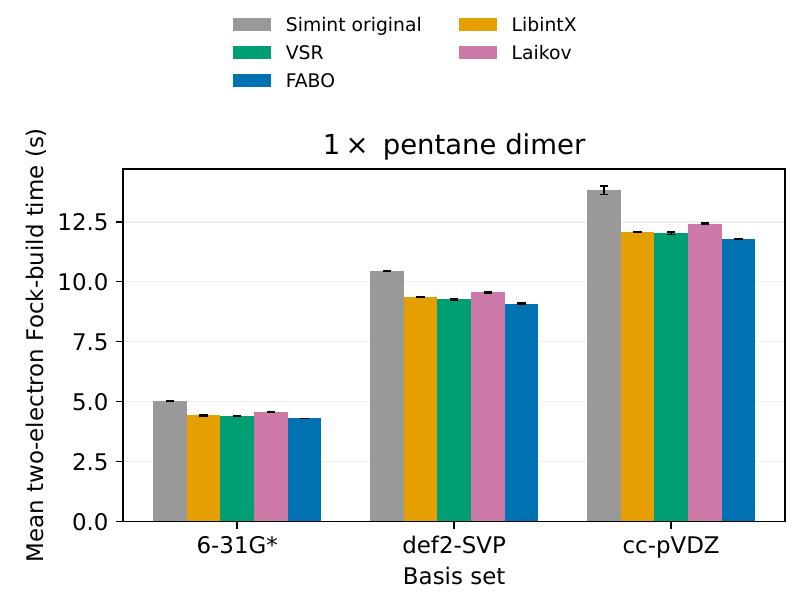}
\par (d)
\end{minipage}
\caption{Two-electron Fock-build times in NWChem--Simint for (a) water, (b) uracil, (c) benzene, and (d) pentane dimers. Bars show the mean of five measurements, and error bars show one sample standard deviation. The panels use independent vertical scales.}
\label{fig:cpu-fock-performance}
\end{figure*}

Fig.~\ref{fig:cpu-fock-performance} and Table~\ref{tab:cpu-fock-times} show the corresponding NWChem--Simint results. The CPU and \gls{gpu} measurements use different workloads, so their absolute times are not directly comparable.

\begin{table*}[tbp]
\caption{Mean two-electron Fock-build times in seconds for the NWChem--Simint benchmarks shown in Fig.~\ref{fig:cpu-fock-performance}.}
\label{tab:cpu-fock-times}
\begin{ruledtabular}
\begin{tabular}{llrrrrr}
Molecule & Basis & Simint original & LibintX & VSR & Laikov & FABO \\
\hline
 & 6-31G* & 1.466 & 1.405 & 1.430 & 1.424 & 1.429 \\
$8\times$ water dimer & def2-SVP & 2.765 & 2.649 & 2.687 & 2.680 & 2.656 \\
 & cc-pVDZ & 3.389 & 3.232 & 3.281 & 3.269 & 3.283 \\
\hline
 & 6-31G* & 6.162 & 5.770 & 5.728 & 5.827 & 5.717 \\
$1\times$ uracil dimer & def2-SVP & 7.075 & 6.635 & 6.594 & 6.700 & 6.543 \\
 & cc-pVDZ & 10.581 & 9.752 & 9.718 & 9.885 & 9.693 \\
\hline
 & 6-31G* & 5.496 & 4.957 & 4.923 & 5.038 & 4.825 \\
$1\times$ benzene dimer & def2-SVP & 7.830 & 7.122 & 7.066 & 7.210 & 6.918 \\
 & cc-pVDZ & 11.378 & 10.131 & 10.114 & 10.420 & 9.945 \\
\hline
 & 6-31G* & 5.011 & 4.436 & 4.396 & 4.552 & 4.296 \\
$1\times$ pentane dimer & def2-SVP & 10.452 & 9.344 & 9.257 & 9.564 & 9.079 \\
 & cc-pVDZ & 13.819 & 12.072 & 12.021 & 12.423 & 11.776 \\
\end{tabular}
\end{ruledtabular}
\end{table*}

\gls{fabo} reduces the mean two-electron build time relative to Simint's original evaluator for every tested system and basis set. Its performance relative to the other alternatives depends on the workload. LibintX gives the lowest mean time for the water calculations, while \gls{fabo} gives the lowest mean for the uracil, benzene, and pentane calculations. For the latter systems, \gls{vsr} also performs closely to \gls{fabo}.

The relatively favorable performance of \gls{vsr} for the water calculations may be related to its region-dependent evaluation strategy. \gls{vsr} combines rational minimax approximations and recurrence relations in two finite argument regions with an asymptotic treatment for large arguments. Its \gls{simd} implementation can skip the other evaluation regions when all lanes in a vector belong to the same region. This behavior may be particularly advantageous on the \gls{cpu}, where an \gls{avx512} vector accommodates only eight double-precision values, compared with the 32 threads in a \gls{gpu} warp. For a similar distribution of input arguments, a smaller group of inputs is more likely to fall entirely within the large-argument region, allowing \gls{vsr} to bypass the evaluation of the small-argument regions. This possibility is relevant to the water calculations, where the spatial separation between different dimers can give rise to relatively large Boys arguments. Consequently, the combination of relatively large arguments and narrower \gls{simd} vectors may increase the frequency with which the asymptotic evaluation path can be used exclusively.

Apart from the favorable performance of \gls{vsr} for the water calculations, the advantage of \gls{fabo} over the other optimized evaluators is also relatively modest in the \gls{cpu} benchmarks. This may partly reflect differences in how their large-argument calculations are executed on the two architectures. As discussed in Sec.~\ref{subsec:large-argument-limitations}, the conventional asymptotic algorithms evaluate a reciprocal square root, as in Eq.~\eqref{eq:asymptotic-f0}. On the \gls{gpu}, double-precision reciprocal square root is typically obtained from an approximate initial value followed by refinement, with the refinement operations competing with the surrounding calculation for floating-point execution resources. On the \gls{cpu}, the same operation is typically implemented using a square root and a division. Although these operations have relatively long latencies, their latency can be hidden by independent calculations in the surrounding integral kernel, as discussed in the Intel optimization manual.\cite{IntelOptimizationManual} Consequently, the relative cost of conventional large-argument evaluation can be lower on the \gls{cpu}, reducing the performance advantage of \gls{fabo}.

Taken together, the QUICK and NWChem--Simint benchmarks demonstrate that \gls{fabo} provides consistent application-level performance improvements over the original Boys evaluators on both \gls{gpu} and \gls{cpu} platforms. In QUICK, \gls{fabo} achieves the lowest mean Fock-build time in all tested cases, while in NWChem--Simint it achieves the lowest mean time in 75\% of cases. In particular, the results show that the improved numerical accuracy of \gls{fabo} can be achieved without a substantial loss of application-level performance. The \gls{cpu} and \gls{gpu} benchmarks therefore demonstrate the practical applicability of \gls{fabo} across distinct computational architectures.

\subsection{Effects on Molecular Calculation Results}
\label{subsec:molecular-accuracy-results}

\begin{figure*}[tbp]
\centering
\includegraphics[width=\textwidth]{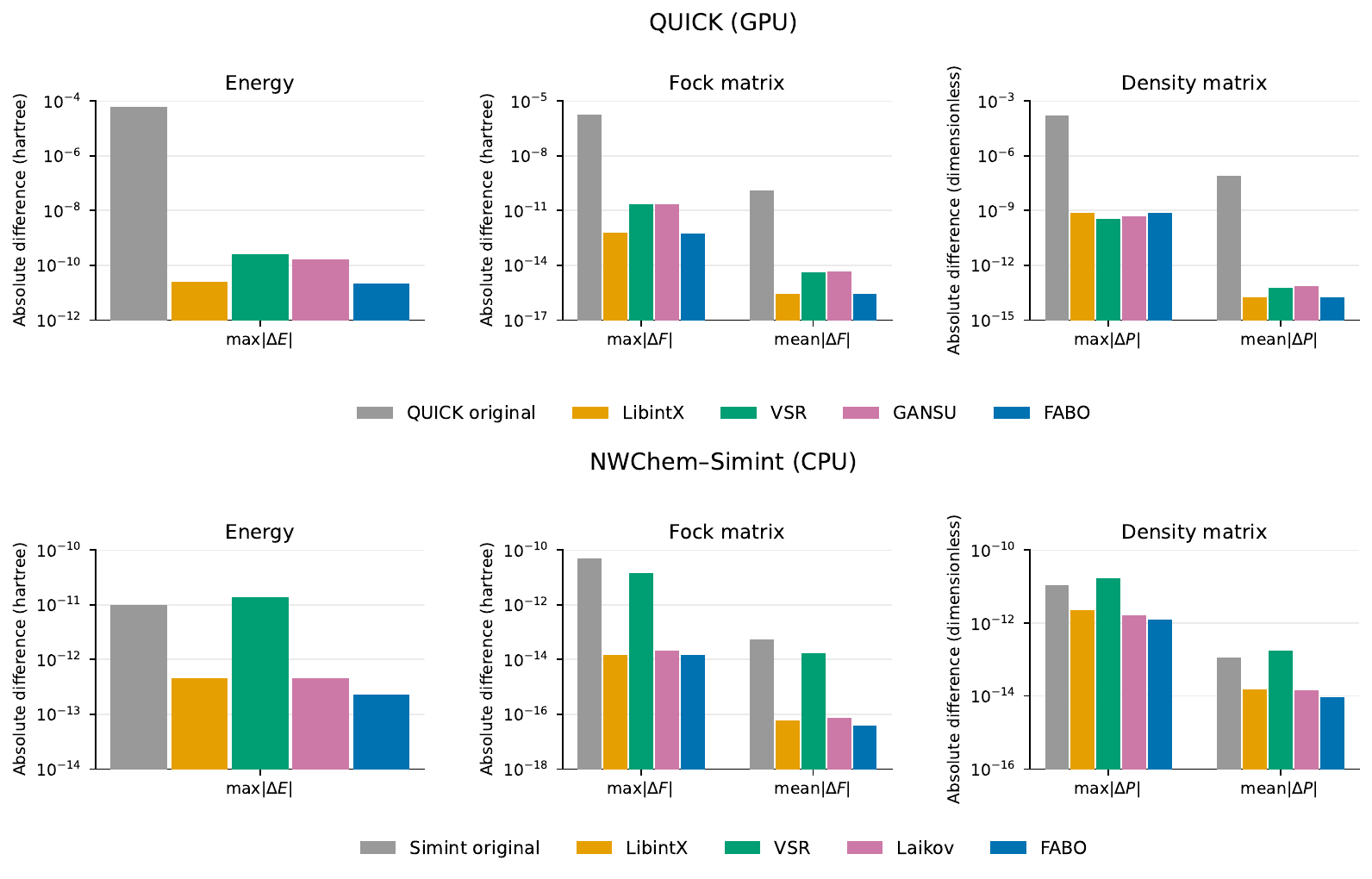}
\caption{Largest casewise differences in the energy, Fock matrix, and updated density matrix relative to the correctly rounded Boys reference. The bars correspond to the values in Table~\ref{tab:molecular-differences}. Each panel uses an independent logarithmic scale.}
\label{fig:molecular-differences}
\end{figure*}

Table~\ref{tab:molecular-differences} and Fig.~\ref{fig:molecular-differences} compare the molecular results obtained with each Boys evaluator against those obtained using correctly rounded Boys values. The comparison covers the energy, Fock matrix, and density matrix after one update from the fixed initial state. For each matrix, the maximum and mean are taken over its elements before selecting the largest value across the tested cases. The maximizing case may differ between columns. The complete casewise results are provided in the supplementary material (Tables~S8 and~S9).

\begin{table*}[tbp]
\caption{Largest casewise absolute differences in molecular results relative to the correctly rounded Boys reference.}
\label{tab:molecular-differences}
\begin{ruledtabular}
\begin{tabular}{llrrrrr}
Application & Boys backend & $\max|\Delta E|$ & $\max|\Delta F|$ & $\operatorname{mean}|\Delta F|$ & $\max|\Delta P|$ & $\operatorname{mean}|\Delta P|$ \\
\hline
\multirow{5}{*}{QUICK (GPU)} & QUICK original & $5.98\times10^{-5}$ & $1.78\times10^{-6}$ & $1.30\times10^{-10}$ & $1.55\times10^{-4}$ & $8.00\times10^{-8}$ \\
 & LibintX & $2.47\times10^{-11}$ & $5.80\times10^{-13}$ & $2.58\times10^{-16}$ & $7.36\times10^{-10}$ & $1.83\times10^{-14}$ \\
 & VSR & $2.44\times10^{-10}$ & $2.27\times10^{-11}$ & $3.97\times10^{-15}$ & $3.32\times10^{-10}$ & $5.51\times10^{-14}$ \\
 & GANSU & $1.69\times10^{-10}$ & $2.27\times10^{-11}$ & $4.75\times10^{-15}$ & $4.63\times10^{-10}$ & $6.87\times10^{-14}$ \\
 & FABO & $2.18\times10^{-11}$ & $5.74\times10^{-13}$ & $2.58\times10^{-16}$ & $7.11\times10^{-10}$ & $1.79\times10^{-14}$ \\
\hline
\multirow{5}{*}{NWChem--Simint (CPU)} & Simint original & $9.78\times10^{-12}$ & $5.10\times10^{-11}$ & $5.56\times10^{-14}$ & $1.07\times10^{-11}$ & $1.11\times10^{-13}$ \\
 & LibintX & $4.55\times10^{-13}$ & $1.42\times10^{-14}$ & $5.95\times10^{-17}$ & $2.27\times10^{-12}$ & $1.49\times10^{-14}$ \\
 & VSR & $1.36\times10^{-11}$ & $1.41\times10^{-11}$ & $1.66\times10^{-14}$ & $1.64\times10^{-11}$ & $1.76\times10^{-13}$ \\
 & Laikov & $4.55\times10^{-13}$ & $2.13\times10^{-14}$ & $7.55\times10^{-17}$ & $1.60\times10^{-12}$ & $1.42\times10^{-14}$ \\
 & FABO & $2.27\times10^{-13}$ & $1.42\times10^{-14}$ & $3.78\times10^{-17}$ & $1.22\times10^{-12}$ & $9.27\times10^{-15}$ \\
\end{tabular}
\end{ruledtabular}
\end{table*}

In QUICK, all four alternative evaluators give substantially smaller deviations from the reference than the original Boys evaluator. Among them, \gls{fabo} and LibintX give the closest energy and Fock-matrix results to the reference, with similar differences in both the maximum and mean absolute matrix errors. Their maximum density-matrix differences, however, are somewhat larger than those obtained with \gls{vsr} and GANSU. This ordering differs from that observed in the function-level accuracy tests. In contrast, the mean density-matrix differences of \gls{fabo} and LibintX remain smaller than those of \gls{vsr} and GANSU, consistent with their better function-level accuracy.

This difference may arise from the run-to-run variation in floating-point accumulation caused by dynamic \gls{gpu} scheduling. Since floating-point addition is not associative, different accumulation orders can produce different matrix elements even when the same Boys evaluator is used. This run-to-run variation is also observed for the correctly rounded reference calculation. It may therefore contribute to the larger maximum density-matrix differences observed for \gls{fabo} and LibintX, despite their smaller mean differences. The five-run averaging used in the comparison reduces the influence of individual runs but does not eliminate this variation. Thus, the maximum density-matrix differences in QUICK do not necessarily reflect the relative accuracy of the Boys evaluators alone.

The NWChem--Simint results provide a useful comparison because the single-threaded \gls{cpu} calculations are reproducible across repetitions. Here, \gls{fabo} gives the smallest maximum density-matrix difference, followed by Laikov's approximation, LibintX, and \gls{vsr}. The mean density-matrix differences show the same ordering. In particular, the relative ordering of \gls{fabo}, LibintX, and \gls{vsr} is consistent with their function-level accuracy results. This agreement supports the interpretation that the different ordering of maximum density-matrix errors in QUICK is associated with the variability of \gls{gpu} accumulation rather than the intrinsic accuracy of the Boys evaluators.

The errors in the energy, Fock matrix, and density matrix do not necessarily follow the same ordering. The energy is a scalar quantity obtained from the input density and the corresponding Fock matrix, whereas the density matrix is obtained after diagonalization and occupation of the resulting orbitals. Consequently, differences in the Fock matrix can propagate differently into the energy and density matrix, and the latter can also depend on the sensitivity of the orbital eigenvectors to small perturbations. The maximum and mean elementwise matrix differences further characterize different aspects of these deviations. Therefore, no single molecular quantity is expected to reproduce the function-level accuracy ranking in every case.

Overall, the molecular calculations demonstrate that \gls{fabo} combines high Boys-function accuracy with competitive application-level performance. In both QUICK and NWChem--Simint, the results obtained with \gls{fabo} remain close to those obtained using correctly rounded Boys values, with particularly small deviations in the energy and Fock matrix. Its density-matrix differences are also among the smallest in the reproducible \gls{cpu} calculations. The comparatively larger maximum density-matrix differences observed in QUICK may result from nondeterministic \gls{gpu} scheduling and the non-associativity of floating-point accumulation. Together with the function-level accuracy results, these comparisons show that the accuracy improvements achieved by \gls{fabo} are retained in practical molecular integral calculations. Moreover, \gls{fabo} provides modest application-level speedups in most of the tested cases, demonstrating that its improved numerical accuracy does not come at the expense of computational efficiency.

\section{Conclusions}
\label{sec:conclusions}

We have developed a gridded Taylor method for evaluating Boys functions in molecular integral calculations. A nonuniform reference grid, constructed from a shifted floating-point representation, provides fine spacing near the origin and progressively wider spacing at larger arguments. This reduces the growth of reference storage while allowing the reference point to be selected directly from the input. A compact low-part correction retains additional information about each output's constant Taylor coefficient.

To extend the method to large arguments, we introduced a normalized
asymptotic representation based on an auxiliary function whose
derivatives follow the same arithmetic relation as those of the
Boys functions. The main-domain and normalized large-argument expansions can therefore use a common sequence of coefficient accesses and Taylor operations. We constructed their reference tables jointly, selecting shared, order-dependent polynomial degrees under a combined storage budget. The selected models met the one-\gls{ulp} target at the prescribed qualification inputs; independent tests also found small exceedances of this target, consistent with the limitations of sampled qualification.

We implemented the method in \gls{fabo} and integrated it into QUICK and NWChem--Simint. In the reproducible NWChem--Simint calculations, \gls{fabo} gave the smallest deviation from the correctly rounded Boys reference for every energy and matrix measure considered. In QUICK, where accumulation order can vary between runs, \gls{fabo} also gave among the smallest deviations overall, including the smallest energy and Fock-matrix differences. For a single Fock build, \gls{fabo} achieves geometric mean speedups of 1.22\%--10.67\% over the other tested Boys implementations, depending on the hardware and the implementation used for comparison.

\section*{Supplementary Material}

The supplementary material PDF describes the construction and numerical qualification of the reference tables, the floating-point evaluation algorithm, and the independent Boys function accuracy tests. It also gives the QUICK and NWChem--Simint benchmark protocols, individual Fock-build timings, and casewise differences in energies, Fock matrices, and density matrices relative to calculations using correctly rounded Boys values.

The data package available from \href{https://github.com/LXYan2333/fabo/releases/tag/paper-si-v1}{FABO release \texttt{paper-si-v1}} contains machine-readable construction and benchmark records, application inputs, and instructions for reproducing the numerical summaries and plots. Table~S11 in Sec.~S7.3 of the supplementary material indexes the data and reporting tools.

\begin{acknowledgements}
This work was supported by the National Natural Science Foundation of China (Nos. 22473092, and 22373077) and the Fundamental Research Funds for the Central Universities (No. 20720260032)	
\end{acknowledgements}

\section*{AUTHOR DECLARATIONS}
\subsection*{Conflict of Interest}
The authors have no conflict of interest to disclose.

\section*{Data Availability Statement}

The data supporting the findings of this study are provided in the article and its supplementary material. The compact records and reproduction instructions are available from \href{https://github.com/LXYan2333/fabo/releases/tag/paper-si-v1}{FABO release \texttt{paper-si-v1}}. The \gls{fabo} source code is available at \url{https://github.com/LXYan2333/fabo}. The numerical summaries and plots can be reproduced from the supplementary records using the reporting tools at \href{https://github.com/LXYan2333/fabo/tree/6dff8737080dc14b18ae7641b1cf5c889bd6e3b6}{commit \texttt{6dff873}}.

\appendix

\nocite{*}
\makeatletter
\def\@bibstyle{fabo-aipnum}
\makeatother
\bibliography{aipsamp}

\end{document}


\maketitle

This supplement describes numerical construction, floating-point evaluation, and the function and molecular tests of FABO (Fast and Accurate BOys). Numerical qualification uses selected inputs, not an exhaustive enumeration of binary64 arguments. Directed reference enclosures control reference uncertainty at tested inputs; they do not establish a continuous-domain bound on implementation error. The accompanying data retain complete configurations and measurements.

\tableofcontents
\section{Numerical representation and implementation details}
\label{sec:implementation}
\subsection{Arithmetic contract}
Inputs are finite binary64 values $x\geq0$, including negative zero. Requested orders lie within the selected model's range. The arithmetic contract is round-to-nearest, ties-to-even, with gradual underflow; users should not enable floating-point traps. NaNs, infinities and negative nonzero inputs are outside the contract. Explicit fused multiply-add operations remain fused in strict builds: disabling contraction prevents the compiler from combining otherwise separate operations, not the use of explicit FMA.

Let $U$ map a positive binary64 value to its unsigned 64-bit representation and let $U^{-1}$ be the inverse bit reinterpretation. The main manuscript's $r_i$, $h=r_i-x$, $p_n$, and $G_{i,n}=\RN(F_n(r_i))$ are used throughout. Binary64 reference coefficients are distinguished from the mathematical values used to construct them.

\subsection{Main-grid indexing and displacement}
\label{subsec:grid-indexing}
For retained fraction-bit count $b_g$, put $d=52-b_g$. The main grid uses
\begin{align}
 z&=\RN(x+s),&K&=(U(z)+2^{d-1})\mathbin{\gg}d,\\
 i&=K-K_{\min},&r_i&=\RN(U^{-1}(K\mathbin{\ll}d)-s),\\
 h&=\RN(r_i-x).
\end{align}
The integer quantizer rounds halfway cases toward the larger key. It is therefore distinct from the ties-to-even floating-point additions and subtractions. The SIMD implementation adds the rounding bit and then shifts, as in the expression for $K$ above. In CUDA, when $d>32$, the high 32-bit word alone contains all bits needed for this quantization. This changes instruction width, not the selected key.

The main domain includes $x=C$; the tail is selected for $x>C$. Define $K_{\min}=K(0)$ and $K_{\max}=K(C)$, giving $K_{\max}-K_{\min}+1$ rows. The offset code is $U(s)\gg d$; $s$ is on the retained lattice, giving a node at zero. The last main node need not equal $C$. During construction, monotone searches in nonnegative binary64 bit order locate the first and last representable arguments assigned to each key by the complete rounded mapping. Qualification uses these actual cells, clipped to the admitted domain, rather than ideal real-valued midpoints.

\subsection{Correction encoding and final assembly}
\label{subsec:correction}
Write $G=\RN(F_n(r_i))$, $G^+=U^{-1}(U(G)+1)$ and $u_G=\ulp(G)$. The encoder forms $\delta=(G^+-F_n(r_i))/u_G$ and chooses
\begin{equation}
 c=\begin{cases}\operatorname{RN}_{\rm int}(256(\delta-1/2)),&\delta\leq1,\\128+\operatorname{RN}_{\rm int}(128(\delta-1)),&\delta>1,\end{cases}
\end{equation}
where integer rounding is nearest, ties-to-even. Codes are restricted to $0\leq c\leq192$. Directed bounds must agree on both the rounded coefficient and the code; unresolved cases are not accepted. The construction also checks positivity, a biased coefficient exponent between 54 and 2046, and a finite $G^+$.

The decoded correction is assembled without floating-point conversion:
\begin{equation}
 \widehat\Delta=U^{-1}\!\left[(U(G)\mathbin{\&}\mathtt{0x7ff0000000000000})-(53\mathbin{\ll}52)+(c\mathbin{\ll}45)\right].
\end{equation}
The additions are integer additions, not bitwise OR: the high code bit contributes to the exponent. The exponent is taken from the ordinary $G$, including when $G^+$ crosses a binade boundary. The lowest stored order is prebiased offline to $G^+$ because it cannot occur as a nonconstant coefficient of any supported output. Its original bits are recovered by subtracting one for decoding. Other coefficients retain ordinary rounded values; their final high parts are incremented at evaluation.

For each needed $j$, form $h_j=\RN(h\,\RN(1/j))$. The nonconstant Horner chain uses ordinary rounded coefficients and explicit FMAs. If its final accumulator is $q_n$, output assembly is
\begin{equation}
 t_n=\FMA(h,q_n,-\widehat\Delta_n),\qquad v_n=\RN(G_n^++t_n).
\end{equation}
The outer addition is a separate rounding. Arithmetic fences and explicit CUDA rounding intrinsics preserve this organization; it must not be replaced by a single FMA with $G_n^+$ as the addend.

\subsection{Tail normalization and scaling}
For a tail argument, let $e$ be its unbiased binary exponent, $k=\lfloor e/2\rfloor$, and retain the significand with exponent $e\bmod2$. This gives $x=4^ku$ with $1\leq u<4$. The tail grid retains $b_t$ fraction bits of $u$ using the same half-up quantizer. Its local key origin is $1023\,2^{b_t}$ and its reference nodes include 4: arguments immediately below 4 can round to that terminal node. The displacement is $\RN(u_i-u)$.

The shared polynomial approximates $A_n(u)$ in the tail and is scaled by $2^{-(2n+1)k}$. Tail correction encoding uses exactly the main-grid rule with $A_n(u_i)$ substituted for $F_n(r_i)$. For a positive normal polynomial result with biased exponent $E_v$, if $(2n+1)k<E_v$, integer subtraction of $((2n+1)k)\ll52$ from its representation performs exact normal scaling. CUDA otherwise uses \texttt{ldexp} for subnormal and zero results. The SIMD helper \texttt{Scale} in \path{include/fast_boys/detail/simd/ops-inl.h} returns immediately if all lanes remain normal; otherwise it uses Highway's \texttt{MulByPow2} with the signed negative scaling exponent. Gradual underflow and nearest rounding are essential on that path. Main-domain preparation sets $k=0$, making scaling an identity.

\subsection{Shared table and output stages}
Main rows and tail rows occupy one contiguous allocation with the same row width. Tail indices are offset by the main-row count. Each correction word contains up to eight bytes, with lower output orders in lower bits, and precedes the associated coefficient group. Coefficients needed only for nonconstant Taylor terms follow without additional correction words. For the $[0,4]$ model, one word holds five correction bytes and three padding bytes before the coefficient sequence. The 1 MiB limit applies to the combined main and tail allocation of each model, including padding, not to each domain separately or all models collectively.

CUDA emits straight-line stages in ascending output order, loading each coefficient at first use and returning when the thread's requested endpoint is reached. There is no warp-wide endpoint reduction or dynamically indexed coefficient array. SIMD output bounds are compile-time constants and results are written through restricted vector references. Native and software gather select the same reference data and numerical algorithm.

\section{Reference construction and numerical qualification}
\label{sec:construction}
\subsection{Reference arithmetic}
The frozen producer records Python 3.12.15, gmpy2 2.3.2, GMP 6.3.0 and MPFR 4.2.2. MPFR directed rounding supplies enclosures rather than relying on agreement of two floating-point approximations. Binary64 coefficients are accepted only when enclosure endpoints round to the same value; correction codes require the analogous endpoint agreement. Precision is increased from 256 through 4096 bits for unresolved decisions. Failure to resolve a decision is distinguished from a demonstrated numerical failure.

For nonnegative $x$, the positive-term representation used by the producer is
\begin{equation}
 F_n(x)=e^{-x}\sum_{j=0}^{\infty}t_j,\qquad t_0=\frac1{2n+1},\qquad t_{j+1}=t_j\frac{2x}{2n+2j+3}.
\end{equation}
After summing through $t_J$, set $r_J=2x/(2n+2J+3)$. Subsequent ratios decrease, so, when $r_J<1$, the unsummed series is bounded by $t_Jr_J/(1-r_J)$. Lower and upper terms and partial sums are propagated with MPFR rounding toward minus and plus infinity, respectively. The upper ratio, a downward-rounded $1-r_J$, and upward division give a conservative remainder $R_J$. At precision $q$, summation stops when $R_J\leq S_J^+2^{-q+16}$, with the right-hand side rounded downward. Multiplying the lower partial sum and the upper partial sum plus $R_J$ by directed bounds on $e^{-x}$ encloses $F_n(x)$. Reaching the cap of 100000 term-recurrence steps without satisfying this test fails the reference calculation; it does not accept the last partial sum. The implementation is \path{data/records/numerical/paired-1ulp/producer/workflow/additive/reference.py}, functions \texttt{enclosure} and \path{reference_sequence}. At $x\geq\max(n+1,q+2048)$, the direct helper instead bounds the complete integral and subtracts $2^{-(q+2048)}$ from its lower endpoint.

This mathematical enclosure is separate from the candidate evaluator. The producer's \texttt{rn64}, \texttt{add64}, \texttt{mul64} and \texttt{fma64} functions use exact rational operands followed by one ties-to-even binary64 rounding per specified operation; FMA rounds the exact product plus addend only once. Thus qualification reproduces the candidate's rounding sequence rather than evaluating its polynomial in MPFR and rounding only the final answer.

Local Taylor interval references accelerate qualification. The helper \path{taylor_reference} uses 32 terms (degree 31) for the highest required order $m$. At center $r$, its truncation error is at most
\begin{equation}
 \frac{|x-r|^{32}}{32!}F_{m+32}(\min(x,r))
 \leq\frac{|x-r|^{32}}{32!}\min\left\{\frac1{2(m+32)+1},A_{m+32}(\min(x,r))\right\}.
\end{equation}
At a zero left endpoint the first bound suffices. Directed coefficient intervals and directed Horner operations include reference-data and evaluation uncertainty, and the truncation bound is added outward. Lower orders are obtained by directed downward recurrence with an exponential enclosure. Ambiguous main-domain and near-switch error decisions return to direct MPFR refinement. The auxiliary function is constructed from $A_0(u)=\sqrt{\pi}/(2\sqrt u)$ and $A_{n+1}(u)=(2n+1)A_n(u)/(2u)$ with directed bounds; these are offline reference operations, not runtime coefficient reconstruction.

Tail qualification compares the scaled polynomial to the Boys function, not merely to $A_n$. Let $a_n=\max(n-1/2,0)$. For $y\geq1$, $y^{n-1/2}\leq\exp(a_n(y-1))$, by $\log y\leq y-1$. Consequently,
\begin{equation}
 B_n(x)=A_n(x)-F_n(x)=\frac12\int_1^\infty y^{n-1/2}e^{-xy}\,dy\leq\frac{e^{-x}}{2(x-a_n)},\qquad x>a_n.
\end{equation}
The implementation's condition $x\geq n+1$ is sufficient. For its supported far-tail reference range $0\leq n\leq128$ and original input $x>10^7$, the denominator exceeds one and $e^{-x}<2^{-x}<2^{-10000000}$, justifying the fixed conservative bound. This threshold concerns $x$, not the normalized coordinate $u\in[1,4)$.

If $x=4^ku$, $P=(2n+1)k$, and $[A_L(u),A_U(u)]$ encloses $A_n(u)$, the reference enclosure is
\begin{equation}
 \left[\max\{0,2^{-P}A_L(u)-B_{\rm up}(x)\},\;2^{-P}A_U(u)\right].
\end{equation}
The auxiliary interval is scaled first, then the original-variable omitted-integral bound is subtracted with outward rounding. Near the switch the producer can instead use direct Boys enclosures. The omitted-integral bound controls one contribution, not the complete floating-point error.

\subsection{Joint search and retained choices}
The completed configuration considers main-grid bit counts $b_g=4,\ldots,12$ and common polynomial ceilings $p=1,\ldots,12$. Tail-bit candidates are enumerated from zero through 20, stopping when even the minimum-degree tail plus one main row exceeds the budget. Each tail grid has $2^{b_t+1}+1$ nodes. The stored metadata identify the selected candidates; nominal combinations are not reported as evaluated-candidate counts.

For each grid pair, degree ceilings are tried in increasing order. Sparse tail screening precedes complete tail qualification and switch determination. The maximum of the per-output switches defines $C$. After accounting for the tail allocation, the main grid takes the smallest lattice offset whose row count fits the remaining capacity. This first-fit operation concerns storage: the producer checks monotonic row-count geometry before bisection and has an exact scanning fallback when that check fails.

The main candidate is screened and then qualified at the common ceiling. Individual degrees are then tested from one upward, requiring acceptance in both domains. The initial main grid, tail grid and common switch remain fixed during this reduction; a smaller tail degree is accepted only if its qualified switch is no later than $C$.

With these degrees fixed, numerical coarsening increases the main offset code by exponentially growing distances until a tested failure brackets a passing point. Bisection narrows this bracket, and a neighborhood extending four codes around the final bracket is tested. The largest tested passing offset is retained. Numerical acceptance is not assumed monotone, so this is a bounded heuristic, not proof of a globally largest admissible offset. The illustrative main-text workflow summarizes this strategy rather than these individual probes.

Candidates are ranked lexicographically by maximum degree, sum of degrees, combined bytes, largest recorded main qualification error, degree vector, tail bits, main bits and offset code. Earlier numerical decisions are not retuned using molecular performance. The $10^7$ value is a tail-qualification anchor, not the final main-domain endpoint.

\subsection{Sampling, refinement and acceptance}
\label{subsec:qualification}
Cheap main screening selects up to 17 evenly distributed row indices and uses four subdivisions per selected cell. Full main qualification visits every rounded-index cell $[a,z]$ with 16 subdivisions. Interior samples are formed as the binary64 expression $a+(z-a)(j/16)$, $j=1,\ldots,15$, rounding each subtraction, division, multiplication and addition separately. Endpoints, the reference when inside the cell, and neighbors of endpoints and the reference in both representable directions are included subject to the domain limits; duplicates are removed. At refinement level $\ell=0,\ldots,3$, the current output's global worst witness $w$ is probed at $w\pm(z-a)/(16\,2^{\ell+1})$ and at \texttt{nextafter}$(w,0)$ and \texttt{nextafter}$(w,+\infty)$, retaining admissible inputs. This is not refinement in every main cell. The retained witness is reevaluated directly at at least 512 bits.

Tail screening uses row indices $0$, $M-1$, $\lfloor(M-1)/2\rfloor$ and $\lfloor(M-1)i/16\rfloor$, $i=0,\ldots,16$, where $M$ is the tail-row count, with duplicates removed. Five binary64 points $a+(z-a)j/4$, $j=0,\ldots,4$, are tested at $x=2^{24}u$ (normalization exponent $k=12$). Full tail sampling includes endpoints, reference nodes below 4, and $\RN(a+(z-a)j/17)$, $j=1,\ldots,16$, with each interior expression formed as an exact rational before its single rounding. Normalized forms of the anchor, its neighbors and the largest finite input are also included. Four refinement rounds bisect the adjacent sample intervals around each cell's worst point, again rounding the exact rational midpoint once.

For each normalized $u$ and output $n$, let $k_{\min}$ be the smallest exponent admitting $x=4^ku$ above the qualification minimum. The far-tail checks include $k_{\min}$, $511$, and all $k\in[k_{\min},511]$ for which $-1077\leq\lfloor\log_2 A_U(u)\rfloor-(2n+1)k\leq-1020$. These classes cover normal/subnormal and rounded-zero transitions; other normal classes use scaling equivalence. Exact sample-generation functions are supplied in \texttt{qualify.py}, \texttt{tail.py} and \texttt{paired.py} under the frozen producer directory.

Switch discovery tests normalized samples scaled through $k=0,\ldots,11$ over $[1,10^7]$. After finding the rightmost sampled failure, four rounds add 16 rational subdivision points between that failure and its next passing point. The retained switch and its immediate successor must pass. The main branch includes the switch; the tail branch starts strictly above it. A passing suffix of this sampled set is not a certificate for every intervening binary64 argument.

Main-domain and near-switch decisions accept an error upper bound at most one ULP, reject a lower bound above one ULP, and adaptively refine ambiguous references. The far-tail qualifier instead screens conservative 256-bit upper bounds and rejects a candidate when a sampled upper bound exceeds the target; its retained worst witness is separately reevaluated at 512 bits. Such a conservative rejection alone does not prove that the true error exceeds one ULP. Unresolved decisions do not replace accepted candidates. Per-model records retain sample counts, witnesses and the largest reported upper bound on sampled ULP error.

\Needspace{32\baselineskip}
\subsection{Selected parameters}
\Needspace{32\baselineskip}
\begin{longtable}{llllllll}
\caption{Selected models: $H=\max_{0\leq n\leq N}(n+p_n)$ is the highest stored reference order. Storage includes coefficients, corrections and padding; full vectors and hexadecimal parameters are in data/models.csv.}\label{tab:models}\\
\toprule
$N$ & $H$ & $b_g$ & $b_t$ & Main rows & Tail rows & Row bytes & Total bytes \\ \midrule
\endfirsthead
\toprule
$N$ & $H$ & $b_g$ & $b_t$ & Main rows & Tail rows & Row bytes & Total bytes \\ \midrule
\endhead
\midrule\multicolumn{8}{r}{Continued on next page}\\\endfoot
\bottomrule\endlastfoot
0 & 4 & 10 & 10 & 3272 & 2049 & 48 & 255408 \\
1 & 5 & 11 & 11 & 4931 & 4097 & 56 & 505568 \\
2 & 6 & 11 & 11 & 6512 & 4097 & 64 & 678976 \\
3 & 7 & 11 & 11 & 8324 & 4097 & 72 & 894312 \\
4 & 9 & 11 & 11 & 6861 & 4097 & 88 & 964304 \\
5 & 10 & 11 & 11 & 5659 & 4097 & 96 & 936576 \\
6 & 11 & 11 & 11 & 5849 & 4097 & 104 & 1034384 \\
7 & 12 & 10 & 10 & 3941 & 2049 & 112 & 670880 \\
8 & 13 & 10 & 10 & 4123 & 2049 & 128 & 790016 \\
9 & 14 & 11 & 10 & 5365 & 2049 & 136 & 1008304 \\
10 & 16 & 10 & 10 & 4174 & 2049 & 152 & 945896 \\
11 & 17 & 10 & 10 & 4207 & 2049 & 160 & 1000960 \\
12 & 18 & 10 & 10 & 4188 & 2049 & 168 & 1047816 \\
13 & 19 & 10 & 10 & 3685 & 2049 & 176 & 1009184 \\
14 & 20 & 10 & 10 & 3254 & 2049 & 184 & 975752 \\
15 & 21 & 10 & 10 & 3408 & 2049 & 192 & 1047744 \\
16 & 22 & 10 & 10 & 2991 & 2049 & 208 & 1048320 \\
17 & 23 & 10 & 10 & 2712 & 2049 & 216 & 1028376 \\
18 & 24 & 10 & 10 & 2166 & 2049 & 224 & 944160 \\
19 & 25 & 10 & 10 & 2323 & 2049 & 232 & 1014304 \\
20 & 26 & 11 & 10 & 2027 & 2049 & 240 & 978240 \\
21 & 27 & 12 & 10 & 2080 & 2049 & 248 & 1023992 \\
22 & 29 & 9 & 9 & 2731 & 1025 & 264 & 991584 \\
23 & 30 & 9 & 9 & 2741 & 1025 & 272 & 1024352 \\
24 & 31 & 10 & 9 & 2341 & 1025 & 288 & 969408 \\
\end{longtable}

The accompanying \path{data/models.csv} supplies degree vectors, exact hexadecimal $C$ and $s$, both resolutions, row counts, padding and domain-specific byte totals. The summary under \path{data/records/numerical/paired-1ulp/} retains qualification witnesses and sample counts. Canonical numerical payloads and emitted arrays have separate hashes because their serialized layouts differ; both already contain the lowest-order prebias. The README identifies the corresponding serializers and manifests.

\section{Details of the numerical illustrations}
\label{sec:illustrations}
\subsection{Matched reference grids}
\label{subsec:grid-illustration}
The comparison uses the $[0,0]$ main model, degree four and 3272 references. Its retained fraction count is 10, offset is \texttt{0x1.0300000000000p+2}, and endpoint is \texttt{0x1.146b4b4e3f86dp+5}. The uniform comparison places exact rational nodes at $r_i=iC/3271$, including both endpoints. Both grids use the same number of coefficient rows and degree; the tail is excluded.

At each exact adjacent-center midpoint, both neighboring mathematical Taylor polynomials are evaluated and the larger absolute relative remainder is retained. Domain endpoints are also evaluated. This illustration uses ideal midpoint cells, unlike the rounded-index cells used for model qualification. Coefficients are mathematical Boys values, not rounded table entries, and no correction decoding or runtime evaluation rounding is included.

The calculation uses MPFR analytic $F_0$ and high-precision upward recurrence for its Taylor coefficients. Results at 256 and 512 bits agree to relative tolerance $10^{-10}$; this precision-consistency test is distinct from a directed enclosure proof. The recorded plotting environment is gmpy2 2.3.1 with MPFR 4.2.2. The mathematical derivative remainder bound provides an additional check. Maximum relative remainders are $6.48463556794868\times10^{-17}$ and $3.11676804346875\times10^{-15}$ for nonuniform and uniform grids, respectively.

\begin{figure}[!htbp]\centering
\includegraphics[width=.9\linewidth]{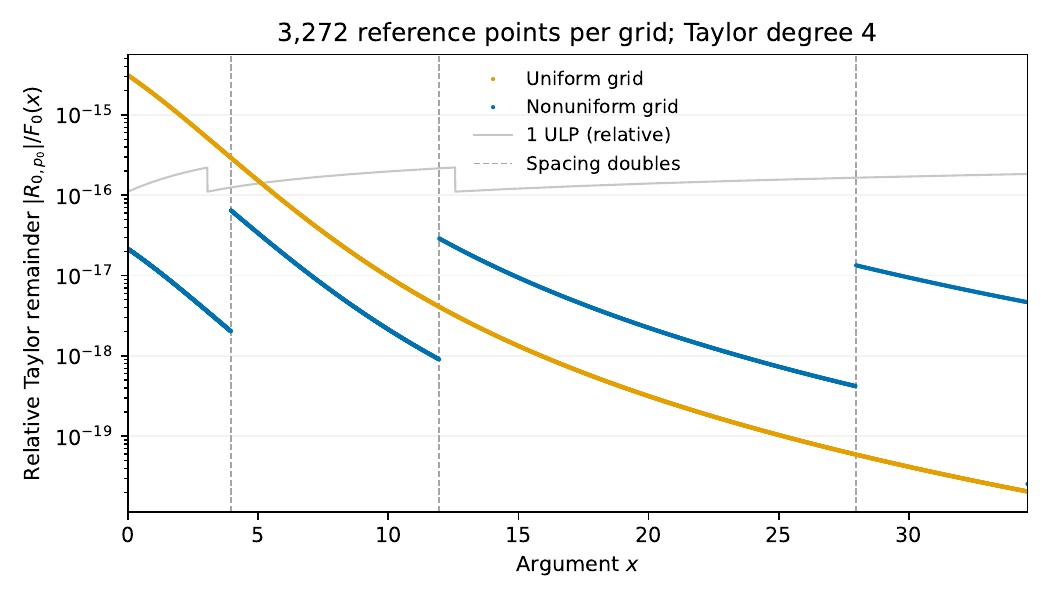}
\caption{Reproduction of the main-text matched-grid illustration. Gray denotes $2^{\lfloor\log_2F_0(x)\rfloor-52}/F_0(x)$; dashed lines mark $x+s=8,16,32$, or $x=3.953125,11.953125,27.953125$. Exact zero remainders are omitted on the logarithmic axis. This is not a measurement of evaluator error or performance.}
\end{figure}

\subsection{Upward-recurrence passing windows}
Correctly rounded binary64 $F_0$ and $E=e^{-x}$ initialize the independent experiment. It forms $t=\RN(x+x)$, $\rho=\RN(1/t)$ and $q=\RN(E\rho)$ once, then evaluates
\begin{equation}
 a_n=\RN((2n-1)\rho),\qquad \widehat F_n=\FMA(a_n,\widehat F_{n-1},-q).
\end{equation}
The exponential may correctly underflow to zero. No approximate reciprocal or asymptotic replacement is used. Reference intervals use MPFR-generated degree-12 Taylor expansions of width $1/16$, binary128 evaluation and direct MPFR refinement through 4096 bits for ambiguous rounding or gates.

Inputs are $x_i=\RN(i/10^6)$, $i=1,\ldots,10^9$. For $N=1,\ldots,12$ and tolerances 1, 2, 4, 8 and 16 ULP, every output $1,\ldots,N$ must pass at one million consecutive inputs. The window spans 0.999999 in $x$. Shards of 250000 inputs retain failure boundaries; joining passing prefixes and suffixes locates the earliest complete window. No complete one-ULP window was found. Missing windows remain explicit in the machine-readable records.

\begin{figure}[H]\centering
\includegraphics[width=.85\linewidth]{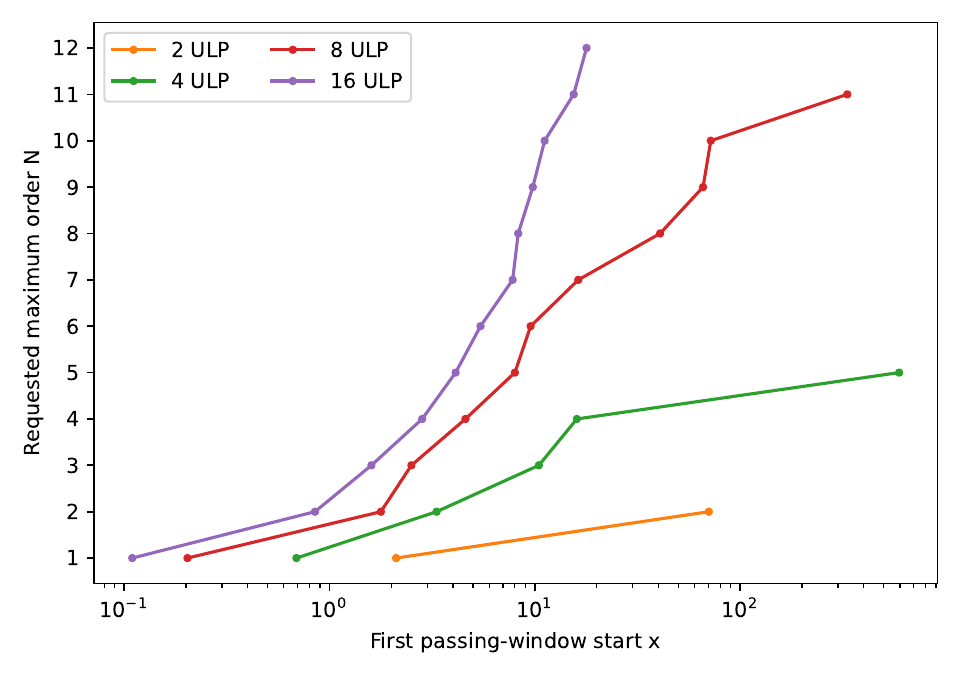}
\caption{Reproduction of the main-text upward-recurrence illustration: the first complete passing-window start, not a boundary guaranteeing acceptance at every larger argument. Full tolerance/order coverage, including missing windows, is retained in the accompanying diagnostic records.}
\end{figure}

\section{Independent function-level accuracy assessment}
\label{sec:accuracy}
For each requested range $[0,N]$, $N=0,\ldots,12$, the completed scans evaluate all outputs on $x_i=\RN(i/10^6)$, $i=0,\ldots,10^9$. Inputs are formed from the integer index and denominator, not by repeated addition of a rounded step. References are evaluated at these actual binary64 inputs. This validation range is separate from the constructed range $N\leq24$.

Throughout this supplement, spacing is defined by
\begin{equation}
\ulp(y)=\begin{cases}2^{-1074},&0\leq y<2^{-1022},\\2^{\lfloor\log_2y\rfloor-52},&y\geq2^{-1022},\end{cases}
\end{equation}
for finite nonnegative $y$. Accuracy measurements use $u=\ulp(\RN(F_n(x)))$, although the reference is not rounded before subtracting the computed value. For $L>0$ and $[L,U]\ni F_n(x)$, a finite computed $v$ has error enclosures
\begin{align}
a_-&=\max(0,L-v,v-U),&a_+&=\max(|v-L|,|v-U|),\\
e_{\rm abs}&\in[a_-,a_+],&e_{\rm rel}&\in[a_-/U,a_+/L],&e_{\rm ULP}&\in[a_-/u,a_+/u].
\end{align}
All bounds are evaluated outward. The scanner rounds the interval endpoints to binary64 and checks agreement of their binary exponents (\texttt{ilogb}); a disagreement triggers a direct 512-bit reference and a repeated check. It requires a positive lower bound and resolved spacing, not agreement of every rounded significand bit. Thus the dense reference intervals need not certify a unique binary64 value, unlike the molecular reference backend.

The scan reference uses 4000 cells of width $1/4$ on $[0,1000]$, with midpoint centers $r=a+1/8$. Each order stores 25 coefficient intervals for the degree-24 Taylor polynomial, constructed by directed 256-bit MPFR evaluation and outward conversion to binary128. With $\delta=x-r$, the remainder is bounded by
\begin{equation}
 |R_{n,24}(x)|\leq F_{n+25}(a)\frac{|\delta|^{25}}{25!},
\end{equation}
where $a$ is the cell's left endpoint. The upper factor is separately tabulated with outward rounding. Binary128 interval Horner evaluation propagates coefficient and arithmetic uncertainty; the remainder is added in both directions. The final endpoint uses the last cell, and $x=0$ uses directed bounds on $1/(2n+1)$. This configuration is independent of the degree-12 reference used for the recurrence illustration. Source and build identities are indexed in Sec.~\ref{sec:artifact-map}.

Retained witnesses are reevaluated directly at 512 bits. Per-shard sums are bounded in binary128; merging uses directed 80-digit decimal arithmetic. The scan has 1000 million-input shards and a final one-input shard, processed in batches of up to 4096. Coverage checks reject missing, duplicated or incompatible shards. The retained lower and upper bounds on maxima and sums are provided separately; the largest reported upper bound need not identify the unique true maximizer when reference intervals overlap.

Each returned value has equal weight. Pooling all requested ranges counts $91\times1\,000\,000\,001$ values per backend and mode. Repeated physical orders in different requested-range tests are not deduplicated. Pooled means are sums divided by counts, not unweighted averages of the 13 range means. Complete per-order and pooled records retain both bounds, counts and exact witness inputs and results.

CPU strict adapters use \texttt{-fno-fast-math} and \texttt{-ffp-contract=off}; fast adapters use \texttt{-ffast-math} and \texttt{-ffp-contract=fast}. CUDA strict adapters use \path{--fmad=false}, and fast adapters use \path{--use_fast_math}. Explicit FMAs remain fused. The collector is strict, links without fast-math startup objects, and checks RN-even, disabled FTZ/DAZ and masked traps after adapter calls. These checks concern the host environment; CUDA device arithmetic follows its adapter compilation mode. The mixed-layout pilot exercises partial vectors/blocks and mixed-domain inputs separately from the dense adjacent-layout scan. These adapters are not the molecular timing executables.

CPU coverage includes original Simint, FABO, native LibintX SIMD, and Highway ports of VSR and Laikov. CUDA coverage includes original QUICK, FABO, LibintX, VSR and GANSU. Original QUICK and Simint routines are included directly. LibintX CPU output storage is initialized as required by its accumulation interface. The accompanying build records identify exact dependency revisions, adapter commands and source hashes; each library is assessed under those settings rather than assigned FABO's construction target.

FABO's largest reported upper bound on sampled ULP error is 1.0669248898976357 at $N=n=3$, $x=$\texttt{0x1.6b0ff5c6c11a1p+5}, for both platforms and both arithmetic modes. This slightly exceeds the construction target. Equal reported maxima do not imply bitwise equality at all other inputs. The QUICK $N=12$ witness near $x=0.100019$ belongs to a function test extending beyond the $N\leq8$ molecular workloads.

\Needspace{29\baselineskip}
\begin{longtable}{lllllll}
\caption{Largest reported upper bounds on sampled ULP error by implementation and arithmetic mode. Exact witnesses and all error measures are in data/maximum-witnesses.csv.}\label{tab:witnesses}\\
\toprule
Platform & Mode & Backend & $N$ & $n$ & $x$ & ULP upper \\ \midrule
\endfirsthead
\toprule
Platform & Mode & Backend & $N$ & $n$ & $x$ & ULP upper \\ \midrule
\endhead
\midrule\multicolumn{7}{r}{Continued on next page}\\\endfoot
\bottomrule\endlastfoot
cpu & strict & FABO & 3 & 3 & 45.382793 & 1.066925 \\
cpu & strict & FABO (software) & 3 & 3 & 45.382793 & 1.066925 \\
cpu & strict & Laikov & 12 & 12 & 0.269971 & 329.2376 \\
cpu & strict & LibintX & 8 & 8 & 36.000000 & 7.804764e+07 \\
cpu & strict & Simint & 12 & 7 & 36.500000 & 1.185666e+10 \\
cpu & strict & VSR & 12 & 12 & 29.491339 & 1.298784e+12 \\
cpu & fast & FABO & 3 & 3 & 45.382793 & 1.066925 \\
cpu & fast & FABO (software) & 3 & 3 & 45.382793 & 1.066925 \\
cpu & fast & Laikov & 12 & 12 & 0.269971 & 329.2376 \\
cpu & fast & LibintX & 8 & 8 & 36.000000 & 7.804764e+07 \\
cpu & fast & Simint & 12 & 7 & 36.500000 & 1.185666e+10 \\
cpu & fast & VSR & 12 & 12 & 29.491339 & 1.298784e+12 \\
cuda & strict & FABO & 3 & 3 & 45.382793 & 1.066925 \\
cuda & strict & GANSU & 12 & 12 & 29.491339 & 1.298784e+12 \\
cuda & strict & LibintX & 11 & 11 & 18.581440 & 39.33279 \\
cuda & strict & QUICK & 12 & 12 & 0.100019 & 3.407894e+22 \\
cuda & strict & VSR & 12 & 12 & 29.491339 & 1.298784e+12 \\
cuda & fast & FABO & 3 & 3 & 45.382793 & 1.066925 \\
cuda & fast & GANSU & 12 & 12 & 29.491339 & 1.298784e+12 \\
cuda & fast & LibintX & 11 & 11 & 18.583692 & 39.44849 \\
cuda & fast & QUICK & 12 & 12 & 0.100055 & 3.376171e+22 \\
cuda & fast & VSR & 12 & 12 & 29.491339 & 1.298784e+12 \\
\end{longtable}

\begin{figure}[H]\centering
\includegraphics[width=\linewidth]{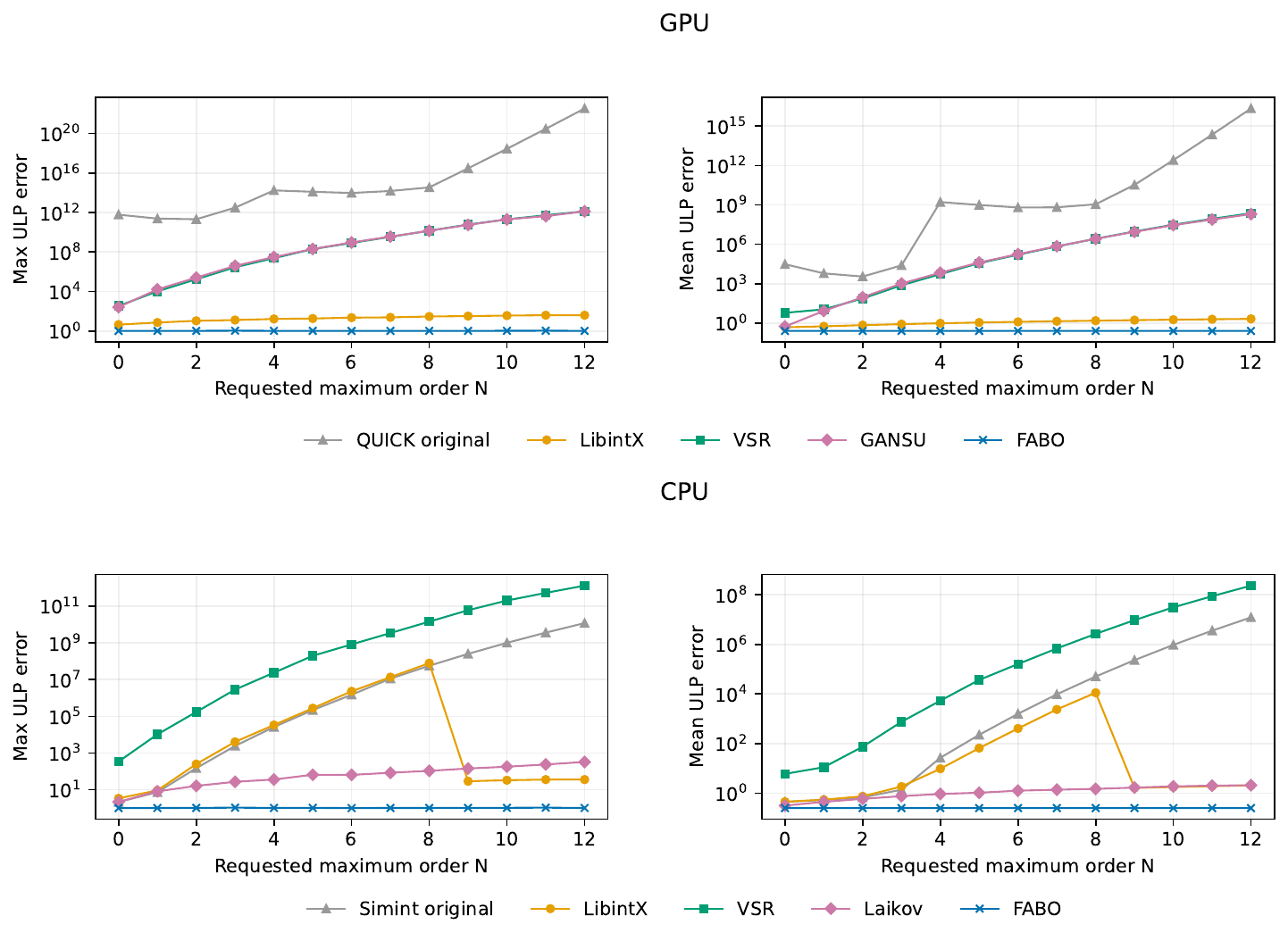}
\caption{Supplementary fast-math accuracy results from the completed scans. Values are upper bounds on sampled errors, not continuous-domain bounds. Native and software gather are policies of the same FABO algorithm; the complete CPU records include both.}
\end{figure}
\FloatBarrier

\section{Application integrations and benchmark setup}
\subsection{Platforms and compilation}
\label{subsec:platforms}
The GPU measurements use an NVIDIA GeForce RTX 5090 D v2, NVCC 13.4.92 and target \texttt{sm\_120}, retaining QUICK's \texttt{-O2} and \path{--use_fast_math} build options. The CPU measurements use one thread on an Intel Xeon Gold 6526Y, GCC 14.4.0, \texttt{-O3 -ffast-math -march=sapphirerapids}, and AVX-512. Intel MKL operates with one thread and automatic dispatch. These application flags are distinct from the standalone validation modes in Sec.~\ref{sec:accuracy}; instruction-inspection probes do not define application build settings.

QUICK's two-electron energy-integral Boys call sites are replaced in separately compiled executables. Screening, contraction, integral recurrences and accumulation are unchanged. The ordinary s/p and s/p/d paths select FABO $[0,4]$ and $[0,8]$ models, and each thread stops at its actual requested endpoint. \texttt{ENABLEF} is off. NWChem substitutions occur in Simint's generated integral kernels, whose requested endpoint is known at compile time and selects the corresponding model. The comparative algorithms and applications are described in their original publications.\cite{Cruzeiro2021AMBERQUICK,Pritchard2016Simint,Asadchev2023LibintX,VikhamarSandberg2026Boys,Laikov2025Boys,Ito2026GANSU}

\begin{table}[htbp]
\centering
\caption{Recorded compilation and execution distinctions. Full commands and loaded-library identities accompany the data.}
\label{tab:build-settings}
\begin{tabular}{lll}\toprule
Artifact & Target/compiler & Arithmetic policy \\\midrule
QUICK timing & SM120 / NVCC 13.4.92 & O2, CUDA fast math \\
NWChem integral timing & AVX-512 / GCC 14.4.0 & O3, fast math \\
CPU accuracy adapters & Sapphire Rapids / GCC & strict and fast separately \\
CUDA accuracy adapters & SM120 / NVCC & contraction off / fast math \\
Accuracy collector & host CPU & strict, RN-even, FTZ/DAZ off \\\bottomrule
\end{tabular}
\end{table}
The CPU application flags also contain \texttt{-fopenmp-simd}; this enables SIMD directives and does not establish multithreaded execution. The archived linkage uses sequential MKL. The recorded runtime is Intel oneAPI MKL 2026.1 (build 20260612), one thread, with automatic dispatch loading \path{libmkl_avx512.so.3}; \path{MKL_CBWR} is not used to force a lower instruction set. CUDA validation ran with driver 615.71.09 on a Ryzen 9 7950X host; the host reference arithmetic and device adapter arithmetic are distinct parts of that experiment.

\subsubsection{Source and build mapping}
\label{sec:artifact-map}
QUICK 26.03 is pinned to revision \texttt{730af1533da8}; NWChem 7.3.1 to \texttt{e2869a2c8144}; and the Simint generator to \texttt{38b24bb040c6}. The CPU comparison records LibintX \texttt{dd5bb8b9536c} and Highway \texttt{9ca61efcab7f}, with the VSR/Laikov SIMD ports from snapshot \texttt{728ec72}. These are dependency identities, not a single revision assigned to all observations. QUICK links its bundled BLAS and LAPACK; NWChem links sequential ILP64 MKL. Full compiler commands, dependency hashes and loaded-library records are retained in the following package-relative locations.

The GPU GANSU adapter uses release \texttt{v2026.8.10}; the VSR coefficients and supplied implementation are attributed to the authors' supplementary collection, DOI \texttt{10.60893/figshare.jcp.c.8390929}. Their source provenance is retained under \path{data/application-inputs/comparators/}. The adapter and coefficient headers match the hashes in the completed CUDA accuracy build; these identities distinguish them from the separately archived CPU SIMD ports.

\begin{description}
\item[QUICK timing binaries.] \path{data/application-inputs/quick/variants/} contains variant recipes and the FABO \path{build.json} and \path{compile_commands.json}; \path{data/records/benchmarks/quick-fixed-density-20261106/reuse.json} maps the final run to executable, fixture and reference identities.
\item[NWChem timing binaries.] \path{data/application-inputs/nwchem/built.json} records the six numerical binaries (including the untimed reference), flags, compiler versions and linkage; \path{inputs.json} in that directory hashes their source inputs. \path{data/records/benchmarks/nwchem-fixed-density-20261106/config.json} records pinned dependencies and execution settings.
\item[Function-accuracy adapters and collector.] Under \path{data/records/benchmarks/function-accuracy-20261107/}, each of \path{cpu/} and \path{cuda/} contains \path{built.json}, \path{run.json} and \path{scan.build.json}. The adapter commands and hashes are separate from the strict host collector's command and reference-table hash. Matching archived source files are supplied under \path{data/methods/function-accuracy/}.
\item[Fixtures, references and reductions.] Each application's \path{reuse.json} identifies reused initial states and accepted reference outputs. The QUICK \path{molecular-means.manifest.json} and NWChem \path{matrix-absolute.manifest.json} authenticate the later reductions. \path{data/method-sources.csv} indexes the reference and reporting code supplied with this supplement.
\end{description}
All final timed observations were newly measured. The previously verified sizing decisions, initial states and correctly rounded reference outputs were reused as identified by those manifests. Reporting scripts subsequently reduced these observations; their identities are kept separate from application builds and numerical-model generation.

\subsection{Frozen initial states and workloads}
\label{subsec:workloads}
QUICK 26.03 prepares its default superposition-of-atomic-densities guess once per case. Saved $P_0$, $H_{\rm core}$, overlap and orthogonalizer are replayed before each update. NWChem prepares an atomic guess, replays orbitals, occupations, overlap and orthogonalizer, and verifies reconstructed input-density and core-Hamiltonian hashes. Initialization and fixture loading are untimed. Every run builds $F[P_0]$, diagonalizes, and constructs $P_1$; it is not a convergence trajectory. Normal integral screening is retained, and the inputs to it are fixed across backends.

Workloads use equilibrium S66 systems 1, 17, 24 and 34: water, uracil, benzene and pentane dimers.\cite{Rezac2011S66} Bases are 6-31G*, def2-SVP and cc-pVDZ.\cite{Hehre1972Basis,Hariharan1973Polarization,Weigend2005Basis,Dunning1989Basis} Both applications perform neutral, closed-shell singlet RHF energy calculations, without density fitting. QUICK uses Cartesian basis functions and direct GPU integrals without molecular-symmetry reduction. NWChem uses spherical, segmented bases, direct integrals and explicit $C_1$ symmetry (\texttt{noautosym noautoz}); its saved inputs disable the alternative Texas and native integral routes and require replicated Fock construction with Simint. The patched single-update entry point rejects a reduced-rank NWChem state. The accepted dimensions are listed below; they are not inferred from basis names.

The QUICK input decks specify a density RMS criterion of $10^{-6}$ but the fixed-state hook bypasses the convergence loop. In the pinned source, these decks leave the two-electron cutoff at $10^{-7}$, primitive cutoff at $10^{-8}$ and density cutoff at $10^{-10}$; neither coarse nor tight integral options are selected. These values are uploaded by \texttt{scf\_operator}. The one-electron prescreening cutoff is $10^{-12}$ and overlap cutoff is $10^{-5}$. The supplied source excerpts retain the associated initialization and keyword parsing. NWChem retains its density-dependent screening, with \texttt{tol2e}$=10^{-7}$ and \texttt{dentolmax}$=10^{-6}$ for these full-rank inputs. Its replicated build computes the integral-only threshold as $\min(10^{-6},10^{-7}/D,10^{-7}/D^2)$ for nonzero maximum shell-block RMS density $D$, and $10^{-7}$ when $D=0$. Screening therefore depends on the fixed density, not just one universal threshold.

Copies are translated along the shortest bounding-box axis by the dimer extent on that axis plus 3.4 Angstrom. Internal geometry is unchanged. Workload sizing uses the original backend and def2-SVP, selecting the first configured copy count reaching a two-second Fock build. The frozen GPU copy counts are 128, 8, 16 and 16; CPU counts are 8, 1, 1 and 1. The resulting geometry is reused across all three bases and backends, with a distinct initial-state fixture per basis. This sizing is not Boys-parameter tuning.

\Needspace{31\baselineskip}
\begin{longtable}{lllll}
\caption{Accepted workload dimensions. QUICK uses Cartesian functions; NWChem uses spherical functions. The orbital rank equals the basis dimension in these cases.}\label{tab:workloads}\\
\toprule
Platform & Molecule & Basis & AO dimension & Electrons \\ \midrule
\endfirsthead
\toprule
Platform & Molecule & Basis & AO dimension & Electrons \\ \midrule
\endhead
\midrule\multicolumn{5}{r}{Continued on next page}\\\endfoot
\bottomrule\endlastfoot
GPU & $128\times$ water dimer & 6-31G* & 4864 & 2560 \\
GPU & $128\times$ water dimer & def2-SVP & 6400 & 2560 \\
GPU & $128\times$ water dimer & cc-pVDZ & 6400 & 2560 \\
GPU & $8\times$ uracil dimer & 6-31G* & 2048 & 928 \\
GPU & $8\times$ uracil dimer & def2-SVP & 2240 & 928 \\
GPU & $8\times$ uracil dimer & cc-pVDZ & 2240 & 928 \\
GPU & $16\times$ benzene dimer & 6-31G* & 3264 & 1344 \\
GPU & $16\times$ benzene dimer & def2-SVP & 3840 & 1344 \\
GPU & $16\times$ benzene dimer & cc-pVDZ & 3840 & 1344 \\
GPU & $16\times$ pentane dimer & 6-31G* & 3168 & 1344 \\
GPU & $16\times$ pentane dimer & def2-SVP & 4320 & 1344 \\
GPU & $16\times$ pentane dimer & cc-pVDZ & 4320 & 1344 \\
CPU & $8\times$ water dimer & 6-31G* & 288 & 160 \\
CPU & $8\times$ water dimer & def2-SVP & 384 & 160 \\
CPU & $8\times$ water dimer & cc-pVDZ & 384 & 160 \\
CPU & $1\times$ uracil dimer & 6-31G* & 240 & 116 \\
CPU & $1\times$ uracil dimer & def2-SVP & 264 & 116 \\
CPU & $1\times$ uracil dimer & cc-pVDZ & 264 & 116 \\
CPU & $1\times$ benzene dimer & 6-31G* & 192 & 84 \\
CPU & $1\times$ benzene dimer & def2-SVP & 228 & 84 \\
CPU & $1\times$ benzene dimer & cc-pVDZ & 228 & 84 \\
CPU & $1\times$ pentane dimer & 6-31G* & 188 & 84 \\
CPU & $1\times$ pentane dimer & def2-SVP & 260 & 84 \\
CPU & $1\times$ pentane dimer & cc-pVDZ & 260 & 84 \\
\end{longtable}

\path{data/workloads.csv} maps every case to its geometry, basis and dimension; \path{data/application-inputs/} supplies the XYZ files in Angstrom, saved input decks, basis records and sizing decisions. Fixture and replay hashes are retained with each accepted observation. The input decks' nominal SCF iteration limits do not override the patched single-update protocol.

\subsection{Timed boundaries}
QUICK Fock timing brackets \texttt{scf\_operator(.false.)} with \texttt{clock\_gettime(CLOCK\_MONOTONIC)} and CUDA synchronization before each timestamp. It includes density-dependent screening preparation, associated transfers, GPU two-electron work and Fock assembly; the one-electron matrix is prepared beforehand. NWChem's \texttt{jk\_seconds} brackets \texttt{fock\_2e} with \texttt{util\_wallsec()}, isolating the two-electron Coulomb/exchange build. These scopes must not be treated as identical. Boys-table allocation, initialization and host-to-device upload are completed before timing. Matrix serialization and energy diagnostics follow timing. Complete-update timings are archived but are not substituted for either Fock metric.

\section{Individual timings and speedup statistics}
\label{sec:timings}
Each application has 12 cases, five timed backends and five fresh-process repetitions, giving 300 observations and 60 case/backend groups. There is no molecular warmup. Backend order rotates cyclically across repetitions. The accompanying raw timing CSV retains repetition indices; original records and execution provenance are retained separately. A reference-backend runtime is never included as a performance observation.

For times $t_1,\ldots,t_5$, the reported mean is $\bar t=\sum_jt_j/5$ and the sample standard deviation is $[\sum_j(t_j-\bar t)^2/4]^{1/2}$. Both are calculated from the individual observations. Table values are in seconds with three decimal places; machine-readable files preserve full precision.

For a fixed comparator $b$ and case $c$, define $S_{b,c}=\bar t_{b,c}/\bar t_{\mathrm{FABO},c}$. The overall ratio is
\begin{equation}
G_b=\exp\left[\frac1{12}\sum_c\log S_{b,c}\right].
\end{equation}
Percentage speedup is $100(G_b-1)$, not percentage time reduction $100(1-1/G_b)$. The comparator is held fixed across cases; it is not the fastest alternative chosen separately for each case. Recomputed ratios give the main manuscript's 1.22--10.67 percent range.
\Needspace{15\baselineskip}
\begin{longtable}{llll}
\caption{Geometric means of comparator/FABO mean-time ratios over 12 cases.}\label{tab:speedups}\\
\toprule
Platform & Comparator & Ratio & Speedup (percent) \\ \midrule
\endfirsthead
\toprule
Platform & Comparator & Ratio & Speedup (percent) \\ \midrule
\endhead
\midrule\multicolumn{4}{r}{Continued on next page}\\\endfoot
\bottomrule\endlastfoot
GPU & GANSU & 1.080227 & 8.02 \\
GPU & LibintX & 1.016422 & 1.64 \\
GPU & QUICK & 1.106705 & 10.67 \\
GPU & VSR & 1.061879 & 6.19 \\
CPU & Laikov & 1.030355 & 3.04 \\
CPU & LibintX & 1.013004 & 1.30 \\
CPU & Simint & 1.103530 & 10.35 \\
CPU & VSR & 1.012189 & 1.22 \\
\end{longtable}

\begin{longtable}{lllll}
\caption{GPU timings in seconds; arithmetic means and sample standard deviations of five observations.}\label{tab:timings-gpu}\\
\toprule
Molecule & Basis & Backend & Mean (s) & SD (s) \\ \midrule
\endfirsthead
\toprule
Molecule & Basis & Backend & Mean (s) & SD (s) \\ \midrule
\endhead
\midrule\multicolumn{5}{r}{Continued on next page}\\\endfoot
\bottomrule\endlastfoot
$128\times$ water dimer & 6-31G* & QUICK & 2.347 & 0.025 \\*
$128\times$ water dimer & 6-31G* & LibintX & 2.122 & 0.012 \\*
$128\times$ water dimer & 6-31G* & VSR & 2.221 & 0.004 \\*
$128\times$ water dimer & 6-31G* & GANSU & 2.237 & 0.008 \\*
$128\times$ water dimer & 6-31G* & FABO & 2.103 & 0.012 \\
$128\times$ water dimer & def2-SVP & QUICK & 2.553 & 0.018 \\*
$128\times$ water dimer & def2-SVP & LibintX & 2.410 & 0.016 \\*
$128\times$ water dimer & def2-SVP & VSR & 2.463 & 0.020 \\*
$128\times$ water dimer & def2-SVP & GANSU & 2.490 & 0.024 \\*
$128\times$ water dimer & def2-SVP & FABO & 2.408 & 0.017 \\
$128\times$ water dimer & cc-pVDZ & QUICK & 3.703 & 0.016 \\*
$128\times$ water dimer & cc-pVDZ & LibintX & 3.350 & 0.022 \\*
$128\times$ water dimer & cc-pVDZ & VSR & 3.462 & 0.014 \\*
$128\times$ water dimer & cc-pVDZ & GANSU & 3.530 & 0.014 \\*
$128\times$ water dimer & cc-pVDZ & FABO & 3.297 & 0.019 \\
$8\times$ uracil dimer & 6-31G* & QUICK & 3.674 & 0.005 \\*
$8\times$ uracil dimer & 6-31G* & LibintX & 3.439 & 0.001 \\*
$8\times$ uracil dimer & 6-31G* & VSR & 3.624 & 0.003 \\*
$8\times$ uracil dimer & 6-31G* & GANSU & 3.669 & 0.005 \\*
$8\times$ uracil dimer & 6-31G* & FABO & 3.348 & 0.004 \\
$8\times$ uracil dimer & def2-SVP & QUICK & 2.237 & 0.002 \\*
$8\times$ uracil dimer & def2-SVP & LibintX & 2.128 & 0.003 \\*
$8\times$ uracil dimer & def2-SVP & VSR & 2.212 & 0.002 \\*
$8\times$ uracil dimer & def2-SVP & GANSU & 2.251 & 0.001 \\*
$8\times$ uracil dimer & def2-SVP & FABO & 2.091 & 0.004 \\
$8\times$ uracil dimer & cc-pVDZ & QUICK & 3.818 & 0.003 \\*
$8\times$ uracil dimer & cc-pVDZ & LibintX & 3.473 & 0.004 \\*
$8\times$ uracil dimer & cc-pVDZ & VSR & 3.667 & 0.002 \\*
$8\times$ uracil dimer & cc-pVDZ & GANSU & 3.760 & 0.002 \\*
$8\times$ uracil dimer & cc-pVDZ & FABO & 3.370 & 0.004 \\
$16\times$ benzene dimer & 6-31G* & QUICK & 5.918 & 0.007 \\*
$16\times$ benzene dimer & 6-31G* & LibintX & 5.380 & 0.004 \\*
$16\times$ benzene dimer & 6-31G* & VSR & 5.701 & 0.003 \\*
$16\times$ benzene dimer & 6-31G* & GANSU & 5.765 & 0.003 \\*
$16\times$ benzene dimer & 6-31G* & FABO & 5.290 & 0.004 \\
$16\times$ benzene dimer & def2-SVP & QUICK & 4.540 & 0.005 \\*
$16\times$ benzene dimer & def2-SVP & LibintX & 4.264 & 0.006 \\*
$16\times$ benzene dimer & def2-SVP & VSR & 4.424 & 0.009 \\*
$16\times$ benzene dimer & def2-SVP & GANSU & 4.473 & 0.007 \\*
$16\times$ benzene dimer & def2-SVP & FABO & 4.234 & 0.010 \\
$16\times$ benzene dimer & cc-pVDZ & QUICK & 7.955 & 0.012 \\*
$16\times$ benzene dimer & cc-pVDZ & LibintX & 7.113 & 0.009 \\*
$16\times$ benzene dimer & cc-pVDZ & VSR & 7.503 & 0.004 \\*
$16\times$ benzene dimer & cc-pVDZ & GANSU & 7.665 & 0.006 \\*
$16\times$ benzene dimer & cc-pVDZ & FABO & 6.984 & 0.009 \\
$16\times$ pentane dimer & 6-31G* & QUICK & 5.192 & 0.004 \\*
$16\times$ pentane dimer & 6-31G* & LibintX & 4.668 & 0.003 \\*
$16\times$ pentane dimer & 6-31G* & VSR & 4.916 & 0.004 \\*
$16\times$ pentane dimer & 6-31G* & GANSU & 5.037 & 0.002 \\*
$16\times$ pentane dimer & 6-31G* & FABO & 4.567 & 0.005 \\
$16\times$ pentane dimer & def2-SVP & QUICK & 6.192 & 0.005 \\*
$16\times$ pentane dimer & def2-SVP & LibintX & 5.765 & 0.007 \\*
$16\times$ pentane dimer & def2-SVP & VSR & 5.977 & 0.007 \\*
$16\times$ pentane dimer & def2-SVP & GANSU & 6.082 & 0.006 \\*
$16\times$ pentane dimer & def2-SVP & FABO & 5.699 & 0.007 \\
$16\times$ pentane dimer & cc-pVDZ & QUICK & 8.823 & 0.008 \\*
$16\times$ pentane dimer & cc-pVDZ & LibintX & 7.947 & 0.006 \\*
$16\times$ pentane dimer & cc-pVDZ & VSR & 8.298 & 0.010 \\*
$16\times$ pentane dimer & cc-pVDZ & GANSU & 8.535 & 0.009 \\*
$16\times$ pentane dimer & cc-pVDZ & FABO & 7.800 & 0.008 \\
\end{longtable}

\begin{longtable}{lllll}
\caption{CPU timings in seconds; arithmetic means and sample standard deviations of five observations.}\label{tab:timings-cpu}\\
\toprule
Molecule & Basis & Backend & Mean (s) & SD (s) \\ \midrule
\endfirsthead
\toprule
Molecule & Basis & Backend & Mean (s) & SD (s) \\ \midrule
\endhead
\midrule\multicolumn{5}{r}{Continued on next page}\\\endfoot
\bottomrule\endlastfoot
$8\times$ water dimer & 6-31G* & Simint & 1.466 & 0.003 \\*
$8\times$ water dimer & 6-31G* & LibintX & 1.405 & 0.003 \\*
$8\times$ water dimer & 6-31G* & VSR & 1.430 & 0.013 \\*
$8\times$ water dimer & 6-31G* & Laikov & 1.424 & 0.002 \\*
$8\times$ water dimer & 6-31G* & FABO & 1.429 & 0.029 \\
$8\times$ water dimer & def2-SVP & Simint & 2.765 & 0.022 \\*
$8\times$ water dimer & def2-SVP & LibintX & 2.649 & 0.001 \\*
$8\times$ water dimer & def2-SVP & VSR & 2.687 & 0.002 \\*
$8\times$ water dimer & def2-SVP & Laikov & 2.680 & 0.002 \\*
$8\times$ water dimer & def2-SVP & FABO & 2.656 & 0.005 \\
$8\times$ water dimer & cc-pVDZ & Simint & 3.389 & 0.008 \\*
$8\times$ water dimer & cc-pVDZ & LibintX & 3.232 & 0.009 \\*
$8\times$ water dimer & cc-pVDZ & VSR & 3.281 & 0.006 \\*
$8\times$ water dimer & cc-pVDZ & Laikov & 3.269 & 0.002 \\*
$8\times$ water dimer & cc-pVDZ & FABO & 3.283 & 0.001 \\
$1\times$ uracil dimer & 6-31G* & Simint & 6.162 & 0.011 \\*
$1\times$ uracil dimer & 6-31G* & LibintX & 5.770 & 0.024 \\*
$1\times$ uracil dimer & 6-31G* & VSR & 5.728 & 0.004 \\*
$1\times$ uracil dimer & 6-31G* & Laikov & 5.827 & 0.004 \\*
$1\times$ uracil dimer & 6-31G* & FABO & 5.717 & 0.046 \\
$1\times$ uracil dimer & def2-SVP & Simint & 7.075 & 0.008 \\*
$1\times$ uracil dimer & def2-SVP & LibintX & 6.635 & 0.002 \\*
$1\times$ uracil dimer & def2-SVP & VSR & 6.594 & 0.010 \\*
$1\times$ uracil dimer & def2-SVP & Laikov & 6.700 & 0.003 \\*
$1\times$ uracil dimer & def2-SVP & FABO & 6.543 & 0.006 \\
$1\times$ uracil dimer & cc-pVDZ & Simint & 10.581 & 0.094 \\*
$1\times$ uracil dimer & cc-pVDZ & LibintX & 9.752 & 0.059 \\*
$1\times$ uracil dimer & cc-pVDZ & VSR & 9.718 & 0.005 \\*
$1\times$ uracil dimer & cc-pVDZ & Laikov & 9.885 & 0.007 \\*
$1\times$ uracil dimer & cc-pVDZ & FABO & 9.693 & 0.010 \\
$1\times$ benzene dimer & 6-31G* & Simint & 5.496 & 0.003 \\*
$1\times$ benzene dimer & 6-31G* & LibintX & 4.957 & 0.006 \\*
$1\times$ benzene dimer & 6-31G* & VSR & 4.923 & 0.003 \\*
$1\times$ benzene dimer & 6-31G* & Laikov & 5.038 & 0.004 \\*
$1\times$ benzene dimer & 6-31G* & FABO & 4.825 & 0.003 \\
$1\times$ benzene dimer & def2-SVP & Simint & 7.830 & 0.007 \\*
$1\times$ benzene dimer & def2-SVP & LibintX & 7.122 & 0.022 \\*
$1\times$ benzene dimer & def2-SVP & VSR & 7.066 & 0.050 \\*
$1\times$ benzene dimer & def2-SVP & Laikov & 7.210 & 0.018 \\*
$1\times$ benzene dimer & def2-SVP & FABO & 6.918 & 0.003 \\
$1\times$ benzene dimer & cc-pVDZ & Simint & 11.378 & 0.005 \\*
$1\times$ benzene dimer & cc-pVDZ & LibintX & 10.131 & 0.014 \\*
$1\times$ benzene dimer & cc-pVDZ & VSR & 10.114 & 0.025 \\*
$1\times$ benzene dimer & cc-pVDZ & Laikov & 10.420 & 0.092 \\*
$1\times$ benzene dimer & cc-pVDZ & FABO & 9.945 & 0.025 \\
$1\times$ pentane dimer & 6-31G* & Simint & 5.011 & 0.003 \\*
$1\times$ pentane dimer & 6-31G* & LibintX & 4.436 & 0.032 \\*
$1\times$ pentane dimer & 6-31G* & VSR & 4.396 & 0.005 \\*
$1\times$ pentane dimer & 6-31G* & Laikov & 4.552 & 0.002 \\*
$1\times$ pentane dimer & 6-31G* & FABO & 4.296 & 0.006 \\
$1\times$ pentane dimer & def2-SVP & Simint & 10.452 & 0.004 \\*
$1\times$ pentane dimer & def2-SVP & LibintX & 9.344 & 0.009 \\*
$1\times$ pentane dimer & def2-SVP & VSR & 9.257 & 0.010 \\*
$1\times$ pentane dimer & def2-SVP & Laikov & 9.564 & 0.027 \\*
$1\times$ pentane dimer & def2-SVP & FABO & 9.079 & 0.029 \\
$1\times$ pentane dimer & cc-pVDZ & Simint & 13.819 & 0.181 \\*
$1\times$ pentane dimer & cc-pVDZ & LibintX & 12.072 & 0.021 \\*
$1\times$ pentane dimer & cc-pVDZ & VSR & 12.021 & 0.051 \\*
$1\times$ pentane dimer & cc-pVDZ & Laikov & 12.423 & 0.028 \\*
$1\times$ pentane dimer & cc-pVDZ & FABO & 11.776 & 0.023 \\
\end{longtable}

\section{Molecular-result comparisons and accompanying data}
\label{sec:molecular}
\subsection{Correctly rounded Boys backends}
The reference changes Boys evaluation, not the subsequent application arithmetic. QUICK evaluates MPFR-generated Taylor intervals in binary128 and accepts a value only when both endpoints round to the same binary64 bits. Ambiguous calls are recorded for adaptive directed-MPFR resolution; the same saved fixture is replayed in a fresh process. Unresolved attempts do not count toward the five accepted reference updates. Queue overflow, invalid inputs and exhausted replay limits fail rather than supply guessed values. Reference timings are excluded.

The NWChem oracle uses the shared MPFR-generated table and 113-bit MPFR arithmetic for its interval evaluator, with directed refinement through 8192 bits when needed. Two untimed fresh-process reference updates are checked for reproducibility. The oracle temporarily establishes the required MXCSR controls and restores the caller's complete register. Correct rounding of Boys values does not make integral recurrence, matrix accumulation or eigensolution exact.

\subsection{Diagnostics and repeated-output reduction}
Using the spin-summed input density, the energy diagnostic is
\begin{equation}
 E[P_0]=E_{\rm nuc}+\frac12\sum_{\mu\nu}(P_0)_{\mu\nu}\bigl(H_{\rm core}+F[P_0]\bigr)_{\mu\nu}.
\end{equation}
Energy and Fock differences are in hartree; density entries are in the application's dimensionless atomic-orbital representation. The energy is associated with $P_0$ and its Fock matrix, not the updated $P_1$.

For QUICK, each backend's five accepted energies and each element of its five Fock and updated-density matrices are averaged first. Reference outputs are averaged separately. Differences are then taken between these means. Centering sums on an observation and retaining high/low mean parts preserves sub-ULP offsets; this is neither an average of absolute errors nor a density reconstructed from an averaged Fock matrix. GPU accumulation order can vary between runs, so reference dispersion remains relevant even for correctly rounded Boys calls.

For NWChem, all repetitions are checked for identical Fock, input/output density, core-Hamiltonian and energy bytes. The verified first output is used for the matrix reduction. This establishes repeatability for the saved binary/node/environment, not across arbitrary systems.

For either matrix $M$, statistics are $\max_{\mu\nu}|\Delta M_{\mu\nu}|$ and $d^{-2}\sum_{\mu\nu}|\Delta M_{\mu\nu}|$, where $d$ is its dimension. All $d^2$ entries are included, including zeros and both symmetric halves. No relative matrix errors are reported. Each main-text summary column independently takes its largest casewise statistic, so the maximizing cases need not coincide. Reference-run sample standard deviations measure dispersion, not rigorous errors in the reference mean.

{\setlength{\tabcolsep}{3pt}
\begin{longtable}{llllllll}
\caption{GPU casewise differences from the correctly rounded Boys backend. Energy and Fock quantities are in hartree; density entries are dimensionless. Matrix means include every element.}\label{tab:molecular-gpu}\\
\toprule
Molecule & Basis & Backend & $|\Delta E|$ & $\max|\Delta F|$ & $\mathrm{mean}|\Delta F|$ & $\max|\Delta P|$ & $\mathrm{mean}|\Delta P|$ \\ \midrule
\endfirsthead
\toprule
Molecule & Basis & Backend & $|\Delta E|$ & $\max|\Delta F|$ & $\mathrm{mean}|\Delta F|$ & $\max|\Delta P|$ & $\mathrm{mean}|\Delta P|$ \\ \midrule
\endhead
\midrule\multicolumn{8}{r}{Continued on next page}\\\endfoot
\bottomrule\endlastfoot
$128\times$ water dimer & 6-31G* & QUICK & 3.07e-07 & 4.92e-07 & 4.89e-11 & 3.31e-05 & 4.76e-10 \\*
$128\times$ water dimer & 6-31G* & LibintX & 1.16e-11 & 1.94e-13 & 2.75e-17 & 7.36e-10 & 1.93e-15 \\*
$128\times$ water dimer & 6-31G* & VSR & 2.44e-10 & 1.65e-12 & 1.62e-16 & 3.32e-10 & 2.88e-15 \\*
$128\times$ water dimer & 6-31G* & GANSU & 1.69e-10 & 1.66e-12 & 1.52e-16 & 4.63e-10 & 2.29e-15 \\*
$128\times$ water dimer & 6-31G* & FABO & 1.46e-11 & 2.83e-13 & 2.75e-17 & 7.11e-10 & 1.89e-15 \\
$128\times$ water dimer & def2-SVP & QUICK & 2.78e-05 & 3.60e-07 & 3.01e-11 & 1.55e-04 & 1.16e-09 \\*
$128\times$ water dimer & def2-SVP & LibintX & 0.00e+00 & 1.78e-13 & 1.92e-17 & 1.44e-10 & 8.23e-16 \\*
$128\times$ water dimer & def2-SVP & VSR & 5.82e-11 & 8.92e-13 & 6.14e-17 & 1.47e-10 & 1.59e-15 \\*
$128\times$ water dimer & def2-SVP & GANSU & 6.69e-11 & 1.06e-12 & 6.04e-17 & 3.69e-10 & 1.84e-15 \\*
$128\times$ water dimer & def2-SVP & FABO & 0.00e+00 & 1.95e-13 & 1.92e-17 & 1.49e-10 & 8.61e-16 \\
$128\times$ water dimer & cc-pVDZ & QUICK & 4.31e-07 & 7.98e-07 & 3.50e-11 & 7.93e-06 & 2.25e-10 \\*
$128\times$ water dimer & cc-pVDZ & LibintX & 1.46e-12 & 1.54e-13 & 2.00e-17 & 8.16e-11 & 6.86e-16 \\*
$128\times$ water dimer & cc-pVDZ & VSR & 9.90e-11 & 1.25e-12 & 6.42e-17 & 1.24e-10 & 1.75e-15 \\*
$128\times$ water dimer & cc-pVDZ & GANSU & 6.84e-11 & 1.17e-12 & 5.68e-17 & 1.07e-10 & 1.53e-15 \\*
$128\times$ water dimer & cc-pVDZ & FABO & 5.82e-12 & 1.50e-13 & 2.01e-17 & 5.60e-11 & 6.60e-16 \\
$8\times$ uracil dimer & 6-31G* & QUICK & 2.35e-07 & 7.00e-07 & 9.42e-11 & 6.20e-06 & 8.98e-09 \\*
$8\times$ uracil dimer & 6-31G* & LibintX & 1.75e-11 & 5.80e-13 & 2.58e-16 & 2.74e-11 & 1.52e-14 \\*
$8\times$ uracil dimer & 6-31G* & VSR & 1.46e-10 & 2.27e-11 & 3.97e-15 & 5.20e-11 & 4.66e-14 \\*
$8\times$ uracil dimer & 6-31G* & GANSU & 1.60e-10 & 2.27e-11 & 4.75e-15 & 4.25e-11 & 5.10e-14 \\*
$8\times$ uracil dimer & 6-31G* & FABO & 2.18e-11 & 5.66e-13 & 2.58e-16 & 2.46e-11 & 1.50e-14 \\
$8\times$ uracil dimer & def2-SVP & QUICK & 5.98e-05 & 1.78e-06 & 1.30e-10 & 9.72e-05 & 8.00e-08 \\*
$8\times$ uracil dimer & def2-SVP & LibintX & 1.46e-12 & 5.63e-13 & 1.79e-16 & 4.71e-11 & 1.83e-14 \\*
$8\times$ uracil dimer & def2-SVP & VSR & 1.41e-10 & 1.41e-11 & 2.55e-15 & 8.32e-11 & 5.51e-14 \\*
$8\times$ uracil dimer & def2-SVP & GANSU & 1.34e-10 & 1.41e-11 & 3.02e-15 & 1.54e-10 & 6.87e-14 \\*
$8\times$ uracil dimer & def2-SVP & FABO & 7.28e-12 & 5.74e-13 & 1.78e-16 & 4.76e-11 & 1.73e-14 \\
$8\times$ uracil dimer & cc-pVDZ & QUICK & 4.81e-05 & 1.32e-06 & 1.07e-10 & 5.62e-05 & 3.14e-08 \\*
$8\times$ uracil dimer & cc-pVDZ & LibintX & 4.37e-12 & 4.52e-13 & 2.13e-16 & 2.53e-11 & 1.15e-14 \\*
$8\times$ uracil dimer & cc-pVDZ & VSR & 1.21e-10 & 1.18e-11 & 3.02e-15 & 3.64e-11 & 4.25e-14 \\*
$8\times$ uracil dimer & cc-pVDZ & GANSU & 1.34e-10 & 1.18e-11 & 3.47e-15 & 4.22e-11 & 4.39e-14 \\*
$8\times$ uracil dimer & cc-pVDZ & FABO & 2.91e-12 & 4.46e-13 & 2.12e-16 & 2.39e-11 & 1.12e-14 \\
$16\times$ benzene dimer & 6-31G* & QUICK & 4.75e-07 & 4.84e-07 & 1.02e-10 & 5.64e-06 & 4.04e-09 \\*
$16\times$ benzene dimer & 6-31G* & LibintX & 1.16e-11 & 2.54e-13 & 1.17e-16 & 2.35e-11 & 8.13e-15 \\*
$16\times$ benzene dimer & 6-31G* & VSR & 6.98e-11 & 1.22e-11 & 1.50e-15 & 3.01e-11 & 1.54e-14 \\*
$16\times$ benzene dimer & 6-31G* & GANSU & 6.98e-11 & 1.21e-11 & 1.61e-15 & 2.73e-11 & 1.39e-14 \\*
$16\times$ benzene dimer & 6-31G* & FABO & 1.16e-11 & 2.98e-13 & 1.16e-16 & 1.96e-11 & 7.96e-15 \\
$16\times$ benzene dimer & def2-SVP & QUICK & 9.75e-06 & 2.10e-07 & 6.20e-11 & 2.30e-05 & 9.76e-09 \\*
$16\times$ benzene dimer & def2-SVP & LibintX & 1.89e-11 & 3.95e-13 & 9.56e-17 & 8.14e-11 & 1.79e-14 \\*
$16\times$ benzene dimer & def2-SVP & VSR & 8.88e-11 & 1.37e-11 & 1.18e-15 & 1.06e-10 & 2.55e-14 \\*
$16\times$ benzene dimer & def2-SVP & GANSU & 1.46e-10 & 1.36e-11 & 1.40e-15 & 8.65e-11 & 2.46e-14 \\*
$16\times$ benzene dimer & def2-SVP & FABO & 1.60e-11 & 4.02e-13 & 9.51e-17 & 8.08e-11 & 1.79e-14 \\
$16\times$ benzene dimer & cc-pVDZ & QUICK & 3.11e-07 & 7.01e-07 & 5.42e-11 & 6.18e-06 & 4.16e-09 \\*
$16\times$ benzene dimer & cc-pVDZ & LibintX & 1.46e-12 & 3.31e-13 & 1.05e-16 & 4.07e-11 & 1.05e-14 \\*
$16\times$ benzene dimer & cc-pVDZ & VSR & 9.31e-11 & 5.67e-12 & 1.25e-15 & 6.86e-11 & 2.73e-14 \\*
$16\times$ benzene dimer & cc-pVDZ & GANSU & 6.69e-11 & 1.20e-11 & 1.49e-15 & 4.39e-11 & 1.78e-14 \\*
$16\times$ benzene dimer & cc-pVDZ & FABO & 1.46e-12 & 3.21e-13 & 1.06e-16 & 4.11e-11 & 1.02e-14 \\
$16\times$ pentane dimer & 6-31G* & QUICK & 6.22e-07 & 4.86e-07 & 9.88e-11 & 4.84e-06 & 1.80e-09 \\*
$16\times$ pentane dimer & 6-31G* & LibintX & 4.37e-12 & 2.84e-13 & 1.19e-16 & 1.58e-11 & 2.15e-15 \\*
$16\times$ pentane dimer & 6-31G* & VSR & 6.40e-11 & 8.02e-12 & 8.72e-16 & 1.94e-11 & 6.28e-15 \\*
$16\times$ pentane dimer & 6-31G* & GANSU & 5.97e-11 & 1.31e-11 & 9.89e-16 & 1.76e-11 & 5.91e-15 \\*
$16\times$ pentane dimer & 6-31G* & FABO & 5.82e-12 & 2.77e-13 & 1.18e-16 & 1.53e-11 & 2.10e-15 \\
$16\times$ pentane dimer & def2-SVP & QUICK & 3.92e-06 & 1.53e-07 & 4.49e-11 & 1.01e-05 & 1.88e-09 \\*
$16\times$ pentane dimer & def2-SVP & LibintX & 1.46e-12 & 3.25e-13 & 7.76e-17 & 2.66e-11 & 3.16e-15 \\*
$16\times$ pentane dimer & def2-SVP & VSR & 9.17e-11 & 1.87e-12 & 4.53e-16 & 4.67e-11 & 9.19e-15 \\*
$16\times$ pentane dimer & def2-SVP & GANSU & 1.03e-10 & 2.10e-12 & 5.11e-16 & 6.44e-11 & 1.25e-14 \\*
$16\times$ pentane dimer & def2-SVP & FABO & 1.16e-11 & 3.14e-13 & 7.73e-17 & 2.54e-11 & 3.10e-15 \\
$16\times$ pentane dimer & cc-pVDZ & QUICK & 4.24e-07 & 7.08e-07 & 6.31e-11 & 6.17e-06 & 1.89e-09 \\*
$16\times$ pentane dimer & cc-pVDZ & LibintX & 2.47e-11 & 2.96e-13 & 8.80e-17 & 9.69e-12 & 2.00e-15 \\*
$16\times$ pentane dimer & cc-pVDZ & VSR & 6.26e-11 & 8.33e-12 & 7.09e-16 & 1.92e-11 & 5.56e-15 \\*
$16\times$ pentane dimer & cc-pVDZ & GANSU & 9.46e-11 & 1.12e-11 & 8.26e-16 & 1.32e-11 & 6.15e-15 \\*
$16\times$ pentane dimer & cc-pVDZ & FABO & 1.31e-11 & 2.77e-13 & 8.79e-17 & 8.46e-12 & 1.97e-15 \\
\end{longtable}

\begin{longtable}{llllllll}
\caption{CPU casewise differences from the correctly rounded Boys backend. Energy and Fock quantities are in hartree; density entries are dimensionless. Matrix means include every element.}\label{tab:molecular-cpu}\\
\toprule
Molecule & Basis & Backend & $|\Delta E|$ & $\max|\Delta F|$ & $\mathrm{mean}|\Delta F|$ & $\max|\Delta P|$ & $\mathrm{mean}|\Delta P|$ \\ \midrule
\endfirsthead
\toprule
Molecule & Basis & Backend & $|\Delta E|$ & $\max|\Delta F|$ & $\mathrm{mean}|\Delta F|$ & $\max|\Delta P|$ & $\mathrm{mean}|\Delta P|$ \\ \midrule
\endhead
\midrule\multicolumn{8}{r}{Continued on next page}\\\endfoot
\bottomrule\endlastfoot
$8\times$ water dimer & 6-31G* & Simint & 5.00e-12 & 1.22e-11 & 1.21e-14 & 4.92e-13 & 4.22e-15 \\*
$8\times$ water dimer & 6-31G* & FABO & 0.00e+00 & 3.55e-15 & 6.33e-18 & 3.28e-14 & 4.92e-16 \\*
$8\times$ water dimer & 6-31G* & LibintX & 0.00e+00 & 7.11e-15 & 1.04e-17 & 3.42e-14 & 4.83e-16 \\*
$8\times$ water dimer & 6-31G* & VSR & 1.36e-11 & 1.86e-12 & 3.04e-15 & 2.40e-12 & 1.06e-14 \\*
$8\times$ water dimer & 6-31G* & Laikov & 0.00e+00 & 7.11e-15 & 1.07e-17 & 3.86e-14 & 5.04e-16 \\
$8\times$ water dimer & def2-SVP & Simint & 1.36e-12 & 1.97e-12 & 2.13e-15 & 2.29e-13 & 9.47e-16 \\*
$8\times$ water dimer & def2-SVP & FABO & 0.00e+00 & 3.55e-15 & 6.95e-18 & 2.72e-14 & 2.39e-16 \\*
$8\times$ water dimer & def2-SVP & LibintX & 4.55e-13 & 1.07e-14 & 1.02e-17 & 4.35e-14 & 2.41e-16 \\*
$8\times$ water dimer & def2-SVP & VSR & 5.91e-12 & 8.60e-13 & 1.05e-15 & 9.41e-13 & 3.51e-15 \\*
$8\times$ water dimer & def2-SVP & Laikov & 4.55e-13 & 7.11e-15 & 1.18e-17 & 4.94e-14 & 2.41e-16 \\
$8\times$ water dimer & cc-pVDZ & Simint & 9.09e-13 & 1.18e-12 & 1.07e-15 & 2.62e-13 & 1.00e-15 \\*
$8\times$ water dimer & cc-pVDZ & FABO & 0.00e+00 & 7.11e-15 & 7.21e-18 & 2.64e-14 & 2.21e-16 \\*
$8\times$ water dimer & cc-pVDZ & LibintX & 0.00e+00 & 7.11e-15 & 1.12e-17 & 3.25e-14 & 2.31e-16 \\*
$8\times$ water dimer & cc-pVDZ & VSR & 5.46e-12 & 9.08e-13 & 9.90e-16 & 1.30e-12 & 3.92e-15 \\*
$8\times$ water dimer & cc-pVDZ & Laikov & 0.00e+00 & 7.11e-15 & 1.16e-17 & 3.04e-14 & 2.27e-16 \\
$1\times$ uracil dimer & 6-31G* & Simint & 9.78e-12 & 4.77e-11 & 4.91e-14 & 8.82e-12 & 7.58e-14 \\*
$1\times$ uracil dimer & 6-31G* & FABO & 2.27e-13 & 1.07e-14 & 3.42e-17 & 3.94e-13 & 2.80e-15 \\*
$1\times$ uracil dimer & 6-31G* & LibintX & 2.27e-13 & 1.07e-14 & 5.03e-17 & 4.13e-13 & 2.55e-15 \\*
$1\times$ uracil dimer & 6-31G* & VSR & 6.37e-12 & 9.05e-12 & 1.39e-14 & 1.35e-11 & 1.07e-13 \\*
$1\times$ uracil dimer & 6-31G* & Laikov & 0.00e+00 & 2.13e-14 & 6.28e-17 & 3.36e-13 & 2.82e-15 \\
$1\times$ uracil dimer & def2-SVP & Simint & 8.19e-12 & 3.50e-11 & 3.37e-14 & 1.07e-11 & 5.49e-14 \\*
$1\times$ uracil dimer & def2-SVP & FABO & 2.27e-13 & 1.42e-14 & 2.85e-17 & 1.22e-12 & 2.88e-15 \\*
$1\times$ uracil dimer & def2-SVP & LibintX & 4.55e-13 & 1.07e-14 & 4.51e-17 & 8.24e-13 & 3.18e-15 \\*
$1\times$ uracil dimer & def2-SVP & VSR & 6.82e-12 & 1.41e-11 & 9.01e-15 & 1.09e-11 & 1.07e-13 \\*
$1\times$ uracil dimer & def2-SVP & Laikov & 2.27e-13 & 1.42e-14 & 5.57e-17 & 8.13e-13 & 3.56e-15 \\
$1\times$ uracil dimer & cc-pVDZ & Simint & 7.05e-12 & 5.10e-11 & 3.45e-14 & 7.27e-12 & 6.47e-14 \\*
$1\times$ uracil dimer & cc-pVDZ & FABO & 2.27e-13 & 1.07e-14 & 3.23e-17 & 4.32e-13 & 2.07e-15 \\*
$1\times$ uracil dimer & cc-pVDZ & LibintX & 0.00e+00 & 1.42e-14 & 4.61e-17 & 1.17e-12 & 2.65e-15 \\*
$1\times$ uracil dimer & cc-pVDZ & VSR & 6.14e-12 & 1.03e-11 & 9.59e-15 & 1.01e-11 & 9.15e-14 \\*
$1\times$ uracil dimer & cc-pVDZ & Laikov & 0.00e+00 & 1.42e-14 & 5.99e-17 & 4.86e-13 & 2.26e-15 \\
$1\times$ benzene dimer & 6-31G* & Simint & 4.09e-12 & 1.31e-11 & 3.84e-14 & 2.82e-12 & 4.79e-14 \\*
$1\times$ benzene dimer & 6-31G* & FABO & 0.00e+00 & 1.07e-14 & 3.78e-17 & 2.51e-13 & 5.63e-15 \\*
$1\times$ benzene dimer & 6-31G* & LibintX & 0.00e+00 & 7.11e-15 & 5.78e-17 & 3.40e-13 & 5.47e-15 \\*
$1\times$ benzene dimer & 6-31G* & VSR & 2.96e-12 & 7.89e-12 & 1.66e-14 & 4.92e-12 & 1.27e-13 \\*
$1\times$ benzene dimer & 6-31G* & Laikov & 0.00e+00 & 1.78e-14 & 7.55e-17 & 4.93e-13 & 6.29e-15 \\
$1\times$ benzene dimer & def2-SVP & Simint & 5.00e-12 & 1.21e-11 & 3.14e-14 & 8.70e-12 & 7.34e-14 \\*
$1\times$ benzene dimer & def2-SVP & FABO & 0.00e+00 & 7.11e-15 & 3.50e-17 & 9.74e-13 & 9.27e-15 \\*
$1\times$ benzene dimer & def2-SVP & LibintX & 0.00e+00 & 7.11e-15 & 5.28e-17 & 2.27e-12 & 1.49e-14 \\*
$1\times$ benzene dimer & def2-SVP & VSR & 3.18e-12 & 7.19e-12 & 1.18e-14 & 1.64e-11 & 1.58e-13 \\*
$1\times$ benzene dimer & def2-SVP & Laikov & 0.00e+00 & 1.07e-14 & 6.70e-17 & 1.60e-12 & 1.42e-14 \\
$1\times$ benzene dimer & cc-pVDZ & Simint & 6.59e-12 & 3.11e-11 & 5.56e-14 & 1.00e-11 & 1.11e-13 \\*
$1\times$ benzene dimer & cc-pVDZ & FABO & 0.00e+00 & 1.07e-14 & 3.71e-17 & 5.34e-13 & 7.05e-15 \\*
$1\times$ benzene dimer & cc-pVDZ & LibintX & 0.00e+00 & 1.07e-14 & 5.95e-17 & 7.50e-13 & 7.22e-15 \\*
$1\times$ benzene dimer & cc-pVDZ & VSR & 2.96e-12 & 3.55e-12 & 1.41e-14 & 1.24e-11 & 1.76e-13 \\*
$1\times$ benzene dimer & cc-pVDZ & Laikov & 0.00e+00 & 1.42e-14 & 7.36e-17 & 6.68e-13 & 6.84e-15 \\
$1\times$ pentane dimer & 6-31G* & Simint & 2.16e-12 & 2.50e-11 & 4.19e-14 & 1.80e-12 & 2.35e-14 \\*
$1\times$ pentane dimer & 6-31G* & FABO & 0.00e+00 & 7.11e-15 & 3.00e-17 & 1.35e-13 & 1.30e-15 \\*
$1\times$ pentane dimer & 6-31G* & LibintX & 0.00e+00 & 7.11e-15 & 4.47e-17 & 1.81e-13 & 1.30e-15 \\*
$1\times$ pentane dimer & 6-31G* & VSR & 3.30e-12 & 6.59e-12 & 1.09e-14 & 2.00e-12 & 3.37e-14 \\*
$1\times$ pentane dimer & 6-31G* & Laikov & 1.14e-13 & 1.07e-14 & 5.56e-17 & 1.86e-13 & 1.56e-15 \\
$1\times$ pentane dimer & def2-SVP & Simint & 2.27e-12 & 2.71e-11 & 2.51e-14 & 2.42e-12 & 2.33e-14 \\*
$1\times$ pentane dimer & def2-SVP & FABO & 0.00e+00 & 7.11e-15 & 2.54e-17 & 2.60e-13 & 1.73e-15 \\*
$1\times$ pentane dimer & def2-SVP & LibintX & 0.00e+00 & 1.07e-14 & 3.51e-17 & 3.31e-13 & 2.01e-15 \\*
$1\times$ pentane dimer & def2-SVP & VSR & 2.84e-12 & 1.38e-12 & 5.27e-15 & 6.18e-12 & 3.84e-14 \\*
$1\times$ pentane dimer & def2-SVP & Laikov & 0.00e+00 & 1.42e-14 & 4.55e-17 & 3.02e-13 & 1.79e-15 \\
$1\times$ pentane dimer & cc-pVDZ & Simint & 1.59e-12 & 6.99e-12 & 1.46e-14 & 2.88e-12 & 2.37e-14 \\*
$1\times$ pentane dimer & cc-pVDZ & FABO & 0.00e+00 & 7.11e-15 & 2.68e-17 & 1.07e-13 & 9.78e-16 \\*
$1\times$ pentane dimer & cc-pVDZ & LibintX & 0.00e+00 & 7.11e-15 & 3.72e-17 & 1.08e-13 & 1.10e-15 \\*
$1\times$ pentane dimer & cc-pVDZ & VSR & 2.61e-12 & 7.59e-12 & 9.43e-15 & 5.06e-12 & 3.58e-14 \\*
$1\times$ pentane dimer & cc-pVDZ & Laikov & 0.00e+00 & 1.42e-14 & 5.10e-17 & 1.07e-13 & 1.08e-15 \\
\end{longtable}
}

\Needspace{18\baselineskip}
\Needspace{19\baselineskip}
\begin{longtable}{lllll}
\caption{QUICK reference-repeat dispersion: energy sample SD and largest elementwise sample SD of each matrix.}\label{tab:reference-variability}\\
\toprule
Molecule & Basis & Energy SD (hartree) & Max Fock SD (hartree) & Max density SD \\ \midrule
\endfirsthead
\toprule
Molecule & Basis & Energy SD (hartree) & Max Fock SD (hartree) & Max density SD \\ \midrule
\endhead
\midrule\multicolumn{5}{r}{Continued on next page}\\\endfoot
\bottomrule\endlastfoot
$128\times$ water dimer & 6-31G* & 0.000e+00 & 1.863e-14 & 1.674e-11 \\
$128\times$ water dimer & def2-SVP & 0.000e+00 & 1.577e-14 & 5.682e-12 \\
$128\times$ water dimer & cc-pVDZ & 3.985e-12 & 1.577e-14 & 6.045e-12 \\
$8\times$ uracil dimer & 6-31G* & 0.000e+00 & 4.426e-14 & 3.171e-12 \\
$8\times$ uracil dimer & def2-SVP & 0.000e+00 & 2.947e-14 & 4.081e-12 \\
$8\times$ uracil dimer & cc-pVDZ & 6.508e-12 & 4.094e-14 & 4.028e-12 \\
$16\times$ benzene dimer & 6-31G* & 7.970e-12 & 3.495e-14 & 1.989e-12 \\
$16\times$ benzene dimer & def2-SVP & 3.254e-12 & 3.178e-14 & 5.409e-12 \\
$16\times$ benzene dimer & cc-pVDZ & 0.000e+00 & 3.495e-14 & 4.333e-12 \\
$16\times$ pentane dimer & 6-31G* & 0.000e+00 & 3.846e-14 & 1.126e-12 \\
$16\times$ pentane dimer & def2-SVP & 0.000e+00 & 4.713e-14 & 2.427e-12 \\
$16\times$ pentane dimer & cc-pVDZ & 7.970e-12 & 3.233e-14 & 9.160e-13 \\
\end{longtable}

\subsection{Accompanying-data guide}
The supplementary data and reproduction instructions are available from \href{https://github.com/LXYan2333/fabo/releases/tag/paper-si-v1}{FABO release \texttt{paper-si-v1}}. Extract \path{si-companion.zip} for the manifest, figures and tables. The archive \path{data.zip} restores the directory \path{data/}; \path{data-manifest.json} identifies its ordinary records and the separately supplied figures and generated tables. The following index uses paths relative to \path{data/}. Regeneration uses the main FABO repository's reporting tools and Python environment, not the data archive alone. The package-root \path{README.md}, under ``Repository and environment prerequisites'', ``Input preparation and integrity'', and ``Regeneration commands'', specifies the reporting commit, setup, input locations and commands; ``Full archives'' identifies the separate matrix and executable archives.

{\small
\begin{longtable}{>{\raggedright\arraybackslash}p{.20\linewidth}>{\raggedright\arraybackslash}p{.46\linewidth}>{\raggedright\arraybackslash}p{.24\linewidth}}
\caption{Accompanying-data index and repository-tool entry points. Short tool names refer to \texttt{workflow.scripts} modules, except \texttt{workflow.upward\_recurrence\_limits}; full commands are in the package-root README.}\label{tab:data-index}\\
\toprule Purpose & Files below \texttt{data/} & Section / entry point \\\midrule\endfirsthead
\toprule Purpose & Files below \texttt{data/} & Section / entry point \\\midrule\endhead
\bottomrule\endfoot
Selected models and qualification & \path{models.csv}; \path{records/numerical/paired-1ulp/summary.json}, \path{config.json}, \path{identity.json} and \path{producer/} in that directory & Secs.~\ref{sec:implementation}--\ref{sec:construction}; \path{export_supporting_information} \\
Function errors and witnesses & \path{function-accuracy.csv}, \path{maximum-witnesses.csv}; \path{records/benchmarks/function-accuracy-20261107/} & Sec.~\ref{sec:accuracy}; \texttt{report\_function\_accuracy} \\
Numerical illustrations & \path{nonuniform-grid-remainder.json}, \path{nonuniform-grid-remainder.csv}; \path{records/diagnostics/upward-recurrence-limits/} & Sec.~\ref{sec:illustrations}; \texttt{plot\_grid\_remainders}, \texttt{upward\_recurrence\_limits plot} \\
Workloads and fixtures & \path{workloads.csv}; \path{application-inputs/quick/} and \path{application-inputs/nwchem/}; each run's \path{reuse.json} & Sec.~\ref{sec:artifact-map}; README, ``Application records'' \\
Sources and builds & \path{method-sources.csv}, \path{methods/}, \path{manifest.json}, \path{simd-manifest.json}; build records listed in Sec.~\ref{sec:artifact-map} & README, ``Source and payload identities'' \\
Repeated timings and speedups & \path{timing-observations.csv}, \path{timing-summary.csv}, \path{geometric-speedups.csv} & Sec.~\ref{sec:timings}; \texttt{plot\_quick\_performance}, \texttt{plot\_nwchem\_performance} \\
Molecular comparisons & \path{molecular-differences.csv}; QUICK \path{molecular-means.json} and NWChem \path{matrix-absolute.json} under \path{records/benchmarks/} & Sec.~\ref{sec:molecular}; \path{report_molecular_accuracy} \\
Reference variability & \path{reference-variability.csv}; QUICK summary and mean-matrix records & Sec.~\ref{sec:molecular}; \texttt{quick\_molecular\_means} (full matrix archive) \\
\end{longtable}}

Raw backend identifiers are preserved in the data; \path{backend-names.csv} maps them to the displayed names. In particular, \texttt{official}/\texttt{quick} denote QUICK, \texttt{original}/\texttt{simint} denote Simint, \texttt{fabo\_additive}/\texttt{fabo} denote FABO, and \texttt{libintx\_gpu}/\texttt{libintx} denote LibintX. \texttt{fabo\_software} is FABO with software gather, not another approximation. Each matrix statistic uses the case and backend named in its row; units and pooled-row conventions are defined in the README.

The compact records suffice to regenerate the reported reductions and plots with the repository tools and environment. The full execution archives (approximately 94\,GiB in total) are too large to distribute with the supplement and are retained on Slater and Heitler. The README lists their locations and access requirements under ``Full archives''. Recomputing differences from the original matrices requires those archives; to generate new matrices or scan records from the inputs, follow Sec.~\ref{sec:rerun-experiments}.

\subsubsection{Rerunning the calculations}
\label{sec:rerun-experiments}
The package-root \path{README.md}, under ``Rebuilding and rerunning'', points to the repository instructions for the standalone Boys scans (\path{workflow/library_accuracy/README.md}) and the fixed-density molecular calculations (\path{workflow/quick/FIXED_DENSITY.md} and \path{workflow/NWCHEM_FIXED_DENSITY.md}). To repeat a molecular case, use its recorded geometry, basis and initial-state protocol. QUICK's matrix elements can vary between repetitions; compare the five-run means as described above. Reconstructing coefficient tables from frozen metadata or regenerating reports from the compact records does not rerun these experiments.

\begingroup\small
\bibliographystyle{unsrt}
\bibliography{references}
\endgroup